\documentclass[rmp,aps,twocolumn,nofootinbib,floatfix]{revtex4}

\usepackage{graphicx}
\usepackage{dcolumn}
\usepackage{bm}

\def\S{Sec.}

\def\Omat{\Omega_{\rm M}}
\def\Olam{\Omega_{\Lambda}}
\def\Orad{\Omega_{\rm R}}
\def\Obar{\Omega_{\rm b}}
\def\Bcg{B_{\rm cg}}
\def\Lx{L_{\rm X}}
\def\Tx{T_{\rm X}}
\def\Tlum{T_{\rm lum}}
\def\Tsp{T_{\rm sp}}
\def\rsh{r_{\rm sh}}
\def\rta{r_{\rm ta}}
\def\tvir{t_{\rm c}}
\def\rvir{r_{\rm v}}
\def\rvt{r_G}
\def\rb{r_{\rm b}}
\def\Pb{P_{\rm b}}
\def\Msh{M_{\rm sh}}
\def\Mvir{M_{\rm v}}
\def\cvir{c_{\rm v}}
\def\Delvir{\Delta_{\rm v}}
\def\rhocr{\rho_{\rm cr}}
\def\rhocro{\rho_{\rm cr0}}
\def\erfc{{\rm erfc}\,}
\def\sigv{\sigma_{\rm 1D}}
\def\keV{{\rm keV}}
\def\kB{k_{\rm B}}
\def\gammaeff{{\gamma_{\rm eff}}}
\def\betafit{\beta_{\rm fit}}
\def\betasp{\beta_{\rm sp}}

\begin{document}


\title{Tracing cosmic evolution with 
       clusters of galaxies}

\author{G. Mark Voit}

\affiliation{%
Department of Physics and Astronomy, Michigan State University,
East Lansing, MI 48824
}%

\date{September 10, 2004}

\begin{abstract}
The most successful cosmological models to date
envision structure formation as a hierarchical
process in which gravity is constantly drawing
lumps of matter together to form increasingly
larger structures.  Clusters of galaxies currently
sit atop this hierarchy as the 
largest objects that have had time to collapse
under the influence of their own gravity.  Thus,
their appearance on the cosmic scene is also
relatively recent.  Two features of clusters 
make them uniquely useful tracers of cosmic 
evolution.  First, clusters are the biggest things 
whose masses we can reliably measure because 
they are the largest objects to have undergone gravitational
relaxation and entered into virial
equilibrium.  Mass measurements of nearby
clusters can therefore be used to determine 
the amount of structure in the universe on scales 
of $10^{14}$-$10^{15} \, M_\odot$, and
comparisons of the present-day cluster mass
distribution with the mass distribution at earlier times
can be used to measure the rate of structure
formation, placing important constraints on
cosmological models.  Second, clusters are essentially
``closed boxes'' that retain all their gaseous 
matter, despite the enormous energy input associated
with supernovae and active galactic nuclei, because 
the gravitational potential wells of clusters are so deep.
The baryonic component of clusters therefore contains
a wealth of information about the processes
associated with galaxy formation, including the
efficiency with which baryons are converted into
stars and the effects of the resulting feedback processes
on galaxy formation.  This article reviews our theoretical
understanding of both the dark-matter component
and the baryonic component of clusters, providing
a context for interpreting the flood of new cluster observations
that are now arriving from the latest generation
of X-ray observatories, large optical surveys, and 
measurements of cluster-induced distortions in the 
spectrum of the cosmic microwave background.
\end{abstract}

\maketitle
\tableofcontents

\section{Introduction}
\label{sec:intro}

Cosmology has recently reached an important milestone.  A wide variety
of cosmological observations now support a single model for the
overall architecture of the observable universe and the development
of galaxies and other structures within it.  According to this so-called
concordance model, the geometry of the observable universe is indistinguishable
from a flat geometry, implying that its total energy density is very close 
to the critical density needed to close the universe.  The two dominant
components of the universe appear to be a non-baryonic form of dark
matter, whose gravity is responsible for structure formation, and a
mysterious form of dark energy, whose pressure is currently causing
the expansion of the universe to accelerate.  The mean density of baryonic
matter is about 15\% of the total amount of matter, 
and we can observe the baryonic matter  only because the
gravitational attraction of non-baryonic dark matter has drawn the baryonic gas into
deep potential wells, where a small fraction of it condenses into stars and galaxies.

This model explains many different features of the observable universe,
but it is not entirely satisfying because the nature of the dark matter and 
the provenance of the dark energy remain unknown.  The implications
of dark energy for fundamental physics are particularly serious, so we
need to be sure that it is absolutely necessary to explain the astronomical
observations.  In addition, many aspects of galaxy formation remain poorly 
understood.  Dark-matter models successfully account for the spatial distribution 
of mass in the universe, as traced by the galaxies, but they do not explain
all the properties of the galaxies themselves.  Dark matter initiates the 
process of galaxy formation, but once stars begin to form, supernova
explosions and disturbances wrought by supermassive black holes 
can inhibit further star formation by pumping thermal energy into the
universe's baryonic gas.

Clusters of galaxies are a particularly rich source of information about
the underlying cosmological model, making possible a number of
critical tests.  According to the concordance model, clusters are the 
largest and most recent gravitationally-relaxed objects to form 
because structure grows hierarchically.  The universe begins in a state 
of rapid expansion whose current manifestation is Hubble's Law 
relating a galaxy's recessional velocity $v_r$ to its distance $d$ through
Hubble's constant $H_0$: $v_r = H_0 d$.   Generalizing this feature of
the local universe to all of observable space links an object's cosmological
redshift $z = (\lambda_{\rm observed}/\lambda_{\rm rest}) - 1$ with a 
unique time $t(z)$ since the Big Bang, enabling us to probe the evolution 
of the universe with observations of distant objects.\footnote{In this definition, 
$\lambda_{\rm rest}$ is the wavelength of a photon emitted by a distant object 
and $\lambda_{\rm observed}$ is the wavelength it is observed to 
have when it reaches Earth.}  Gravity drives structure formation
in this expanding realm because the matter density is nearly equal to
the critical density during much of cosmic history.  Regions whose density
slightly exceeds the mean density are therefore gravitationally bound and
eventually decouple from the expansion, collapse upon themselves,
and enter a state of virial equilibrium in which the mean speeds of the
component particles are approximately half the escape velocity.
Because density perturbations in the concordance model have greater
amplitudes on smaller length scales, small sub-galactic objects are
the first to decouple, collapse, and virialize.  These small objects then
collect into galaxies, and galaxies later collect into clusters of galaxies,
whose masses now top out at roughly $10^{15}$ times that of the
Sun's ($10^{15} \, M_\odot$).  Thus, the growth and development of
clusters directly traces the process of structure formation in the
universe.

Section~\ref{sec:observ} outlines the observable properties of galaxy clusters that
enable us to measure their masses.  Observables in the optical band
include the overall luminosity of a cluster's galaxies, which scales with
the overall mass, the velocity dispersion of a cluster's member galaxies,
which responds to the depth of the cluster's potential well, and gravitational
lensing of background galaxies by the cluster's potential.  Observables
in the X-ray band include the overall X-ray luminosity of a cluster, coming
from the hot gas trapped in the cluster's gravitational potential, the temperature
inferred from the X-ray spectrum of that gas, and the abundances of
various elements inferred from the emission lines in that spectrum.
This hot gas also leaves an imprint on the microwave sky because
its electrons Compton scatter the photons of the cosmic microwave
background radiation.  Microwave observations are therefore an
alternative source of information about the hot gas and its temperature.

Once we have measured the masses of a sample of clusters, we can
use that sample to study cosmology.  Section~\ref{sec:dcomp} explains
how the characteristics of the cluster population relate to cosmological
models.  It begins by summarizing the elements of the concordance 
model and provides a number of useful analytical approximations 
to the results of numerical simulations of cluster formation based on 
this model.  Then it covers the dicey middle ground linking those
simulations with observations, currently the main source of uncertainty 
in deriving cosmological parameters from cluster observations.
The section concludes with a look at the evolution  
observed in the cluster population and the constraints that 
cluster evolution places on cosmological models.

Section~\ref{sec:baryons} takes up the subject of the baryonic component of
clusters, with two purposes in mind.  First, in order to improve the
precision of cosmological measurements with clusters, we need
to know how the process of galaxy formation affects the relations
used to derive cluster masses from observations of a cluster's hot 
gas and galaxies.  Current numerical simulations accurately
reproduce the behavior of the dark component, whose interactions
are purely gravitational, but fail to reproduce with similar accuracy
the observed behavior of the baryonic component, whose interactions 
are also hydrodynamical and thermodynamical.  These
discrepancies between simulations and observations indicate that
galaxy formation alters the state of a cluster's hot gas in a way that
preserves information about the poorly understood feedback
processes that regulated galaxy formation long before the cluster
reached its present state.  Our second purpose is therefore to
try to decipher what the state of the hot gas is saying about the
process of galaxy formation, so as to gain insight into those feedback
processes.  Section~\ref{sec:concl} concludes the review with some brief 
remarks about ongoing and future cluster surveys.  

Despite this article's length, it falls somewhat short of being a 
comprehensive review of cluster physics, which would
require more pages than this journal is inclined to provide.  
Instead, I have tried to assemble a readable introduction to cluster
evolution for non-experts, concentrating on the middle ground 
connecting theory to observations and distilling the key
theoretical results into a set of simple analytical tools useful 
to observers.  For more on the subject of clusters and their
evolution, readers should consult \citet{Sarazin88}, \citet{bmv02}, 
\citet{rbn02}, and \citet{mdo04}.

\section{Observable Properties of Clusters}
\label{sec:observ}

Clusters of galaxies might have been called something different if they had 
first been discovered in a waveband other than visible light, because all of 
the stars in all of a cluster's galaxies represent only a small fraction of a cluster's
overall mass.  Clusters contain substantially more mass in the form of hot gas, 
observable with X-ray and microwave instruments.  This section outlines
how clusters are observed in all three of these wavebands and how those observations
reveal a cluster's total mass, which turns out to be about seven times the 
combined baryonic mass in stars and hot gas \citep{wnef93, djf95, Evrard97, asf02}.

\subsection{Clusters in Optical Light}
\label{sec:clopt}

Optical identification of galaxy clusters has been going on for quite a long time.
By the end of the eighteenth century Charles \citet{Messier1784} and William 
\citet{Herschel1785} had already recognized concentrations of galaxies in the 
constellations Virgo and Coma Berenices.  Today these clusters of galaxies are 
known as the Virgo cluster and the Coma cluster.
Optical discoveries of clusters continued to accumulate as observing power
grew over the next two centuries \citep[see][for a review of the history]{Biviano00},
culminating with the definitive cluster catalogs of George Abell and collaborators
\citep{Abell58, aco89}.  Abell's catalogs contain most of the known nearby galaxy clusters
and are the foundation for much of our modern understanding of clusters.

Abell recognized that projection effects can complicate the identification of clusters 
in optical galaxy surveys and therefore was careful in defining his clusters.  Working
from the Palomar Sky Survey plates he estimated the distance of each cluster 
candidate from the apparent brightness of its tenth brightest member galaxy.   He then
counted all the galaxies lying within a fixed projected radius and brighter than a
magnitude limit two magnitudes fainter than the third brightest member.  The bounding radius, 
which he determined from the distance estimate, is now known to be $\sim 2$~Mpc
and was the same for all clusters.\footnote{The Megaparsec is astronomers' favored unit
of distance on cluster scales: $1 \, {\rm Mpc} = 3.09 \times 10^{24} \, {\rm cm} = 3.26 \times
10^6$ light years.}  In order to compensate for projection effects, 
he subtracted from his galaxy counts a background level equivalent to 
the mean number of galaxies brighter than the magnitude limit for the cluster 
in similarly-sized, cluster-free regions of the plate, and retained all cluster candidates 
with a net excess of 50 galaxies brighter than the limiting magnitude. 

Most of the optical cluster identification techniques used today extend and refine Abell's 
basic approach \citep[e.g.,][]{Lumsden92, Dalton97, Postman96}, often augmenting it with 
information about galaxy colors \citep[e.g.,][]{gy00, bahcall_maxBCG03, Nichol04}.  These 
improvements are necessary because the contrast of clusters 
against the background galaxy counts decreases with cluster distance.
Galaxy colors can help identify distant clusters because many cluster galaxies
are significantly redder than other galaxies at a similar redshift, 
owing to their lack of ongoing star formation.  The colors
of their aging stellar populations therefore place these cluster members on
a narrow and distinctive locus known as the ``red sequence'' in a plot of
galaxy color versus magnitude \citep[e.g.][]{gy00}.

Once suitable cluster candidates are found, their status as true mass concentrations can 
be checked by measuring the underlying mass.  Optical observations offer two 
complementary ways to perform such measurements, through the 
orbital velocities of the member galaxies and through the degree to which galaxies lying 
behind the cluster are lensed by the cluster's gravitational potential.  
We will discuss both of these methods after a few more words about how galaxy counts
relate to the overall optical luminosities of clusters. 

\subsubsection{Optical Richness}
\label{sec:richness}

To the extent that light traces mass in the universe, the total optical luminosity of a cluster
is itself an indicator of a cluster's mass.  Measuring the luminosity of every galaxy in
a cluster is impractical, especially for distant clusters in which only the brightest
galaxies can be observed.  However, because the luminosity distribution function
of cluster galaxies is nearly the same from cluster to cluster, observing the high-luminosity
tip of that distribution allows one to normalize the overall galaxy luminosity function for
the cluster, yielding estimates for both the cluster's total optical luminosity and its mass.

Abell's catalogs encode this information by placing clusters in categories of
``richness" corresponding to the net excess of galaxies brighter than the magnitude 
limit used to define each cluster.  The richest clusters (class 5) contain over 300 galaxies
brighter than the magnitude limit, while the poorest (class 1) contain 
only 50-79 such galaxies.  Clusters not quite making
Abell's cut (30-49 galaxies above the magnitude limit) were assigned to richness 
class zero.  Within this system, the Coma cluster originally ranked as richness class 2.

Invoking assumptions about the shape of the luminosity distribution function 
helps to link richness more directly to a cluster's total luminosity.   Cluster galaxies generally
adhere to a luminosity distribution function following the form proposed by \citet{Schechter76}, 
with the number of galaxies in luminosity range $dL$ about $L$ proportional
to $L^{-\alpha} \exp(-L/L_*)$, with $\alpha \approx 1$ \citep[e.g.,][]{bczz01}.  
Assuming this distribution function, \citet{Postman96} define a richness
parameter $\Lambda_{\rm cl}$ equivalent to the number of cluster galaxies
brighter than the characteristic luminosity $L_*$.  They find that $\Lambda_{\rm cl}$
is highly correlated with Abell's richness measure, but the scatter between
richness and $\Lambda_{\rm cl}$ is large.  

Another richness parameter in current use is $\Bcg$, the amplitude of the correlation 
function between the cluster center and the member galaxies \citep{ls79, yl99}.  
It is derived from the angular correlation function of galaxies 
measured down to a given magnitude limit, after removing the background counts, 
and is normalized by dividing out the expected luminosity distribution function of galaxies
integrated down to that magnitude limit.  This richness parameter also correlates with Abell's 
richness, but again the scatter is broad.  
\citet{ye03} show that $\Bcg$ correlates well with other global properties
of clusters, suggesting that richness observations may become an inexpensive 
way to measure cluster masses, but first the mass-richness relation must be calibrated
and the scatter in that relation must be quantified.

\subsubsection{Galaxy Velocities}
\label{sec:vdisp}

Once a cluster has been optically identified, obtaining the radial velocities $v_r$ 
of the cluster galaxies from their redshifts helps in mitigating projection effects 
and in measuring the cluster's mass.  Because the velocity distribution of a relaxed 
cluster's galaxies is expected to be gaussian in velocity space,
galaxies with velocities falling well outside the best-fitting gaussian envelope are unlikely to 
be cluster members and are generally discarded.  Fitting the velocity distribution 
$\exp [-(v_r - \langle v_r \rangle)^2 / 2 \sigv^2]$ to the remaining galaxies then yields a 
one-dimensional velocity dispersion $\sigv$ for the cluster.  If the velocity distribution
of a cluster candidate is far from gaussian, then it is probably not a real cluster but rather
a chance superposition of smaller structures.  Obviously, the accuracy of $\sigv$ 
depends critically on the number of galaxies with measured velocities and the method
for identifying and eliminating non-members.

\citet{Zwicky33, Zwicky37} was the first to measure a cluster's velocity dispersion, finding
$\sigv \sim 700 \, {\rm km \, s^{-1}}$ for the Coma cluster.  He correctly concluded
from this fact and his estimate of the Coma cluster's overall radius that this cluster's 
mass must be far greater than the observed mass in stars---the first evidence for
dark matter in the universe.  Shortly thereafter, \citet{Smith36} showed that the same
was true of the Virgo cluster.
Zwicky's reasoning involved the virial theorem of classical mechanics, which applies
to steady, gravitationally bound systems.  Differentiating the system's moment of
inertia $I = \sum_i \, m_i {\bf r}_i^2$ twice with respect to time and setting the
result to zero produces the virial relation
\begin{equation}
  \sum_i \, m_i \dot{\bf r}_i^2 = - \sum_i \, m_i \ddot{\bf r}_i \cdot {\bf r}_i \; \; .
\label{eq:vt}
\end{equation}
The left-hand side is twice the total kinetic energy of the cluster's particles, and in a 
spherically symmetric system of mass $M$ with a gaussian velocity distribution, that 
kinetic energy is $3M \sigv^2 / 2$.  If the system is isolated, then the right-hand side 
is equal to the absolute value of the gravitational potential energy, which can be
expressed as $GM^2 / \rvt$, where
\begin{eqnarray}
  \rvt & \equiv & M^2 \left(  \sum_i \sum_{i < j} \, \frac {m_i m_j} {r_{ij}} \right)^{-1}   \\
     ~ & \approx & \frac {\pi} {2} M^2 \left( \sum_i \sum_{i < j} 
              \, \frac {m_i m_j} {r_{\perp,ij}} \right)^{-1}  \nonumber
\end{eqnarray}
and $r_{ij}$ is the separation between particles $i$ and $j$.  The approximation
gives $\rvt$ for a spherically symmetric system in
terms of the projected particle separations $r_{\perp.ij}$ in the plane of the sky
\citep{lm60}.  According to the virial theorem, the mass of 
a spherical, isolated cluster should therefore be $M = 3 \sigv^2 \rvt / G$.

Applying this virial analysis to real clusters is not quite so simple because 
clusters are not isolated systems---there is no clean boundary separating 
a cluster from the rest of the universe.  Segregating the cluster from the outlying
regions with an arbitrary bounding surface alters the interpretation of the
right-hand side of equation (\ref{eq:vt}).  In a steady state, the momentum flux
of particles exiting the boundary is equal to that entering, so the bounding surface
is formally equivalent to a reflecting wall that adds a pressure correction
term, offsetting some of the gravitational potential energy \citep{tw86, cye97}.   
One must also account for objects seen in projection, such as infalling galaxies that have
not yet entered into virial equilibrium and interlopers that are not true cluster
members, problems that have led to the invention of various kinds of projected 
mass estimators \citep{bt81, htb85}. 

Extensive redshift measurements now allow observers to measure much more than just
a cluster's velocity dispersion, enabling detailed studies of a cluster's mass profile
and dynamical state.  Generally a cluster's velocity dispersion declines with 
projected radius, implying that the relationship between projected radius and
the mass enclosed within that radius is somewhat shallower than linear in 
the cluster's outskirts \citep{Rood72, kg82, cye97}.  
Beyond the approximate virial radius of a cluster, the enclosed mass continues 
to increase and the galaxies move primarily on infalling radial trajectories 
\citep{Kaiser87, rg89, dg97, bg03}.  Even farther out is a thin region where galaxies
are nearly stationary with respect to the cluster because there the cluster's gravity
has just succeeded in reducing the outward Hubble flow to a standstill 
\citep{Kaiser87, rgkd03}.
Eventually these galaxies will fall back toward the cluster and become cluster
members.  

Because clusters are dynamical systems that have not quite finished forming and 
equilibrating, the velocity dispersion and virial theorem by themselves do not 
yield an exact cluster mass measurement.  Detailed information on the spatial 
distribution of galaxy velocities is of great help in measuring the masses of large, 
nearby clusters but similar information is very difficult to obtain for the distant 
clusters so interesting to cosmologists.  In lieu of detailed observations,
one can use simulations of cluster formation to calibrate the approximate virial
relationship between velocity dispersion and cluster mass, but we will postpone
discussion of that procedure to the discussion of dark-matter dynamics in
\S~\ref{sec:dcomp}.

\subsubsection{Gravitational Lensing}
\label{sec:glens}

In his remarkable 1937 paper on the Coma cluster, Zwicky also proposed that
cluster masses could be measured through gravitational lensing of background
galaxies.  That technique did not become practical for six more decades but is
now one of the primary methods for measuring cluster mass.  Lensing is sensitive
to the cluster's mass within a given projected radius $r_\perp$ because the
mass within this radius deflects photons toward our line of sight through the 
cluster's center.  When the deflection angle is small compared to a background
galaxy's angular distance from the cluster center, weak lensing shifts each 
point in the galaxy's image to a slightly larger angular distance from the cluster's center,
thereby distorting the image by stretching it tangentially to $r_\perp$.  Measuring 
the weak-lensing distortion of any single galaxy is nearly impossible because the
exact shape of the unlensed galaxy is generally unknown.  Instead, observers
must measure the shear distortion of an entire field of background galaxies,
under the assumption that any intrinsic deviations of galaxy images from
circular symmetry are uncorrelated.

Many excellent articles explain this weak-lensing technique in more detail \citep{twv90, ks93,
hfks98, Mellier99, bs01}.
Here we wish only to give a flavor of how a cluster's mass can be measured
from the lensing it induces.  The deflection angle itself depends on the gradient
of the gravitational potential in the lensing system, meaning that a mass sheet
of constant surface density produces no shear and goes undetected.  Additional
mass that is distributed symmetrically about the line of sight through a cluster's center
bends photon paths by an angle twice that expected from Newtonian physics,
$4GM(<r_\perp)/c^2 r_\perp$, which can be
measured from the shear distortion and redshift distribution of the background
galaxies.  Obtaining a cluster mass from the mass $M(< r_\perp)$ along the 
column bounded by $r_\perp$ requires additional assumptions about how mass 
is distributed within this column.  A particularly simple mass configuration would be 
a singular isothermal sphere, in which $\sigv$ remains constant with radius
and $M(r) = 2 \sigv^2 r / G$ (\S~\ref{sec:massprof}); notice that the boundary pressure term required
in this configuration alters the usual virial relation.  The deflection angle for this
mass distribution is $4 \pi \sigv^2 / c^2$, independent of radius.  In general, however, the
cluster potential will not be precisely isothermal, nor will the cluster be
perfectly spherical.

Simulations of large-scale structure formation suggest that superpositions
of other mass concentrations limit the accuracy of weak-lensing masses, 
at least for clusters defined to be within spherical volumes.  Projected mass 
fluctuations along the line of sight to a distant cluster can be on 
the order of $\sim 10^{14} \, M_\odot$ \citep{mwnl99, mwl01, Hoekstra01}.   
On the other hand, weak-lensing masses are expected to correlate quite well with 
cluster richness, another measure of the mass within a cylindrical 
region, raising the possibility that at least some of the 
projected mass can be accounted for by using galaxy colors to
separate these mass concentrations from the cluster in redshift space.

\subsection{Clusters in X-rays}
\label{sec:clx}

Clusters of galaxies are X-ray sources because galaxy formation
is inefficient.  Only about a tenth of the universe's baryons reside
with stars in galaxies, leaving the vast majority adrift in intergalactic
space.  Most of these intergalactic baryons are extremely difficult
to observe, but the deep potential wells of galaxy clusters compress
the associated baryonic gas and heat it to X-ray emitting temperatures.
The gas temperature inferred from a cluster's X-ray spectrum 
therefore indicates the depth of a cluster's potential well, and the 
emission-line strengths in that spectrum indicate the abundances 
of elements like iron, oxygen, and silicon in the intracluster 
medium (ICM).  Here we outline the primary characteristics of
that X-ray emission.  For a more detailed discussion of the physics, 
see \citet{Sarazin88}.

\subsubsection{X-ray Surface Brightness}
\label{sec:sx}

Extended X-ray emission from clusters of galaxies was first observed
in the early 1970's \citep{Gursky71, kgtgp72, fkgtg72}, but was correctly attributed to thermal
bremsstrahlung several years earlier by \citet{fgsw66}, who were
inspired by a spurious X-ray detection of the Coma cluster.
For typical cluster temperatures ($kT \gtrsim 2 \, \keV$) 
the emissivity of thermal bremsstrahlung dominates that 
from emission lines, but below $\sim 2 \, \keV$ that situation 
reverses, given the typical heavy-element abundances relative
to hydrogen which are $\sim$0.3 times those found in the Sun.  
The rate at which the ICM radiates energy can be expressed
in terms of a cooling function $\Lambda_c (T)$ computed 
assuming that collisional ionization equilibrium determines
the relative abundance of each ion.  Many collisional
ionization codes have been developed to compute the emissivity
and X-ray spectrum of such gas \citep[e.g.,][]{rs77}.  Because these 
cooling processes all involve electrons colliding with ions, 
the resulting cooling function is usually defined so that either
$n_e n_{\rm H} \Lambda_c(T)$ or $n_e n_{\rm ion} \Lambda_c(T)$
is the luminosity per unit volume.  \citet{tn01} give a useful fit 
to the computations of \citet{sd93} for abundances 
equal to 0.3 times their solar values.  For typical
ICM temperatures, $\Lambda_c \sim 10^{-23} \, {\rm erg \, cm^3 \, s^{-1}}$.

In most clusters, the intracluster gas appears to be in approximate
hydrostatic equilibrium.  Assuming spherical symmetry, the equation
of hydrostatic equilibrium can be written
\begin{equation}
 \frac {d \ln \rho_g} {d \ln r} + \frac {d \ln T} {d \ln r} = -2 \frac {T_\phi(r)} {T} \; \; ,
 \label{eq:hydeq}
\end{equation}
where $\rho_g$ is the gas density and $\kB T_\phi(r) = GM(r) \mu m_p/2r$ 
is the characteristic temperature of a singular isothermal sphere with
the same value of $M(r)/r$.  Making the additional assumption that
the gas is isothermal leads to a classic model for the X-ray surface
brightness of clusters known as the beta model 
\citep{cff76}.  If the velocity distribution of
the particles responsible for $M(r)$ is also isothermal with a constant velocity
dispersion $\sigv$, then Poisson's equation implies
\begin{equation}
  \frac {d \ln \rho_g} {dr} = -  \frac {\mu m_p} {kT} \frac {d\phi} {dr}
                                          = \beta \frac {d \ln \rho} {dr} \; \; ,
\end{equation}
where the eponymous $\beta \equiv \mu m_p \sigv^2 / kT$ \citep[e.g.,][]{Sarazin88}.
Given the approximate isothermal potential of \citet{King62}, $\rho(r) \propto 
[ 1 + (r/r_c)^2]^{-3/2}$, in which $r_c$ is a core radius that keeps the profile 
from becoming singular at the origin, the gas density profile becomes
$\rho_g(r) \propto [ 1 + (r/r_c)^2]^{-3 \beta /2}$.  The expected X-ray surface
brightness profile for an isothermal gas is then $\propto [1 + (r/r_c)^2]^{-3 \beta + 1/2}$,
and fitting this model to the observations gives the best-fit parameters 
$r_c$, $\betafit$, and the normalization of the gas-density distribution.   

Beta models generally describe the observed surface-brightness profiles
of clusters quite well in the radial range from $\sim r_c$ to $\sim 3 r_c$, with 
$\betafit \approx 2/3$ and $r_c \sim 0.1 \rvt$ giving the 
best fits for rich clusters \citep{jf84} and a possible 
trend toward lower $\beta$ values in poorer clusters \citep{hms99, hp00, frb01, spflm03}.  
The X-ray luminosity integrated over radius  
converges for $\beta > 0.5$, meaning that most of the observed X-rays come
from a relatively small proportion of the ICM.  However, beta models often 
underestimate the central surface brightness \citep{jf84} 
and tend to overestimate the brightness at $r \gg r_c$ \citep{vfj99}.  
These discrepancies arise in part because the intracluster medium is 
not strictly isothermal (\S~\ref{sec:mtrel}) and because real cluster
potentials differ from the King model (\S~\ref{sec:massprof}).

The centrally concentrated surface-brightness profiles of clusters make X-ray 
surveys very effective at finding cluster candidates.  Because X-ray emission
depends on density squared, clusters of galaxies strongly stand out
against regions of lesser density, minimizing the complications of projection
effects \citep[see][for a recent review]{rbn02}.  Surveys of X-ray selected
clusters currently extend to $z \approx 1.3$ \citep[e.g.][]{Stanford01,Rosati04},
a limit owing to the decline of surface brightness with redshift (\S~\ref{sec:glogeo}).
Unfortunately, X-ray luminosity correlates less well than one would
like with the optical properties of clusters.  Early studies showed that X-ray
luminosity correlates with optical richness but with a large scatter \citep{Bahcall77, 
Mush84}, and that situation has not improved much in the intervening decades
\citep{d02_rox2, Kochanek03, Gilbank03}.   The optical properties of very luminous 
X-ray clusters are well behaved \citep{Lewis99}, but deep optical surveys have found 
distant cluster candidates that appear to have velocity dispersions much larger than 
one would guess from their X-ray luminosity \citep{lmp03}.  These objects may be 
may be superpositions of smaller clusters whose joint velocity distribution seems 
like that of a larger relaxed cluster.
 
\subsubsection{Plasma Temperature}
\label{sec:tx}

Clusters in hydrostatic equilibrium have a plasma temperature that is closely
related to the overall mass.  Measuring that temperature requires higher
quality data than a simple luminosity measurement, because the photons
must be divided among multiple energy bins.  Ideally, one would like enough data
to measure both $T(r)$ and $\rho_g(r)$, in which case equation (\ref{eq:hydeq})
can be solved directly for $M(r)$.  Even with the highest-quality data, the
derived mass is still slightly model dependent because $T(r)$ and $\rho(r)$
must be determined by deprojecting the surface-brightness information 
\citep{fhcg81, kcc83, wjf97, Smaug2003}.

In practice, the quality of the mass measurement depends on what the total number
of observed X-ray photons allows.  With limited information about the temperature gradient, one
can fit a polytropic law\footnote{Note that this is {\em not} an actual equation
of state for the gas but only a fitting formula for $T(r)$ as a function of $\rho_g(r)$.} 
$T \propto \rho_g^{\gammaeff-1}$, 
giving the radial dependence of temperature in terms of 
an effective adiabatic index $\gammaeff$ with 
density as the radial coordinate.  However, data on distant clusters often do not allow
a temperature gradient to be measured and sometimes are even insufficient
to give an accurate temperature.  In those cases, one must rely on scaling laws
that connect X-ray luminosity with temperature and temperature with mass,
calibrated with either high-quality observations or numerical simulations of
cluster formation that include all the relevant physics (\S~\ref{sec:obsmassfcn}). 

Limitations in the measurement of cluster temperature systematically affect the mass
one infers for the cluster.  If only a single temperature can be measured,
then the isothermal beta model implies
\begin{equation}
 \frac {M(r)} {r}  =  \frac {3 \beta \kB T} {G \mu m_p} \frac  {(r/r_c)^2} {1 + (r/r_c)^2} \; \; .
\end{equation}
Note that at large radii this relation approaches the one for isothermal gas in a singular
isothermal potential, $M(r)/r = 2 \kB T / G \mu m_p$, as long as $\beta = 2/3$.
However, single temperatures gleaned from a cluster's overall spectrum need to be
treated with caution.  Global cluster temperatures quoted in the literature are 
generally spectral-fit temperatures ($\Tsp$) obtained by fitting a single-temperature 
emission model to an overall cluster spectrum containing multiple temperature 
components.  These spectral-fit temperatures are similar to, but not identical to,
the cluster's luminosity-weighted temperature $\Tlum$ in which each temperature 
component is weighted by $\rho_g^2$.   Numerical simulations indicate that both 
$\Tsp$ and $\Tlum$ can differ from the mass-weighted gas temperature $T_g$ 
and from one another by $\sim 10$-20\% \citep{me01, mrmt04}. 

A modest amount of spatially resolved temperature information improves 
the mass measurement.  Allowing for a temperature gradient corresponding 
to $T \propto \rho_g^{\gammaeff-1}$ changes the estimated mass to 
\begin{equation}
 \frac {M(r)} {r}  =  \frac {3 \beta \gammaeff \kB T(r)} {G \mu m_p} \frac  {(r/r_c)^2} {1 + (r/r_c)^2} \; \; .
\end{equation}
Observers are still working toward a consensus on the temperature gradients 
of clusters \citep{mfsv98, ib00, pa02, dm02, Mush04}, 
but measured values of $\gammaeff$ often range as high 
as 1.2 \citep{frb01}.  Cluster temperatures are extremely difficult to observe
in the neighborhood of the virial radius, but extrapolating a $\gammaeff = 1.2$
gradient to $10 r_c$ leads to a gas temperature less than half the core temperature.  
Including temperature-gradient information can therefore lower the 
estimated mass for a cluster of temperature $\Tlum$ by up to $\sim$50\%.

\begin{figure*}
\includegraphics[angle = 270 , width=7.0in , trim = 0.2in 0in 0.3in 0in , clip ]
{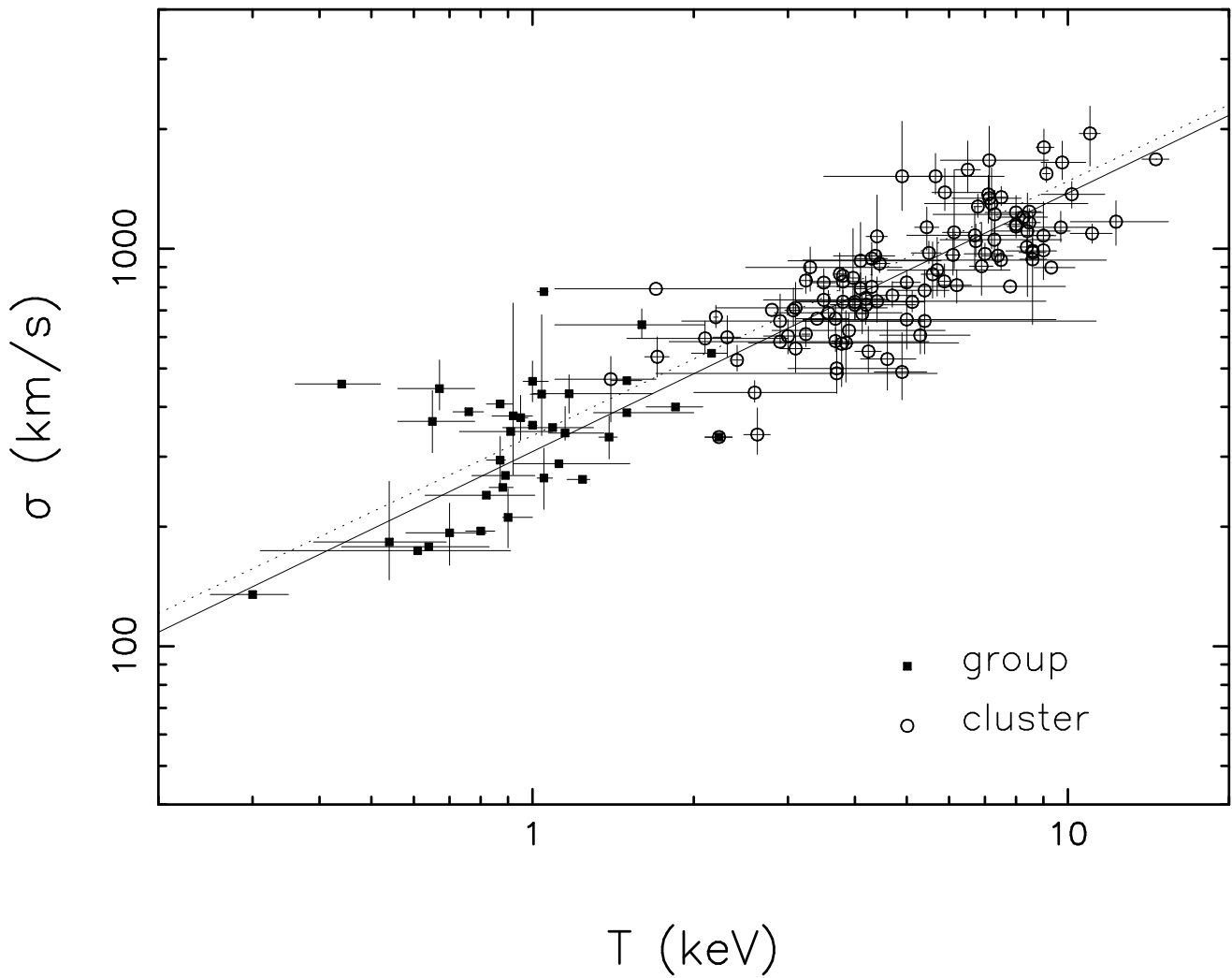}
\caption{Relation between velocity dispersion and temperature for a heterogeneous
sample drawn from the literature.  Solid squares illustrate data on 
galaxy groups and open circles give the cluster data.  The dotted and solid lines 
show the best power-law fits for groups and clusters, respectively.  The best-fitting 
relation to the combined sample is $\sigv = 10^{2.51 \pm 0.01} \, {\rm km \, s^{-1}}
(T / 1 \, \keV)^{0.61 \pm 0.01}$, corresponding to $\betasp = 0.97$ at 6~keV.
 \citep[Figure from][]{xw00}}
 \label{fig:sigt}
\end{figure*}

Despite the potential for systematic uncertainties, the luminosity-weighted
temperatures of clusters correlate well with their velocity dispersions.
Most of the recent comparisons for low-redshift clusters find that $\sigv
\propto \Tsp^{\sim 0.6}$, slightly steeper than expected if both quantities 
track cluster mass \citep{lb93, xw00}.  Those same comparisons find  
normalizations of this relation for rich clusters in the range $\betasp = 
\mu m_p \sigv^2 / k \Tsp = 0.9-1.0$ (Figure~\ref{fig:sigt}).  The discrepancy
between $\betasp$ and $\betafit$ is no cause for concern.  It arises
because the true mass profile is not a King model and because clusters 
are not in perfect hydrostatic equilibrium \citep{Evrard90, bl94}.
More worrisome are recent observations suggesting that the X-ray 
temperatures of distant optically-selected clusters with unusually 
small X-ray luminosities are also considerably cooler than their velocity 
dispersions would indicate \citep{lmp03}.  However, more extensive redshift
measurements have shown that at least one of these systems is composed 
of several smaller systems that have not yet merged to form a single 
large cluster \citep{gl04}. 

\subsubsection{Measuring Abundances}
\label{sec:zx}  

Abundances of elements like iron, oxygen, and silicon in the intracluster
medium are relatively easy to measure from their emission line fluxes,
as long as the temperature of the line-emitting gas is well defined.
Because of the low density of intracluster gas, collisional deexcitation
is negligible, so every collisional excitation produces a photon that
leaves the cluster.  Thus, one can fit the optically-thin spectrum 
of a collisionally-ionized, single-temperature plasma to the observed
spectrum, adjusting the abundances in the model to produce the best fit.
The high spectral resolution of today's X-ray observatories, {\em Chandra}
and {\em XMM-Newton}, allows abundance determinations for individual
elements if enough photons can be gathered.  Otherwise, the solar pattern
of abundance ratios is assumed for elements other than H and He and
the normalization of the overall pattern is fit to the observations.  Because 
the most abundant elements are almost completely ionized in the hottest
clusters, these abundance determinations depend heavily on the strength 
of the K-shell emission lines of iron, sometimes the only lines that are
measurable.

On average, the overall abundances of heavy elements with
respect to hydrogen in clusters are about 0.3 times the solar ratios.  Just as with
temperature, this determination is weighted toward the cluster core because of
the $\rho_g^2$ emissivity.  Spatially resolved observations of Fe $K$-line
emission show that iron abundances, at least, can be higher at the 
cluster's center, particularly when a giant, central-dominant galaxy 
is there.  This iron excess is consistent with being supernova debris
from the giant galaxy's stars \citep{delm03}.  Farther out in clusters, 
these Fe gradients appear to flatten at $\sim$0.3 times the solar level, 
extending to about $\sim 5 r_c$, beyond which point the X-ray surface 
brightness is too low for accurate abundance and temperature measurements.
This abundance level does not seem to have substantially
changed from redshift $z \sim 1$ to the present \citep{dvglhs98, Don99, 
trebmn03}.

The total amount of iron implied by extrapolating this ratio over an
entire cluster is quite impressive, exceeding the total amount of
iron contained within all the stars in the cluster's galaxies \citep{Renzini97}.
Explaining how all that iron got into the intracluster medium is
challenging.  It is comparable to the total amount of iron produced
by all the supernovae thought to have exploded during the history of the
cluster, and according to some estimates, it requires a 
disproportionately large number of massive stars to have formed in order
to produce enough supernovae \citep{dfj91, mg95, gm97, lm96, Loewenstein01,
pmcs03}. 

Presumably all these supernovae could have driven strong 
gaseous outflows known as galactic winds that expelled the 
heavy elements into the intracluster medium \citep{ld75, ham90}.  
However, such powerful galactic winds 
are hard to produce in numerical simulations of galaxies because 
much of the energy released by massive-star (Type II) supernovae 
is transferred to cool gas within the galaxy, where it is radiated away 
before it manages to drive a powerful wind \citep{mf99}.
Alternatively, some of this iron may come from exploding white
dwarfs (Type Ia supernovae), whose iron yields are higher than
those of Type II supernovae.  In either case, the total amount of kinetic
energy released by the supernovae that created these elements
is enormous, corresponding to $\sim 0.3 - 1 \, \keV$ per particle in the
intracluster medium \citep{fad01, pmbb02}.  Yet, the efficiency of energy transfer
from supernovae to the ICM remains an open question \citep{ky00}.

In principle, one can probe the origins of elements in the ICM and
assess whether massive stars were disproportionately common
earlier in time by comparing the abundances of massive-star
products like oxygen to that of iron, which may come largely
from Type Ia supernovae.  No clear answer has yet
emerged from such studies, which depend heavily on a proper
understanding of the gas temperature distribution to get the
correct elemental abundances \citep[e.g.,][]{Buote03}.  Some studies
have concluded that the relative abundance patterns in the
intracluster medium are near solar, implying
that the stellar populations producing those supernovae were
similar to those in our own galaxy \citep{Renzini04}.  Other studies find
an excess of oxygen and other elements of similar atomic
number, suggesting that the cluster's galaxies produced an
unusually large number of massive stars early in the cluster's
history \citep[e.g.,][]{fbb03}. 
  
\subsection{Clusters in Microwaves}
\label{sec:clsz}

Hot gas in clusters can also be observed through its effects 
on the cosmic microwave background.  The background itself 
has a virtually perfect blackbody spectrum \citep{Mather90}.  
Soon after the discovery of this background
radiation, \citet{Weymann65, Weymann66} computed how Compton
scattering would distort its spectrum, slightly shifting some of the 
microwave photons to higher energies as they passed through
hot intergalactic gas.  \citet{sz70, sz72} then predicted
that hot gas in clusters of galaxies would indeed produce such a
distortion, now known as the Sunyaev-Zeldovich (S-Z) effect.  

\subsubsection{The S-Z Effect}
\label{sec:sze}

Two decades after this prediction there were only a few marginal 
detections \citep{Birk91}, but many clusters were detected 
at high significance in the ensuing decade \citep{Birk99, Carlstrom00}. 
With multiple new and highly capable S-Z instruments
coming on line in the next few years, another quantum leap in
this area is poised to happen, enabling wide-field cosmological
studies of clusters to extend through much of the observable 
universe \citep{chr02}.  A number of recent reviews elucidate 
the details of the S-Z effect \citep[e.g.,][]{Birk99, chr02}.  
Here we summarize only a few fundamentals.

To lowest order, the shape of the distorted spectrum depends 
on a single parameter proportional to the product of the probability that
a photon passing through the cluster will Compton scatter and the 
typical amount of energy a scattered photon gains:
\begin{equation}
 y =  \int \frac {\kB T} {m_e c^2} n_e \sigma_{\rm T} \, dl \; \; ,
\end{equation}
where $\sigma_{\rm T}$ is the Thomson cross-section and the
integral is over a line of sight through the cluster.  Because 
the optical depth of the cluster is small, the change in microwave 
intensity at any frequency is linearly proportional to $y \ll 1$, with
reduced intensity at long wavelengths and enhanced intensity
at short wavelengths.  Relativistic corrections in hot clusters add 
a slight frequency dependence to the magnitude of the effect, making 
cluster temperatures measurable with precise observations of the 
microwave distortion at several frequencies \citep[see][for a discussion]{chr02}.  
A cluster's motion with respect to the
microwave background produces additional distortion, known as
the kinetic S-Z effect, but here we will concern ourselves only 
with the thermal S-Z effect.

Cosmological applications of the thermal S-Z effect in clusters
benefit greatly from the fact that the effect is independent of
distance, unlike optical and X-ray surface brightness.  Thus,
a dedicated S-Z cluster survey efficiently finds clusters out
to arbitrarily high redshifts.  Because not all these clusters
will be well resolved, the surveys will be measuring an 
integrated version of the distortion parameter:
\begin{equation}
  Y = \int y \, dA  \propto \int n_e T \, dV \, \, ;
\end{equation}
where the first integral is over a cluster's projected surface
area and the second is over its volume.  The $Y$ parameter
therefore tells us the total thermal energy of the electrons, from which
one easily derives the total gas mass times its mass-weighted temperature
within a given region of space.  If these regions can be chosen so that
the gas mass is always proportional to the cluster's total mass, then the
observable $Y$ can be used a measure of cluster mass, once the relationship
between $Y$ and mass has been calibrated.

The impressive power of the S-Z effect for finding distant clusters
also has a significant drawback, namely sky confusion owing to projection
effects.  Along any line of sight through the entire observable universe,
the probability of passing within the virial radius of a cluster or group
of galaxies is of order unity \citep[e.g.,][]{veb01}.  Because
a cluster's S-Z distortion does not diminish with distance, many of the
objects in a highly sensitive S-Z survey will therefore significantly
overlap. Information on galaxy colors will help to separate
nearby objects from more distant ones, but the implications
of sky confusion for making accurate mass measurements are still
a matter to be reckoned with \citep[e.g.,][]{whs02}.
One way to avoid the problem of sky confusion will be to measure
the statistical S-Z properties of clusters in the angular power spectrum
of the microwave sky instead of analyzing the clusters themselves 
\citep{sbp01, da_silva01, hc01}.  In fact, this statistical signal may already 
have been detected \citep{Pearson03, Kuo04}

\subsubsection{Comparing S-Z with X-ray}
\label{sec:sz-x}

Comparisons between a cluster's X-ray properties and S-Z properties
are useful in several different ways.  X-ray observations are nicely
complementary to S-Z observations of clusters because they give 
the integral of $\rho_g^2$ along lines of sight through a cluster 
in addition to a gas temperature.  Assuming that clusters are
spherical objects with smooth gas distributions, one can 
divide the product of temperature and the line-of-sight
integral of $\rho_g^2$ by the observed $y$ value 
to obtain a cluster's gas density profile.  Combining the data
in this way can be particularly useful in studying the outskirts
of clusters, where the X-ray surface brightness is difficult to
observe but the S-Z signal remains substantial.
With this density profile in hand, one can then derive 
the line-of-sight thickness of the cluster from either the X-ray 
or S-Z observations.  This type of information could help to
solve the S-Z projection problem in fields where there are
high-quality X-ray and S-Z data.

If a cluster is indeed spherical, then a comparison of its cluster's 
physical thickness with its apparent angular size directly gives 
the cluster's distance, which can be used to determine
the scale and geometry of the universe \citep{bha91}.
Deriving the scale of the universe in this way is subject to numerous 
systematic effects.  For example, clusters are not all perfectly spherical.
Many appear slightly ellipsoidal in X-ray images, calling for a sample 
of clusters with random orientations to beat down this systematic
effect, although three-dimensional reconstructions are possible with
the addition of gravitational-lensing data \citep[e.g.][]{Zaroubi01}.  
Note also that comparisons of X-ray images to S-Z images would 
produce nonsensical distances if the intracluster medium were
highly clumpy, owing to the $\rho_g^2$ X-ray emissivity.  The fact that
cluster distances found in this way are consistent with the standard
calibrations of Hubble's Law indicates that the X-ray emitting gas 
is well-behaved and that most clusters are in approximate hydrostatic
equilibrium.

\section{Evolution of the Dark Component}
\label{sec:dcomp}

Cluster masses measured with the techniques outlined in the previous
section range from around $10^{14} \, M_\odot$ to more than $10^{15} \,
M_\odot$, the vast majority of which appears to be dark matter that emits 
no detectable radiation.   Even using alternative theories of gravity, it is 
difficult to explain the cluster observations without dark matter dominating 
the overall mass \citep{Sanders03}.  In contrast, explaining the characteristics 
of clusters and their evolution with redshift is much easier with models 
in which non-baryonic cold dark matter dominates the mass 
density of the universe.   

This section explains how the evolution of the dark component of the universe,
including both dark matter and dark energy, 
is thought to be reflected in the evolution of cluster properties.
It begins with a summary of the concordance model for cosmology and some
closely related alternatives, all of which are predicated on the existence of
non-baryonic cold dark matter.  It then explains how dark matter drives cluster
formation in such models, providing some simple analytical approximations
to the extensive numerical work that has been done on the subject.  These
models do a good job of accounting for the basic properties of observed 
clusters, allowing astronomers to measure several of the parameters in the concordance 
model using cluster observations, most notably the overall mass density 
of the universe and the amplitude of the initial spectrum of density perturbations 
that eventually produces all the structure we observe.  

The accuracy of those parameter measurements is currently limited by
uncertainties in the relationships between cluster masses and the observable
properties that trace those masses.  Numerical simulations of cluster formation
do not yet provide precise calibrations of these relations because they do not
yet account for all of the thermodynamical processes associated with galaxy 
formation.  The third part of this section surveys the mass-observable relations
and how the uncertainties in those relations affect cosmological parameters
derived from them.  The fourth part of this section examines how the properties of clusters 
evolve and how fitting that evolution with cosmological models improves the accuracy
of the derived cosmological parameters.  Even though current surveys of distant
clusters contain relatively few objects, they already place strong constraints on
the overall matter density.   Larger cluster surveys in both the microwave and
X-ray bands have the potential to place much stronger constraints on the overall
cosmological model, measuring both dark matter and dark energy parameters
to 5\% statistical accuracy, independently of other cosmological observations.

\subsection{A Recipe for the Universe}
\label{sec:recipe}

Our current understanding of cluster evolution is an outgrowth of the overall cosmological 
model, whose primary features depend on just a handful of parameters.
One set of parameters specifies the global cosmological model, which describes
the overall geometry of the universe, the mean density of its contents, and 
how its scale changes with time.  The other important set of parameters specifies the
initial spectrum of density perturbations that grew into the galaxies and clusters
of galaxies we see today.  Here we define both sets of parameters and their
roles in the context of the overall model.  More extensive and detailed 
discussions of this recipe for the universe can be found in some of the
excellent books on cosmology \citep[e.g.,][]{Peebles93, Peacock99}.

\subsubsection{Global Dynamics}
\label{sec:glodyn}

The expansion of the universe can be characterized by a time-dependent scale factor 
$a(t)$ proportional to the mean distance between the universe's galaxies.  Hubble's Law
relating the distance $d$ between two galaxies and the speed $v$ at which they appear to move
apart can then be written as $v = H(t) d$, where $H(t) = \dot{a}/a$ is the Hubble parameter.
Many independent measurements indicate that the value of this parameter at the current 
time $t_0$ is $H(t_0) = H_0 = 71 \pm 7 \, {\rm km \, s^{-1} \, Mpc^{-1}}$ \citep{H0KP_01}.   The value of
$H_0$, known as Hubble's constant, is often further distilled in the literature into the
dimensionless quantity $h = H_0 / (100 \, {\rm km \,  \, s^{-1} \, Mpc^{-1}})$.  Sometimes
this review will use the more suitable alternative $h_{70} = H_0 / (70 \, {\rm km \,  \, s^{-1} 
\, Mpc^{-1}})$ when characterizing observable cluster properties.

On very large scales, the universe appears homogenous and isotropic.  Astronomers therefore
assume that the time-dependent behavior of $H(t)$ obeys the Friedmann-Lemaitre model of the 
universe, in which
\begin{equation}
  \frac {\ddot{a}} {a} = - \frac {4} {3} \pi G \left( \rho + \frac {3p} {c^2} \right)  \; \; ,
  \label{eq-fluniv}
\end{equation}
where $\rho(t) c^2$ is the mean density of mass-energy and $p(t)$ is the pressure owing to 
that energy density.  Local energy conservation requires that
\begin{equation}
  \dot{\rho} c^2 = -3 \frac {\dot{a}} {a} (\rho c^2 + p) \; \; ,
\end{equation}
and we can use this expression to integrate the dynamical equation as long as we
know the equation of state linking $\rho$ and $p$.  If the equation of state has the form
$p = w \rho c^2$, then density changes with the expansion as $\rho \propto a^{-3(1+w)}$.
For a single mass-energy component with a constant value of $w$ we therefore obtain 
\begin{equation}
  \dot{a}^2  =  \frac {8 \pi G} {3} \rho_0  a^{-(1+3w)} + {\rm const.} \; \; ,
\end{equation}
where $\rho_0$ is the value of the energy density when $a=1$ and the constant of
integration is related to the global curvature of the universe.

\begin{figure*}
\includegraphics[width=7.0in , trim = 1.1in 1.9in 1.1in 3.1in , clip]
{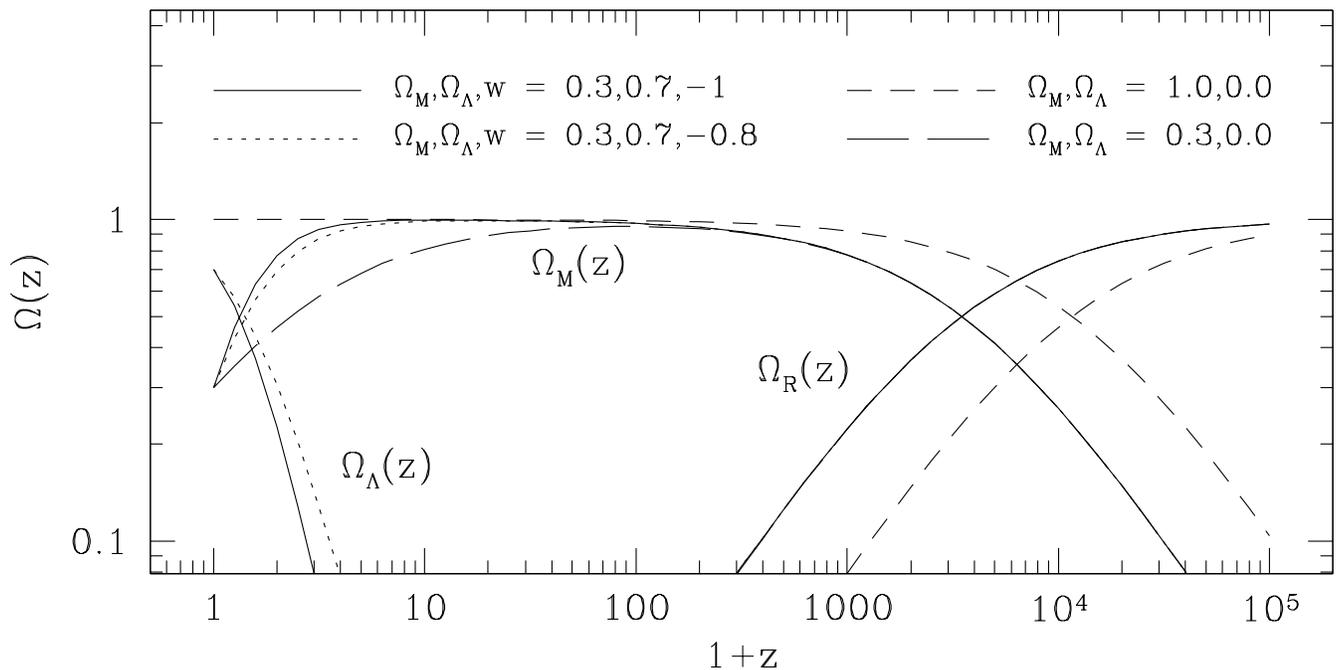}
\caption{Evolution of energy densities with redshift.  The various lines show the dependence
of $\Omat (z)$, $\Olam (z)$, and $\Orad (z)$ on redshift for various sets of present-day 
cosmological parameters.
Structure in the universe grows most rapidly while $\Omat (z) \approx 1$, because positive density
perturbations then exceed the critical density.  This period of time occurs between the redshift
$z_{\rm eq}$ when $\Omat (z_{\rm eq}) = \Orad (z_{\rm eq})$ and the redshift at which $\Omat$ begins
to drop.  Notice that the redshift $z_{\rm eq}$ is earlier for larger present-day values of $\Omat$ and
that the redshift at which $\Omat (z)$ begins to decline depends on the characteristics of dark energy.
Observations of clusters and their evolution provide opportunities to constrain the values of 
$\Omat$, $\Olam$, and $w$ because the timing of both of these epochs influences the 
properties of the cluster population.}
\label{fig:omegas}
\end{figure*}

It is most convenient to normalize the scale factor so that it equals unity at the current 
time.  Then the cosmological redshift $z$ of radiation from distant objects is simply related 
to the scale factor of the universe when that radiation was emitted: $a = (1+z)^{-1}$. 
This definition allows us to link the constant of integration to more familiar
parameters, obtaining
\begin{equation}
  \left(  \frac {\dot{a}} {a} \right)^2 = H_0^2 [ \Omega_0 (1+z)^{3(1+w)} + (1-\Omega_0)(1+z)^2 ] \; \; ,
\end{equation}
where $\Omega_0$ is the current energy density $\rho_0$ in units of the current critical
density $\rhocro = 3 H_0^2 / 8 \pi G$.

Several different components of the universe, each with a different equation of state, 
can influence the overall expansion history.  Non-relativistic particles with a mass density
$\rho_{\rm M}$ contribute negligible pressure, corresponding to $w=0$.  The energy density 
$\rho_{\rm R} c^2$ in photons and other relativistic particles exerts a pressure with $w = 1/3$.
Einstein's cosmological constant acts like an energy density $\rho_\Lambda c^2$ 
that remains constant while the universe expands and therefore exerts a pressure 
corresponding to $w=-1$. Including each of these components yields the dynamical 
equation
\begin{eqnarray}
\label{eq-globaldyn}
 H^2(z)  \;  = \;  \left(  \frac {\dot{a}} {a} \right)^2 & = & H_0^2 [ \Omat (1+z)^3  
                                         + \Orad (1+z)^4  \nonumber     \\
			~ & ~ & \; \; \; \;  +   \,  \Olam  +  (1-\Omega_0) (1+z)^2 ]   \; , 
\end{eqnarray}
where $\Omega_x$ is the current mass-energy density in component $x$ in units of
$\rhocro$ and $\Omega_0 = \Omat + \Orad + \Olam$.   The value of $\Omega_x$ at an arbitrary
redshift is given by $\Omega_x(z) = \Omega_x (1+z)^{3(1+w)} [H(z)/H_0]^{-2}$.

Each of these energy-density parameters can be further articulated.  The matter density 
parameter $\Omat$ consists of a contribution $\Obar$ from baryons and a contribution
$\Omega_{\rm CDM}$ from non-baryonic cold dark matter.
The radiation density parameter includes contributions from the photons of the microwave 
background, $\Omega_{\rm CMB}$, and from relict neutrinos produced in the 
Big Bang, $\Omega_\nu$, as long as they remain relativistic particles.   
Finally, because the physical origin of the $\Olam$ 
term remains mysterious, it may not be correct to assume that the energy density responsible 
for it stays constant with time.  In order to check this possibility observationally, one can replace 
the $\Olam$ term with a generalized dark-energy term $\Olam (1+z)^{3(1+w)}$ and attempt to 
measure the value of $w$ \citep{tw97, ws98}.

Recent observations, including the cluster studies we will discuss later, have provided approximate
values for many of these energy-density parameters, allowing us to estimate when each of
the various energy components dominated the dynamics (Figure~\ref{fig:omegas}).  Dark energy 
with $\Olam \approx 0.7$ seems to be most important at the current epoch, and because 
of the scaling of other terms with redshift, it will grow increasingly dominant 
as time progresses.  Non-relativistic matter appears to have a density corresponding to 
$\Omat \approx 0.3$, implying that matter dominated the dynamics at $z \gtrsim1$.  
The radiation term was most important in the distant past, prior to the redshift
$z_{\rm eq} = \Omat/\Orad -1$ of matter-radiation equality.  Neutrinos with masses less 
than a few eV will be relativistic particles at this epoch, leading to 
\begin{equation}
 z_{\rm eq} = 2.37 \times 10^4 \, \Omat h^2
\end{equation}
for $T_{\rm CMB} = 2.73 \, {\rm K}$ at $z=0$ and three families of neutrinos.  

\subsubsection{Global Geometry}
\label{sec:glogeo}

Geometry in a universe that is homogenous and isotropic has the same radius of curvature
everywhere, but its overall architecture can be either positively curved, flat, or negatively
curved, depending on the value of $\Omega_0$.  Because the scale of the universe is changing
with time, the most sensible coordinate system to use when describing its geometry is one
that expands along with the universe.  In such a comoving coordinate system, a radial interval 
in spherical coordinates has length $a(t) dr$, and the interval corresponding to a small  
transverse angle $d \psi = \sqrt{d\theta^2 + \sin^2 \theta \cdot d\phi^2}$ depends
on the radius of curvature $a(t) R_\kappa$.  For positive curvature, analogous to the surface
of a sphere, the transverse interval is $a(t) R_\kappa \sin (r/R_\kappa) d\psi$, and for negative
curvature it is $a(t) R_\kappa \sinh (r/R_\kappa) d\psi$.  

We can therefore write the Robertson-Walker metric that describes such a universe as
\begin{equation}
 c^2 d\tau^2 = c^2 dt^2 - a^2(t) \left[ dr^2 + R_\kappa^2 S_\kappa^2 (r/R_\kappa) d\psi^2 \right]  \; \; ,
\end{equation}
where $S_\kappa (x) = \sin x$ for positive curvature ($\kappa = 1$), $S_\kappa (x) = \sinh x$
for negative curvature ($\kappa = -1$), and a flat universe ($\kappa = 0$) corresponds to 
$R_\kappa \rightarrow \infty$.  The metric can be written in the more familiar form
\begin{equation}
 c^2 d\tau^2 = c^2 dt^2 - a^2(t)  \left[ \frac {dr_\kappa^2} {1-\kappa r_\kappa^2 / R_\kappa^2}  
                                            + r_\kappa^2 d\psi^2 \right]  \; \; 
\end{equation}
with the definition $r_\kappa \equiv R_\kappa S_\kappa (r/R_\kappa)$.  Plugging this
metric into Einstein's field equations leads to 
\begin{equation}
  H^2(z) = \left(  \frac {\dot{a}} {a} \right)^2 =  \frac {8 \pi G \rho} {3} - 
                \frac {\kappa c^2} {a^2 R_\kappa^2} \; \; ,
\end{equation}
which relates the radius of curvature to other cosmological parameters: 
\begin{equation}
 R_\kappa = \frac {c} {H_0} \sqrt{ \frac {\kappa} {\Omega_0 - 1}  } \; \; .
\end{equation}
Notice that the universe at early times is effectively flat as long as $\Omat + \Orad > 0$ because 
the horizon size of the observable patch is $\sim c/H(z) \ll (1+z)^{-1} R_\kappa$ for observers
at times corresponding to large values of the redshift $z$. 

\begin{figure}
\includegraphics[width=3.2in , trim = 1.0in 1.0in 1.0in 1.0in , clip]
{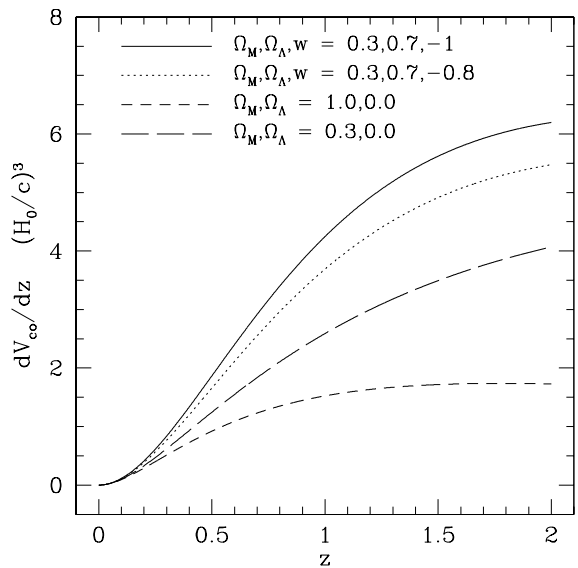}
\caption{Redshift dependence of comoving volume in various cosmologies.
The quantity $dV_{\rm co}/dz$ is the comoving volume of the entire sky between
redshift $z$ and $z + dz$, divided by the redshift interval $dz$.  
If clusters were a non-evolving population 
of objects, one could distinguish between these cosmologies
simply by counting the number of clusters on the sky in each redshift interval. }
\label{fig:Vco}
\end{figure}

The low-redshift universe may also be effectively flat, but that is not guaranteed. Consequently, 
both the expansion of the universe and its curvature need to be taken into account when we 
observe highly redshifted objects like distant clusters of galaxies.  Because the metric relates the 
comoving radial coordinate $r$ to redshift through $dr/dz = -c/H(z)$, the coordinate distance to an 
object with an observed redshift $z$ is
\begin{equation}
 \label{eq-rcoord}
  r(z) = c \int_0^z \frac {dz} {H(z)} \; \; .
\end{equation}
Relations involving the divergence of light paths can then be compactly written in terms
of $r_\kappa(z) = R_\kappa S_\kappa [r(z)/R_\kappa]$, which reduces to $r(z)$ in a flat 
universe.  For example, the angle subtended at coordinate distance $r(z)$
by the transverse length $l$ becomes
\begin{equation}
 \psi = \frac {(1+z) l} {r_\kappa (z)} \; \; .
\end{equation}
In a flat, static universe, an object of physical size $l$ would subtend this same angle
if it were at the distance $d_{\rm A}(z) = r_\kappa (z) / (1+z)$, 
sometimes called the angular-size distance. Likewise, the comoving volume within a solid 
angle $d \Omega$ and a redshift interval $dz$ is given by 
\begin{equation}
  \frac {d^2V_{\rm co}} {d\Omega \, dz}= \frac {c r_\kappa^2 (z)} {H(z)}    \; \; .
\label{eq:vco}
\end{equation}
These formulae are useful to cluster cosmology because they allow us to constrain
$H(z)$ and the cosmological parameters that go into it if we know either the transverse 
sizes of high-redshift clusters or their number density within a given comoving volume.
Figure~\ref{fig:Vco} shows how the comoving volume of the universe depends 
on redshift for several different sets of cosmological parameters. 

When surveying the universe for clusters, we also need to know how the geometry and 
expansion of the universe affect the apparent brightness of a cluster and the galaxies 
within it.  The expansion alone reduces the energy flux received from a distant object
by two factors of $1+z$, with one factor coming from the time dilation of the photon
flux owing to expansion and the other from the redshift of the photons themselves.
The observed energy flux from an object of luminosity $L$ is therefore
\begin{equation}
  F = \frac {L} {4 \pi (1+z)^2 r_\kappa^2 (z)} \; \; .
\end{equation}
We would measure the same flux from an equivalent object in a flat, static universe 
if it were at a distance $d_L (z) = (1+z) r_\kappa (z)$,
sometimes called the luminosity distance.  The consequences for surface brightness, 
equal to flux per unit solid angle, are even more dramatic.  In a flat, static universe, an 
object's surface brightness remains constant, but its surface brightness in an expanding 
universe is reduced by a factor $d_A^2/d_L^2 \propto (1+z)^{-4}$, meaning that extended 
objects like clusters are far less bright at high redshifts.

It would be nice if $r_\kappa(z)$ could be expressed analytically for a general
cosmology, but in most cases it cannot.  However, a useful analytical expression
for the divergence factor does exist for the case in which $\Orad$
and $\Olam$ are negligible (Mattig 1958):
\begin{equation}
 r_\kappa (z) = \frac {2c} {H_0} \frac 
                  {\Omat z + (\Omat - 2) (\sqrt{1+\Omat z} -1)} {\Omat ^2 (1+z)} \; \; .
\end{equation}
Usually one needs to integrate equation (\ref{eq-rcoord}) numerically and then insert the results 
into the $S_\kappa$ function to obtain the rest of the relations.

\subsubsection{Density Perturbations}
\label{sec:dperts}

The very existence of galaxy clusters and the human beings who observe them demonstrates
that the universe is not perfectly homogeneous.  Therefore, the matter density in the early universe 
must have been slightly lumpy.  At some early time these perturbations away from 
the mean density $\langle \rho_M \rangle$ correspond to an overdensity field
\begin{equation}
 \delta({\bf x}) = \frac {\rho_M({\bf x}) - \langle \rho_M \rangle} {\langle \rho_M\rangle}
\end{equation}
with Fourier components
\begin{equation}
 \delta_{\bf k}(k) = \int \delta({\bf x}) e^{i{\bf k}\cdot{\bf x}} \: d^3x \; \; .
\end{equation}
In the plausible case that $\delta({\bf x})$ is isotropic, it can be 
characterized by an isotropic power spectrum
\begin{equation}
 P(k) \equiv \langle | \delta_k | ^2  \rangle \; \; .
\end{equation}
If $\delta({\bf x})$ is also a Gaussian random field, then $P(k)$ is a complete
statistical description of the initial perturbation spectrum.

The physical meaning of $P(k)$ becomes clearer if we assume it has a power-law form, 
with $P(k) \propto k^n$, and consider the variance in mass within identical volume
elements corresponding to the length scale $k^{-1}$.  For example, let $W(r)$ be a spherical window
function that goes quickly to zero outside some characteristic radius $r_W$ and whose
integral over all of space is unity.  The mass perturbation smoothed over the window is
\begin{equation}
  \frac {\delta M} {M} ({\bf r}) = \int \delta({\bf x}) W(|{\bf x} - {\bf r}|) \, d^3x \; \; .
\end{equation}
Using the convolution theorem, we can then write down the variance $\sigma^2 \equiv
\langle | \delta M / M |^2 \rangle$ on this mass scale in terms of $W_k$, the Fourier 
transform of $W(r)$:
\begin{equation}
 \sigma^2 = \frac {1} {(2 \pi)^3} \int P(k) |W_k|^2 d^3k \; \; .
\end{equation}
The variance in mass on scale $k$ for a power-law perturbation spectrum is therefore 
$\sigma^2 \propto k^{n+3}$, because the windowing averages out modes with $k \gg r_W^{-1}$.  
Thus, the typical mass fluctuation on mass scale $M \propto k^{-3}$ is
\begin{equation}
 \frac {\delta M} {M} \propto M^{- \frac {n+3} {6}} \; \; .
\end{equation}
Notice that large-scale homogeneity of the universe requires $n > -3$.

It is also illuminating to consider how $P(k)$ relates to fluctuations in the gravitational
potential, $\delta \Phi \propto k \delta M$.  The potential fluctuations owing to a 
power-law perturbation spectrum scale as $\delta \Phi \propto k^{(n-1)/2}$.  The magnitude
of these fluctuations therefore diverges on either the high-mass end or the low-mass end,
except in the case of $n=1$.
This special property of the $P(k) \propto k$ power spectrum was noted independently
by \citet{Harrison70}, \citet{py70}, and \citet{Zeldovich72}.   
Not only is this the most natural power-law spectrum, it also appears to be a 
good approximation to the true power spectrum of density fluctuations 
in the early universe.  Inflationary models for the seeding of structure in the universe produce
a Gaussian density field with a power-law index close to $n=1$ \citep{gp82},
which is consistent with the observed fluctuations in the cosmic microwave
background \citep[e.g.,][]{WMAP03}.

\subsubsection{Growth of Linear Perturbations}
\label{sec:linperts}

Once the universe has been seeded with density perturbations they begin to grow
because the gravity of slightly overdense regions attracts matter away from neighboring, 
slightly underdense regions.  A complete treatment of perturbation growth is beyond 
the scope of this review, but some key features can be clarified with a simple toy model 
consisting of a uniform-density sphere that is slightly denser than its surroundings.  The
equation of motion for the radius $R$ of an expanding homogeneous sphere is analogous to the one 
governing the universe as whole.  Integrating equation (\ref{eq-fluniv}) with $a = R/R_0$,
where $R_0$ is an arbitrary fiducial radius at which $\rho = \rho_0$, gives
\begin{equation}
  \frac {\dot{R}^2} {2} - \frac {4 \pi G \rho_0 R_0^{3+3w}} {3} R^{-(1+3w)} = \epsilon \; \; .
\end{equation}
The constant of integration $\epsilon$ in this equation is again related to spatial curvature
but can also be interpreted as the net specific energy of the sphere.

Now consider the behavior of two nearly identical spheres that both begin expanding
from $R=0$ at $t=0$ but have specific energies that differ by a small amount
$\delta \epsilon \ll \dot{R}^2 / 2$.  As these two spheres evolve, their radii will become
slightly different by an amount $R_2 - R_1 = \delta R$, which satisfies the equation
\begin{equation}
 \int_0^R \frac {dR_1} {\dot{R}_1} = \int_0^{R+\delta R} \frac {dR_2} {\dot{R}_2} \; \; . 
\end{equation}
In the linear regime, we can make the substitution $\dot{R}_2^{-1} = (1- \dot{R}_1^{-2}
\delta \epsilon) \dot{R}_1^{-1}$.  If we then take the sphere of radius $R_1$ to be 
representative of the universe at large, we obtain
\begin{equation}
 \frac {\delta R} {R} = \frac {\delta \epsilon} {R_0^2} \frac {\dot{a}} {a} \int_0^a \frac {da} {\dot{a}^3} \; \; .
\end{equation}
Because $\delta \rho / \rho = -3 (1+w) \delta R / R$, this model leads to the following 
growth function for linear perturbations:
\begin{equation}
\label{eq-growth}
 D(a) \propto \frac {\delta \rho} {\rho}  \propto \frac{\dot{a}} {a} \int_0^a \frac {da} {\dot{a}^3} \; \; ,
\end{equation}
which is conventionally normalized so that $D(a) = 1$ at $z=0$.  
Notice that the rate of perturbation growth implied by $D(a)$ does not depend on the
scale of the perturbation, implying that density perturbations on all scales grow in unison.

This expression for the growth function is identical to those obtained through more rigorous
arguments \citep[e.g.,][]{Heath77, Peebles93}.  In a matter-dominated universe, perturbation 
amplitudes grow in proportion to the scale factor $a$.  In a radiation-dominated universe, 
they grow $\propto a^2$.  Handy numerical algorithms for computing $D(a)$ can be
found in \citet{Ham01}.  A good approximation for the general case with a constant 
dark-energy density is
\begin{eqnarray}
  D(z) & = & \frac {5 \Omat (z)} {2 (1+z)} \left\{ \Omat (z)^{4/7} - \Olam (z) 
   									\right.   \nonumber \\
           & ~ & \; + \left. \left[ 1 + \frac {\Omat (z)} {2} \right] 
                               \left[ 1 + \frac {\Olam (z)} {70} \right] \right\}^{-1}  
\end{eqnarray}
\citep[see][]{lrlp91, cpt92}.

If the dark-energy density is homogeneous but not constant in time, then the dark-energy
density in the perturbed sphere of radius $R_2$ does not depend on its radius.  In that case, 
one must solve a differential equation to determine the evolution of $\delta \equiv \delta \rho / \rho$
in the linear regime.  Differentiating $R_2 = R_1 (1 - \delta / 3)$ twice with respect to time
and keeping only the lowest order terms leads to
\begin{equation}
  \ddot{\delta} + 2 \frac {\dot{a}} {a} \dot{\delta} = 4 \pi G \rho_{\rm M}(z) \delta 
\label{eq-growthw}
\end{equation}
in a universe with negligible radiation density.  \citet{ws98} derive a useful approximation
to the growth function by defining $\alpha_w$ such that 
\begin{equation}
  \frac {d \ln \delta} {d \ln a} = [\Omat(z)]^{\alpha_w}   \; \; .
\end{equation}
For a slowly varying equation of state ($ | dw/d\Omat(z) | \ll [1- \Omat(z)]^{-1}$), they find
that
\begin{equation}
  \alpha_w = \frac {3} {5 - w/(1-w)} + \frac {3} {125} \frac {(1-w)(1-3w/2)} {{(1-6w/5)}^3} [1-\Omat(z)] \; \;
\end{equation}
to lowest order in $1-\Omat(z)$.   Using this expression for $\alpha_w$ in the integral
\begin{equation}
  D(a) \approx a \exp \left( \int_a^1 \left\{ 1 - [\Omat(z)]^{\alpha_w} \right\}  \frac {da} {a} \right) \; \; 
\end{equation}
reproduces the growth function obtained from numerical integration of equation (\ref{eq-growth})
to better than 1\% for $\Omat(z) > 0.2$.

These growth functions are valid only as long as pressure gradients do not alter the dynamics 
of the perturbation.  Pressure effects are not an issue when the scale of a perturbation is larger 
than the Hubble length $c H^{-1}$.  In that regime the growth functions found by solving 
equations  (\ref{eq-growth}) and (\ref{eq-growthw}) remain valid.  Yet, as the universe ages, 
it encompasses perturbations of increasingly larger scale and additional physical effects 
enter the picture.

The bad news is that a variety of processes alter the scale-free nature of the
original perturbation spectrum.  The good news is that the imprint of these processes 
on $P(k)$ can tell us a great deal about the contents and dynamics of the universe.
During the radiation-dominated era of the universe ($z > z_{\rm eq}$), pressure effects begin 
to alter the growth of a given mode when its wavelength 
is finally contained within the horizon length $\sim cH^{-1}$.
Then radiation pressure can effectively resist gravitational compression, inhibiting
further growth of modes at that wavelength.  Instead, these modes in 
the coupled photon-baryon fluid begin to oscillate as acoustic waves, 
and eventually damp owing to photon diffusion out of higher-density, 
higher-temperature regions.  Perturbation growth in the dark-matter 
component therefore stalls near the amplitude at which the perturbations 
were first contained within the horizon because the gravitationally dominant
photon component no longer spurs mode growth.  These perturbations then
resume growing at $z_{\rm eq}$,  when matter begins to dominate the dynamics.  
The transition from radiation domination to matter domination therefore imprints a 
bend in $P(k)$ on a length scale corresponding to the horizon scale at $z_{\rm eq}$.  

Perturbation growth is scale-independent during the matter-dominated era only insofar 
as the matter can be considered cold on the scale of the perturbation.  If the characteristic 
velocities of the matter particles are not small compared to the escape velocity from 
the perturbation, then both pressure forces and particles streaming out of denser regions
can damp small-scale perturbations.  Each effect of this type imprints its own
characteristic feature on $P(k)$. 

All of these scale-imprinting effects that alter $P(k)$ from the time the primordial
power spectrum is created until the present day are typically subsumed into a single 
quantity known as the transfer function, defined to be
\begin{equation}
  T(k) \equiv \frac {\delta_k (z=0)} {\delta_k (z) D(z)} \; \; ,
\end{equation}
where the symbol $k$ refers to comoving modes with wavenumber 
$(1+z)k$ in physical space, a convention implicit throughout this review.  
The redshift $z$ in this definition is assumed to be large enough that $\delta_k(z)$ 
reflects the original power spectrum imprinted by inflation or some other process.  
The transfer function therefore represents all the alterations of the original power 
spectrum that subsequently occur, except for those involving mode growth in the
non-linear regime.  If the primordial spectrum is a power law of index 
$n_p \approx 1$, then the power spectrum of linear perturbations at $z=0$ 
is $P(k) \propto k^{n_p} T^2(k)$.   

\subsubsection{The CDM Power Spectrum}
\label{sec:cdm}

The most successful models for the formation of large-scale structures like
clusters of galaxies assume that cold dark matter (CDM) is responsible.
Particles that interact only through gravity exert negligible pressure, and if 
their random velocities are small then they will not be able to 
escape from incipient potential wells on the scales of interest.  That is, they
will be too ``cold'' to damp the relevant perturbations by freely streaming out
of them.  Thus, the transfer function for a universe containing only radiation
and cold dark matter has just one feature, corresponding to the wavenumber
of the mode that enters the horizon at the matter-radiation equality 
redshift $z_{\rm eq}$, with a comoving size $l_{\rm eq} \sim c H_0^{-1} 
(\Omat z_{\rm eq})^{-1/2} \sim 20 \, (\Omat h^2)^{-1} \, {\rm Mpc}$.

Growth of modes with smaller comoving wavelengths temporarily 
stalls from the redshift at which they enter the horizon until $z_{\rm eq}$.
Because radiation dominates the universe during this time interval,
the comoving size of the horizon scales as $a$ while the
growth function scales as $a^2$.  Short-wavelength perturbations
therefore miss out on a growth factor $\sim (k l_{\rm eq})^2$, corresponding
to the square of the change in scale factor from the time a perturbation
enters the horizon to the time of matter-radiation equality.  Growth of
long-wavelength modes, on the other hand, does not stall at all.
The behavior of the CDM transfer function in the two extremes is 
$T(k) \approx 1$ for $k \ll l_{\rm eq}^{-1}$ and $T(k) \approx (k l_{\rm eq})^{-2}$
for $k \gg l_{\rm eq}^{-1}$.  For $n_p = 1$, these scalings translate to  $\delta M/M \sim
M^{-2/3}$ on large scales and $\delta M/M \sim {\rm const.}$ on small scales,
meaning that structure formation in a CDM universe is
hierarchical, with small-scale perturbations reaching the non-linear 
regime before larger-scale ones.

Numerical computations are needed to derive the exact CDM transfer 
function, but many authors have provided useful analytical fits to those 
numerical results.  One such expression is
\begin{eqnarray}
 T(k) & = & \frac {\ln (1+2.34q)} {2.34q} [ 1 + 3.89q + (16.1q)^2 \nonumber \\
      ~ & ~ & \; \; \; \; \; \; \; \; \; \; \; \; \; + (5.46q)^3 + (6.71q)^4 ]^{-1/4} \; \; ,
\end{eqnarray}
with $q = k \, (\Omat h^2)^{-1} \, {\rm Mpc}$ \citep{bbks86}.
Allowing for trace populations of baryons and massive neutrinos alters 
the CDM power spectrum in minor but interesting ways.  For example,
a small proportion of baryons lowers the apparent dark-matter density
parameter, causing a shape-preserving shift in the CDM transfer function
\citep{pd94}. This shift can be reproduced by setting $q =
k \, (\Gamma h)^{-1} \, {\rm Mpc}$, so that it includes a shape parameter $\Gamma = 
\Omat h \exp [ - \Obar (1+\sqrt{2h}/\Omat)]$ \citep{Sugi95}. 
Fitting formulae accomodating additional modifications owing
to baryons and massive neutrinos can be found in \citet{eh98, eh99}.

\subsubsection{Power Spectrum Normalization}
\label{sec:pknorm}

The preceding sections give the theoretical expectations for the shape and growth 
rate of the density perturbation spectrum but do not specify its normalization.
Because inflationary theories do not make firm predictions about the
amplitude of the primordial power spectrum, the normalization of 
$P(k)$ must be determined observationally.  For example, measurements of the 
present-day mass distribution of the universe indicate that $\delta M / 
M \approx 1$ within comoving spheres of radius $8 \, h^{-1} \, {\rm Mpc}$ 
(\S~\ref{sec:obsmassfcn}), as suggested by early galaxy surveys showing 
that the variance in galaxy counts was of order unity on this length scale \citep{dp83}.  

This feature of the universe is the motivation for expressing the power-spectrum
normalization in terms of the quantity $\sigma_8$, where
\begin{equation}
 \sigma_8^2 = \frac {1} {(2 \pi)^3} \int P(k) |W_k|^2 d^3k \; \; 
\end{equation} 
is the variance defined with respect to a top-hat window function $W(r)$ having a 
constant value inside a comoving radius of $8 \, h^{-1} \, {\rm Mpc}$ and vanishing 
outside this radius.   When using this formula, one must keep in mind that $P(k)$ 
refers to the power spectrum of linear perturbations evolved to $z=0$ according to the 
growth function $D(z)$, which is valid only for small perturbations.  
There are other ways of characterizing the power-spectrum normalization, but
$\sigma_8$ is the most widely-used parameter.

\subsubsection{Summary of Cosmological Parameters}
\label{sec:cospars}

At the beginning of this recipe, we promised to encapsulate the overall
cosmological model in two small sets of parameters.  The set governing
the global behavior of the universe consists of $H_0$, $\Omat$, $\Obar$,
$\Orad$, $\Olam$, and $w$.  The set governing the initial density 
perturbation spectrum consists of $\sigma_8$ and $n_p$.  The shape
parameter $\Gamma$ is not a free parameter in standard cold dark matter
models but is sometimes treated as a free parameter in order to test
variants of the standard model. 

\begin{figure}
\includegraphics[width=3.2in ] 
{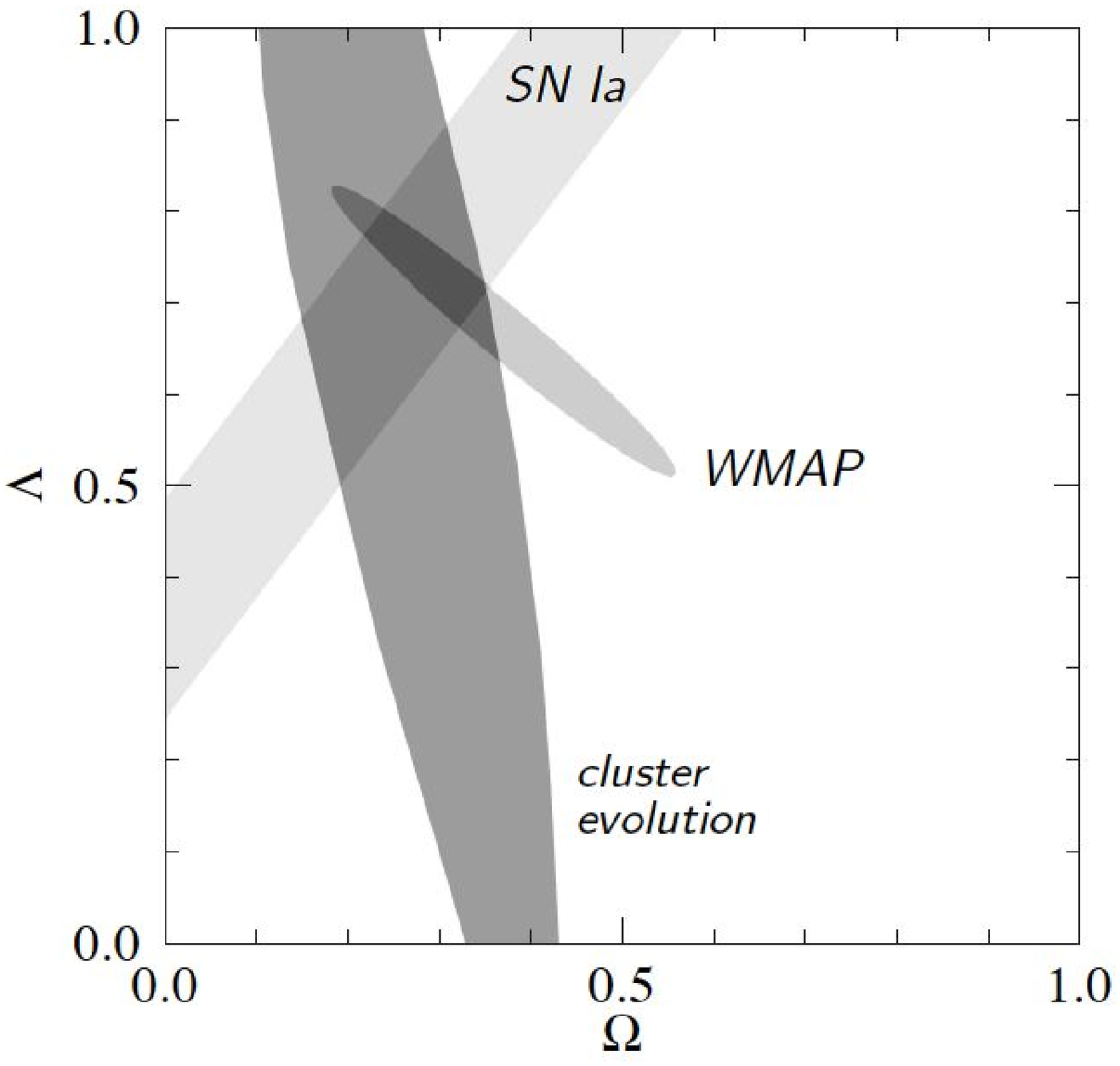}
\caption{Cosmological constraints from cluster evolution \citep{Vikh03}, supernovae
\citep{Riess98, Perl99}, and WMAP observations of the cosmic microwave background
\citep{WMAP03}.  The horizontal axis labeled $\Omega$ gives the value of $\Omat$;
the vertical axis labeled $\Lambda$ gives the value of $\Olam$.  These particular
contraints from cluster evolution are based on the baryonic mass function of clusters
(\S~\ref{sec:omat}), but other measures of cluster evolution give similar results.  The complementarity
of these constraints is evident from the figure, and the common region of overlap near
$\Omat = 0.3$ and $\Olam = 0.7$ is reassuring evidence of consistency in the overall
picture.  (Figure courtesy of Alexey Vikhlinin.)}
\label{fig:constraints_Vikh}
\end{figure}

In the concordance model, also known as the $\Lambda$CDM model, 
to denote cold dark matter with a cosmological constant, these parameters
are all assigned values close to the most likely values implied by observations:
\begin{itemize}

\item {\em Hubble's constant.}  The consensus value of this parameter, measured
          primarily from the expansion rate of the local universe is $H_0 = 71 \pm 7
          \, {\rm km \, s^{-1} \, Mpc^{-1}}$ \citep{H0KP_01}. 
           
\item {\em Matter density.}  Several different methods involving clusters indicate
          that $\Omat \approx 0.3$ (\S~\ref{sec:obsmassfcn}, \S~\ref{sec:omat}).  
          Combining the results of distant 
          supernova observations and observations of temperature patterns in the 
          microwave background gives a similar value for this parameter.  
          Figure~\ref{fig:constraints_Vikh} shows one example of these mutual constraints 
          in the $\Omat$-$\Olam$ plane.

\item {\em Baryon density.}  The abundances of light elements formed during
          primordial nucleosynthesis indicate that $\Obar = 0.02 h^{-2}$, equal to
          $\Obar = 0.04$ for the value of Hubble's constant given above \citep[e.g.,][]{bnt01}.  
          This value is consistent with the baryon density inferred from the fluctuations
          in the cosmic microwave background \citep[e.g.,][]{WMAP03}.
                     
\item {\em Radiation density.}  The energy density $\Orad$ in electromagnetic 
          radiation is simply calculated from the microwave background temperature
          $T_{\rm CMB} = 2.728 \pm 0.004$ \citep{Fixsen96} and Hubble's constant.
          Neutrinos may also contribute to the energy density in relativistic matter, if
          their masses are sufficiently small, but this contribution is currently too small
          to affect the global dynamics.
           
\item {\em Dark energy density.}  Observations of distant supernovae imply that
          the expansion of the universe is accelerating at a rate consistent with a
          constant dark-energy density corresponding to 
          $\Olam \approx 0.7$ \citep{Riess98, Perl99, Riess04}.
          Combining the matter density inferred from clusters with the flat geometry 
          inferred from temperature patterns in the microwave background 
          corroborates this result \citep[e.g.,][see also Figure~\ref{fig:constraints_Vikh}]{bops99}.
                     
\item {\em Dark energy equation of state.}  Observations of microwave
          background patterns, when combined with observations of large-scale 
          structure are consistent with Einstein's cosmological constant ($w = -1.0$)
          but not with $w \gtrsim -0.8$ \citep{WMAP03}.   Alternatively, combining
          cluster surveys with observations of distant supernovae leads to similar
          constraints, $w = -0.95_{-0.35}^{+0.30}$ \citep{Schuecker03w}.   
          However, theoretical arguments suggest that the parameter $w$ may 
          be redshift dependent \citep{pr03}.
           
\item {\em Normalization of density perturbations.}  The cluster observations
          discussed in \S~\ref{sec:obsmassfcn} and \S~\ref{sec:omat} indicate 
          that the power-spectrum normalization falls
          into the range $\sigma_8 \approx 0.7 - 1.0$.  This range is consistent with
          both structures in the cosmic microwave background and other observations
          of large-scale structure.
           
\item {\em Slope of primordial perturbation spectrum.}  All available information
          indicates that $n_p \approx 1$.  Constraints from observations of the microwave
          background, when combined with optical observations of large-scale structure,
          give $n_p = 0.97 \pm 0.03$ \citep{WMAP03}. 
                     
\item {\em Shape parameter of perturbation spectrum.}  The concordance values
          of $\Omat$, $\Obar$, and $H_0$ given above imply $\Gamma \approx 0.2$, 
          which agrees with the value of $\Gamma$ derived from observations of
          large-scale structure \citep[e.g.,][]{pd94, Schuecker01, Szalay03}, 
          an important element of self-consistency in the concordance model.
           
\end{itemize}

Several other closely related models have been pursued during the last two decades,
but none of them have proven as successful in explaining such a large number of observations.
Here are a few variants some of which will be discussed later in connection with the 
parameter constraints derived from cluster surveys:

\begin{itemize}

\item {\em Standard cold dark matter} (SCDM).  In this model, the universe is assumed to be
          flat, with no dark energy, so $\Omat = 1$ and $\Olam = 0$.  For this value of the matter
          density, measurements of large-scale structure, including clusters, imply that $\sigma_8
          = 0.4-0.5$ (\S~\ref{sec:obsmassfcn}).  However, the shape 
          parameter implied for this model, given the observed Hubble constant, is 
          $\Gamma \approx 0.7$, which conflicts with observations of large-scale
          structure \citep[e.g.,][]{Szalay03}.  Also, this value of $\Omat$ leads to a 
          baryon-to-dark-matter ratio $f_b = \Obar/\Omat$ that is inconsistent 
          with cluster observations (\S~\ref{sec:mbaryon}).
           
\item {\em Tilted cold dark matter.}  One way to make models with $\Omat = 1$ more consistent
         with large-scale structure observations is to assume that the primordial perturbation
         spectrum is ``tilted'' so that it is significantly shallower than $n_p = 1$ \citep[e.g.,][]{cgko92}.  	
         However, such models conflict with the strong constraints on $n_p$ inferred
         from microwave background observations.
         
\item {\em Ad hoc power spectrum} ($\tau$CDM).  Another option for making $\Omat = 1$ 
          models more consistent with observations is to arbitrarily adjust the shape of the 
          perturbation spectrum to fit the observations.  One such realization 
          is the $\tau$CDM model \citep{Jenk98}, which sets $\Omat = 1$, $\sigma_8 = 0.5$
          and $\Gamma = 0.2$, even though there is little physical justification for having 
          such a low value of $\Gamma$ in a cosmology with such a large matter density
          \citep[but see][]{wgs95}.
          
\item {\em Open cold dark matter (OCDM).}  There is also the option of accepting the
         evidence that $\Omat \approx 0.3$ but dispensing with dark energy ($\Olam = 0$)
         so that the universe has an open geometry.  In this case, the perturbation spectrum
         can be identical to the one in the $\Lambda$CDM concordance model, but the
         growth rate of those perturbations differs because of the changed expansion rate
         at late times.
                   
\end{itemize}
 
This review does not consider models involving forms of dark matter other than
cold dark matter, but it does consider models with generalized forms of dark energy 
having $w \neq -1$ ($w$CDM).  Current cluster surveys are not yet large enough to place
strong constraints on this equation-of-state parameter but they might provide
much stronger constraints in the coming decade (\S~\ref{sec:olam}).

\subsection{Cluster Formation}
\label{sec:clustform}

Cluster formation from perturbations in the density distribution of cold dark matter
is a hierarchical process.  Small subclumps of matter are the first pieces of the cluster to 
deviate from the Hubble flow and undergo gravitational relaxation 
because the density perturbations have larger amplitudes
on smaller mass scales, These small pieces
then merge and coalesce to form progressively larger structures as perturbations
on larger mass scales reach the non-linear regime.  A full understanding of the
details of this hierarchical merging process requires numerical simulations,
but simplified, spherically symmetric models of cluster formation illustrate
many of the important concepts.  This part of the review shows how a cluster would 
grow from a spherically symmetric mass perturbation and then refines the
details of that simplified approach, based on what we have learned from 
numerical simulations. 

\subsubsection{Spherical Collapse}
\label{sec:sphercoll}

The most basic features of cluster formation can be understood in terms
of a spherically symmetric collapse model \citep[e.g.,][]{gg72, fg84, Bert85}.  
In such a model, the matter that
goes on to form a cluster begins as a low-amplitude density perturbation that
initially expands along with the rest of the universe.   The perturbation's
gravitational pull gradually slows the expansion of that matter, eventually
halting and reversing the expansion.  A cluster of matter then forms at the
center of the perturbation, and the rate at which additional matter accretes
onto the cluster depends on the distribution of density with radius in the
initial perturbation.

In a geometry that is perfectly spherically symmetric, the behavior of an
individual mass shell in the presence of a homogeneous generalized dark-energy field
follows the equation of motion
\begin{equation}
  \ddot{r}_{\rm sh} = - \frac {G \Msh} {\rsh^2} - \frac {1+3w} {2} \Olam H_0^2 (1+z)^{3(1+w)} \rsh \; \; ,
\label{eq-rsh}
\end{equation}
where $\rsh$ is the shell radius and $\Msh$ is the mass enclosed within $\rsh$.
Throughout the early evolution of a spherical perturbation, the value of $\Msh$ within
a given mass shell remains constant.  Thus, if the dark-energy term is negligible,
the radius of a mass shell obeys the parametric solution $\rsh = \rta [(1- \cos \theta_M)/2]$,
$t = \tvir [(\theta_M - \sin \theta_M)/2 \pi]$, with a turnaround radius 
$\rta = [(2G \Msh \tvir^2)/\pi^2]^{1/3}$ for a shell that collapses to the origin 
at time $\tvir$.  The solution for $\Olam \approx 0.7$ and $w \approx -1$ is not 
much different because the dark-energy term remains $\lesssim 15$\% of the 
matter term during the trajectory of all shells that collapse to the origin by
the present time.  If greater accuracy is needed, a shell's trajectory can be 
computed numerically from equation (\ref{eq-rsh}).\footnote{Here we are 
making the standard assumption that the collapsing 
dark matter has no effect on the local dark-energy density \citep[e.g.,][]{ws98, wk03}.
If in fact the dark-matter collapse alters the local properties of dark energy, 
the dynamics could be somewhat altered \citep{mv04}.}

Once a shell collapses, the mass within it no longer remains
constant.  Because the dark matter within a collapsing shell is collisionless,
shells on different trajectories can easily interpenetrate.  The radii 
of collapsed shells in this idealized geometry therefore oscillate symmetrically
about the origin, and the amplitudes of these oscillations modestly decrease
with time as mass associated with other collapsed shells accumulates within
the oscillations' turning points \citep{Gunn77}.

The accretion process in real clusters is not so symmetric.  Instead, gravitational
forces between infalling clumps of matter produce a time-varying gravitational potential
that randomizes the velocities of the infalling particles, yielding a Maxwellian velocity
distribution in which temperature is proportional to the particle mass. 
This process, known as ``violent relaxation" \citep{LB67}, leads to a state of 
virial equilibrium in which the total kinetic energy $E_K$ is related to the 
total gravitational potential energy $E_G$ through the equation
\begin{equation}
 E_G + 2 E_K = 4 \pi \Pb \rb^3
\end{equation}
where $\Pb$ is the effective pressure owing to infalling matter at the boundary 
$\rb$ of the collapsed system (\S~\ref{sec:vdisp}).  Setting $\Pb$ to zero yields the usual form of
the virial theorem for gravitationally bound systems.

A common toy model for estimating the location of a cluster's outer 
boundary is the spherical top-hat model, which assumes that the perturbation leading 
to a cluster is a spherical region of constant density.  All of the mass shells 
in such a perturbation move in unison and collapse to the origin simultaneously. 
The virial theorem therefore suggests that the bounding radius of the cluster 
after it collapses and relaxes should be in the neighborhood of half the turnaround 
radius.  Numerical simulations indeed show that particle velocities within 
this radius are generally isotropic and those outside this radius are generally infalling, 
but the boundary between the isotropic and infalling regions is not particularly 
distinct \citep{Evrard90, nw93}.   

The spherical top-hat model has actually led to several different definitions for the 
virial radius of a cluster.  If one assumes that all the mass in the original top-hat perturbation
ends up within $\rta / 2$, then the mass density in that region is  $6M/\pi \rta^3$.
In a matter-dominated universe with zero dark energy, this density is equal
to $\Delvir = 8\pi^2 / (Ht)^2$ times the critical density $\rhocr \equiv 3 H^2 / 8 \pi G$.
Thus, for a flat, matter-dominated universe in which $Ht = 2/3$, the mean density 
of a perturbation that has just collapsed is taken to be $18 \pi^2 \approx 178$
times the critical density.  A useful approximation for $\Delvir$ in a flat universe
with a non-zero cosmological constant ($w=-1$) is
\begin{equation}
 \Delvir = 18 \pi^2 + 82 \, [\Omat(z) - 1] - 39 \, [\Omat(z) -1]^2
\end{equation}
\citep{bn98}. Because the outer radius of a real cluster is not distinct, 
one pragmatic definition of the virial radius is then the radius $\rvir$ within which 
the mean matter density is $\Delvir \rhocr$ \citep{ecf96}.  However, the numerical value
of $\Delvir$ in a flat, matter-dominated universe has inspired other definitions.
A common alternative is the scale radius $r_{200}$, within which the 
mean matter density is $200 \rhocr$.  Another frequently used scale radius is 
$r_{\rm 180m}$, within which the mean matter density is 180 times the mean background 
density $\Omat (z) \rhocr$.  As long as $\Omat(z) \approx 1$, both of these scale 
radii are nearly identical to $\rvir$, but because $\Omat \approx 0.3$ at the 
present time, these radii are now somewhat different, with $r_{200} < \rvir < r_{\rm 180m}$.  
This multiplicity of definitions for the radius of a cluster is a potential source of 
confusion, but as we will see below, each of these scale radii can be particularly
well suited to certain applications.  

\subsubsection{Cluster Mass Profiles}
\label{sec:massprof}

Observations of galaxy clusters have long indicated that the velocity dispersion
of a cluster's galaxies remains relatively constant with distance from
the cluster's center, implying an underlying mass-density profile 
$\rho_{\rm M} (r) \propto r^{-2}$.  The simplest analytical cluster model 
consistent with such a density profile is the singular isothermal sphere, 
in which the velocity dispersion $\sigma_v$ is constant and isotropic 
at every point and $\rho_{\rm M}(r) = \sigma_v^2 / 2 \pi G r^2$
\citep[e.g.,][]{bt87}.  This model is useful for making 
analytical estimates of cluster properties, but it is incomplete because 
the total mass diverges linearly with radius.  

Numerical simulations of cluster formation produce dark-matter halos 
whose density profiles are shallower than isothermal at small radii and
steeper than isothermal at large radii.  A generic form for representing
these profiles is
\begin{equation}
 \rho_{\rm M}(r) \propto r^{-p} (r + r_s)^{p-q} \; \; ,
\end{equation}
where the parameters $p$ and $q$ describe the inner and outer power-law
slopes and the radius $r_s$ specifies where the profile steepens.
Groups that have fit such profiles to simulated clusters disagree about the
best values of $p$ and $q$ but typically find $1 \lesssim p \lesssim 1.5$
and $2.5 \lesssim q \lesssim 3$.  Specific examples include the NFW
profile, with $p=1$ and $q=3$ \citep{nfw97}, the
Moore profile, with $p=1.5$ and $q=3$, and the \citet{rtm03} 
profile, with $p=1$ and $q=2.5$.  Both optical and X-ray observations 
indicate that density profiles of this sort are good representations of 
the underlying mass profiles of clusters, at least outside of the innermost
regions \citep{CNOC_mass_Profile, pa02, lbs03}.  
Observing the asymptotic inner slope $p$ is currently 
a matter of great observational interest, as the cuspiness of dark-matter 
density profiles at $r=0$ is one of the acid tests of the CDM paradigm 
for structure formation \citep[see][and references therein]{Navarro03}.  
However, we will not discuss that issue here because the global properties 
of clusters depend little on the value of $p$.  In this review, we 
will use the NFW profile when necessary because it remains thee
most widely used fitting formula for representing the results of cosmological
simulations.

The transition of the density profile from shallow to steep can also
be expressed in terms of a concentration parameter $c = \rb/r_s$,
which expresses the bounding radius of the cluster in units of $r_s$.  
Because the concentration parameter depends on 
$\rb$, numerical values of $c$ depend somewhat on whether the
bounding radius is taken to be $\rvir$, $r_{200}$, $r_{\rm 180m}$, or
something similar.  However, these radii are not vastly different
because they are generally several times larger than $r_s$, meaning that the
enclosed mass is not rapidly diverging in the neighborhood of the
virial radius.  Typical concentration parameters for simulated clusters
are in the range $c \sim 4 - 10$, with a scatter in $\ln c$ of $0.2$-$0.35$ \citep{Jing00}.
Also, lower-mass objects tend to have higher halo concentrations
because they formed earlier in time, when the overall density of the
universe was greater \citep{nfw97, Bullock01, ens01}. 

\subsubsection{Defining Cluster Mass}
\label{sec:massdef}

Even with these more sophisticated forms for the density profile, mass 
still diverges with radius.  Thus, a cluster's mass and all the relations
linking that mass to other observable quantities depend on how one 
chooses to define a cluster's outer boundary.  One would like to define
that boundary so as to maximize the simplicity of the relationships 
between cluster mass and other observables, but no single definition 
is best for all applications.

The easiest way to link observations to theoretical models is through
definitions taking the mass of a cluster to be $M_\Delta$, the amount
of matter contained in a spherical region of radius $r_\Delta$ 
whose mean density is $\Delta \cdot \rhocr$.  It is also common for cluster mass 
to be defined with respect to the background mass density, so that the
mean density of matter within the virial radius is $\Delta \cdot \Omat(z) \rhocr$,
but applying this definition to observations requires prior knowledge of $\Omat$.
Spherical top-hat collapse suggests that $\Delvir$ is a good
choice for the density threshold.  However, observers often prefer to
raise that threshold to $\Delta = 200$ or even $\Delta = 500$ for two 
reasons.  The properties of a cluster are easier to observe in regions 
where the density contrast is higher, and simulations show that the 
region within $r_{500}$ is considerably more relaxed than the region 
within $\rvir$.  

As an example of such definitions in action, consider the relation between 
velocity dispersion and the virial mass $\Mvir$ obtained by truncating 
a singular isothermal sphere at the virial radius $\rvir$:
\begin{equation}
 M_v = f_\sigma \frac {4 \sigma_v^3} {GH \Delvir^{1/2}} \; \; ,
\label{eq:mvsigma}
\end{equation}
where the factor $f_\sigma$ is a parameter that can be adjusted to
account for the fact that clusters are not perfect isothermal spheres
\citep{Evrard89, ecf96, bn98}.   The presence of this parameter is
a reminder that the derivation of this relation should not be taken too 
literally.  Truncation of the mass distribution at $r_\Delta$ formally implies
a non-zero boundary pressure that shifts the virial
relation for this configuration so that $E_K =  - 3 E_G / 4$, which is 
inconsistent with the definition of $\rvir$\citep{v00}.  This functional
form for the $\Mvir$-$\sigma_v$ is useful primarily as a fitting formula
that accounts for most of the cosmology-dependent changes in the
normalization of the relation.  However, because the density profiles 
of dark matter halos defined in this way depend on both mass and 
redshift, the correction factor $f_\sigma$ is not a universal constant.

Recent work has shown that defining a cluster's mass using the
threshold $\Delta = 200$ leads to an $M_{200}$-$\sigma_v$
relation that is remarkably independent of cosmology.  \citet{Evrard04}
finds that the relation
\begin{equation}
 M_{200} = \frac {10^{15} \, h^{-1} \, M_\odot} {H/H_0} 
                      \left( \frac {\sigma_v} {1080 \, {\rm km \, s^{-1}}} \right)^3  
\end{equation}
is an excellent fit to a wide range of simulated clusters sampled
over a wide range of redshifts.  This relation
is equivalent to setting $f_\sigma = 1.2$ and $\Delvir = 200$ in
equation ({\ref{eq:mvsigma}).

Conversions between $\Mvir$ and $M_\Delta$ defined with respect 
to an arbitrary $\Delta$ are straightforward as long as a cluster's 
concentration parameter is known.  From the definitions 
of these masses, we have
\begin{equation}
 \frac {M_\Delta} {\Mvir} = \frac {\Delta} {\Delvir} \left( \frac {r_\Delta} {\rvir} \right)^3 \; \; ,
\label{eq:mconv}
\end{equation}
and the halo concentration gives the relationship between $r_\Delta$ and
$\rvir$.  \citet{hk03} have provided a useful approximation for this
relation in the case of an NFW profile.    Recasting their formulae in slightly
different notation, one can write the halo concentration $c_\Delta$ defined
with respect to $r_\Delta$ in terms of the concentration $\cvir$ defined with
respect to $\rvir$:
\begin{equation}
 c_\Delta = \frac {1} {\left[ a_1 f_c^{2p_c} + \left( \frac {3} {4} \right)^2 \right]^{-1/2} + 2f_c} \; \; ,
\end{equation}
with $f_c = (\Delta/\cvir^3 \Delvir) [\ln (1+\cvir) - \cvir / (1 + \cvir ) ]$, 
$p_c = a_2 + a_3 \ln f_c + a_4 (\ln f_c)^2$ and $(a_1, ..., a_4)$ = ($0.5116$, $-0.4283$,
$-3.13 \times 10^{-3}$, $-3.52 \times 10^{-5})$.  Plugging the ratio $r_\Delta/\rvir = c_\Delta/\cvir$
given by this approximation into equation (\ref{eq:mconv}) converts cluster masses with an
accuracy $\sim 0.3$\% for the halo concentrations typical of clusters.
                                          
Some other definitions of cluster mass are useful in certain contexts but are more
difficult to relate to the top-hat collapse model.  For example, observers who measure
cluster mass using gravitational lensing or the total optical luminosity are essentially
measuring cluster mass within a cylinder along the line of sight 
rather than a sphere.  In principle, these
observables can be linked with cluster masses defined with respect to a cylindrical
boundary, but the relationships between those cylindrical masses and models of 
structure formation are not as well understood as their spherical counterparts.
On the theoretical side, the masses of clusters identified in numerical simulations
are sometimes defined using a ``friends-of-friends" algorithm that links neighboring
mass particles \citep[e.g.,][]{defw85}.  However, clusters defined in this way often
have irregular boundaries \citep{White01}, making this sort of definition difficult to 
apply to observations.  Masses defined within spheres also have their shortcomings, 
particularly in cases where two clusters are just beginning to merge, but in general provide
the most direct link between cosmological models and observations.
  
\subsubsection{Cluster Mass Function}
\label{sec:massfcn}

Some of the most powerful constraints on current cosmological models come
from observations of how clusters evolve with time.  Because cosmological
time scales are so long, we cannot observe how individual clusters evolve but
rather observe how the demographics of the entire cluster population changes
with redshift.  A important conceptual tool in this effort is the cluster mass function,
$n_M(M)$ which gives the number density of clusters with mass greater than 
$M$ in a comoving volume element.   Notice that the cluster mass function inevitably 
depends on how one defines cluster mass.

Combining spherical top-hat collapse with the growth function for linear
perturbations has led to a widely used semi-analytical method for expressing
the cluster mass function in terms of cosmological parameters.  \citet{ps74} 
pioneered the basic approach, which was refined and
extended by \citet{bcek91}, \citet{Bower91}, and \citet{lc93}. 
This class of models simplifies the problem of structure formation by assuming 
that all density perturbations continue to grow according to the linear growth 
rate $D(z)$ even when their amplitudes become non-linear.  When perturbations
are treated in this way, their variance on mass scale $M$ as a function of redshift is
\begin{equation}
 \sigma^2(M,z) = \frac {D^2(z)} {(2\pi)^3} \int P(k) | W_k(M) |^2 d^3 k \; \; ,
\end{equation}
where $W_k(M) = 3 ( \sin kr_M - kr_M \cos kr_M)/(kr_M)^3$ with 
$r_M = (3 M / 4 \pi \Omat \rhocro)^{1/3}$ is the Fourier-space representation 
of a top-hat window function that encloses mass $M$.
The normalization of $P(k)$ is set so that $\sigma (M_8,0) = \sigma_8$ for
$M_8 \equiv (8 \, h^{-1} \, {\rm Mpc})^3 H_0^2 \Omat / 2G 
= 6.0 \times 10^{14} \, \Omat h^{-1} \, M_\odot$.   
These perturbations are then assumed to collapse and virialize when their 
density contrast $\delta = \delta \rho / \rho$ exceeds a critical threshold $\delta_c$.

Suppose the initial density perturbations are gaussian with a variance 
$\sigma^2(M,z)$ that declines monotonically with mass.  Then according
to the Press-Schechter approach, the probability that a region of mass $M$ 
exceeds the collapse threshold at redshift $z$ is $\erfc [ \delta_c / \sqrt{2} \sigma(M,z) ]$,
where $\erfc (x)$ is the complementary error function. 
Implicit in this expression is the notion that all the mass in the
universe belongs to collapsed, virialized objects when viewed on
sufficiently small mass scales.  It then follows that the cluster 
mass function on scale $M$ at redshift $z$ is
\begin{equation}
 n_M(M,z) = \frac {\Omat \rhocro} {M} \; \erfc 
                          \left[ \frac {\delta_c} {\sqrt{2} \sigma(M,z)} \right] \; \; .
\end{equation}
This expression implies that the shape of the mass function depends
only on $\sigma(M,z)$ and remains invariant with respect to the characteristic
collapsing mass scale $M_*(z)$ at which $\sigma(M_*,z) = \delta_c$.  
Observers often work with the mass function in a differential form,
such as $d n_M / d \ln M$, but theorists prefer expressing the 
differential form in terms of the shape-governing function $\sigma(M,z)$.  
Then the differential mass function takes the form 
\begin{equation}
 \frac {dn_M} {d \ln \sigma^{-1}} = \sqrt {\frac {2} {\pi}} 
 					\; \frac {\Omat \rhocro} {M}
					\; \frac {\delta_c} {\sigma} 
                                            \; \exp \left[ - \frac {\delta_c^2} {2 \sigma^2} \right] \; \; .
   \label{eq:mfunc_ps}
\end{equation} 
Both of these forms for the mass function can be straightforwardly
extended to cases in which the perturbations are non-gaussian 
\citep[e.g.,][]{rgs00}.

The value of the critical threshold $\delta_c$ was originally inferred
from spherical top-hat collapse.  Expanding the parametric solution
for spherical collapse in powers of $\theta_M$ leads to the following
relation at early times:
\begin{equation}
  r_{\rm sh} \approx \left( \frac {9GMt^2} {2} \right)^{1/3} 
                 \left[ 1 - \frac {1} {20} \left( \frac {12 \pi t} {\tvir} \right)^{2/3} \right] \; \; .
    \label{eq:lingr}
\end{equation}
The leading term in this expression characterizes the behavior of a
critical-density sphere and the second term describes how the evolution
of a slightly overdense sphere deviates from that of a critical-density
sphere.  Assuming that this deviation grows according to equation (\ref{eq:lingr}) 
until the moment of collapse and virialization ($t=\tvir$) gives the value
of the critical threshold in a flat universe with $\Omat = 1$:
$\delta_c = 3(12\pi)^{2/3}/20 \approx 1.686$.  
Generalizing this treatment to cases where $\Omat \neq 1$ produces 
only minor differences in $\delta_c$ for interesting values of the 
cosmological parameters \citep{lc93, ecf96}. 

The preceding derivation of the cluster mass function is not terribly rigorous, 
but it is useful because adopting $\delta_c = 1.686$ leads to 
mass functions that agree reasonably well
with those derived from numerical simulations.  Treating 
perturbation collapse as ellipsoidal rather than spherical improves that
agreement \citep{smt01}. \citet{st99} have shown that the expression
\begin{eqnarray}
 \frac {dn_M} {d \ln \sigma^{-1}} & =  & A_{\rm s} \sqrt {\frac {2a_{\rm s}} {\pi}} 
 			            \; \left[ 1 + \left( \frac {\sigma^2} {a_{\rm s} \delta_c^2} \right)^{p_{\rm s}} \right] 
 					 \; \; \; \; \; \; \; \; \; \nonumber \\ 
 ~ & ~ & \; \; \; \; \; \;         \times \;   \frac {\Omat \rhocro} {M}
 				\frac {\delta_c} {\sigma} 
 			               \; \exp \left[ - \frac {a_{\rm s} \delta_c^2} {2 \sigma^2} \right] \; \; ,
\label{eq:mfunc_st}
\end{eqnarray} 
with $A_{\rm s} = 0.3222$, $a_{\rm s} = 0.707$, and $p_{\rm s} = 0.3$
is quite an accurate representation of the mass functions from several
different numerical simulations.  However, because semi-analytical mass 
functions like these are not rigorously derived, they are essentially just 
fitting formulae that conveniently express the simulation results and should
treated cautiously outside the cosmological models against which they have been
tested.  

\begin{figure*}
\includegraphics[width=6.5in , trim = 0.1in 2.2in 0.1in 0.1in , clip]
{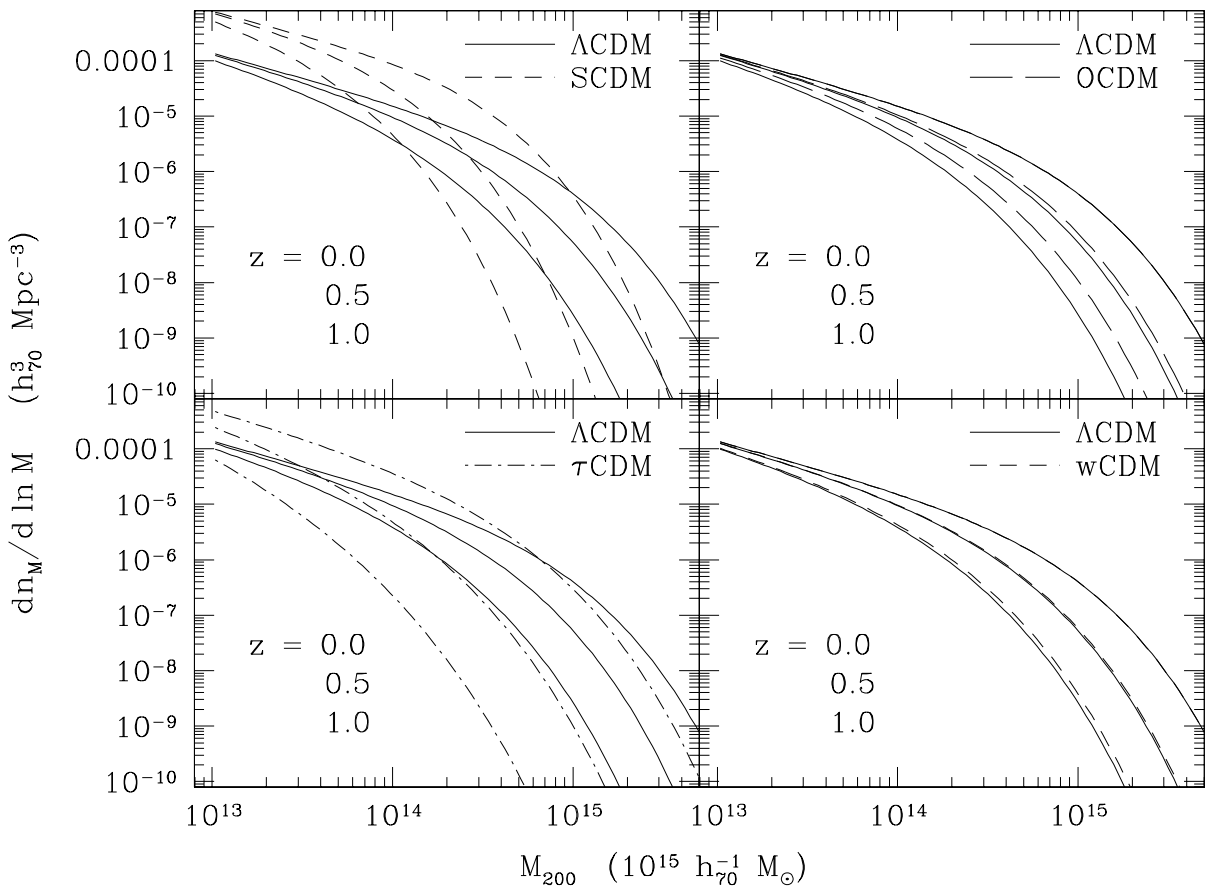}
\caption{Mass-function evolution in five different cosmologies.  The fiducial
model in all cases is the $\Lambda$CDM model with $\Omat = 0.3$, $\Olam = 0.7$,
$w = -1$, and $\sigma_8 = 0.9$.  The upper left panel compares cluster evolution
in the $\Lambda$CDM case with a standard cold-dark-matter model (SCDM) having $\Omat = 1.0$,
$\Olam = 0.0$, and $\sigma_8 = 0.5$.  Evolution in the SCDM case is much more dramatic, 
and the steeper slope of the mass function strongly disagrees with observations of local 
clusters \citep[e.g.,][]{ReiBohr02}.  Retaining $\Omat = 1.0$ and $\Olam = 0.0$ while 
adjusting the power spectrum so that $\Gamma = 0.21$ gives a $\tau$CDM model (lower left) 
in which the slope of the low-redshift mass function is more acceptable, but the evolution
is still very strong.  Dispensing with dark energy while keeping the matter 
density low gives an OCDM model ($\Omat = 0.3$, $\Olam = 0$, $\sigma_8 = 0.9$; upper right)
with less evolution than the $\Lambda$CDM case because structure formation
starts to ramp down earlier in time (see Figure~\ref{fig:omegas}).  Dark energy 
in a $w$CDM model identical to the $\Lambda$CDM model except with $w = -0.8$
(lower right) also slows cluster evolution relative to the $\Lambda$CDM case.}
\label{fig:mfunc}
\end{figure*}

A particularly well-tested fitting formula for cluster mass functions has 
been provided by \citet{Jenkins01}.   Combining results for simulated clusters
spanning a mass range from $< 10^{12} M_\odot$ to $> 10^{15} M_\odot$ 
and sampled at a number of different redshifts, they found that the form of 
$dn_M / d \ln \sigma^{-1}$ was nearly invariant if they 
defined cluster mass to be $M_{\rm 180m}$, the mass within a sphere 
of mean density $180 \, \Omat (z) \rhocr$.  When this definition of 
cluster mass is used, the formula 
\begin{equation}
 \frac {dn_M} {d \ln \sigma^{-1}} =  A_{\rm J}
                                             \frac {\Omat \rhocro} {M}
 			            \; \exp [ - | \ln \sigma^{-1} + B_{\rm J} |^{\epsilon_{\rm J}} ]
 \label{eq:mfunc_jenk}
\end{equation} 
with $A_{\rm J} = 0.301$, $B_{\rm J} = 0.64$, and $\epsilon_{\rm J} = 3.82$,
reproduces the cluster mass function to $\sim$20\% accuracy for all the
cosmologies tested, including $\Lambda$CDM, $\tau$CDM, and OCDM.
In this expression, $A_{\rm J}$ governs the fraction of the total mass in collapsed
objects, $e^{B_{\rm J}}$ functions as a collapse threshold analogous to $\delta_c$,
and $\epsilon_{\rm J}$ stretches the mass function to fit the simulations.
The Sheth-Tormen mass function of equation (\ref{eq:mfunc_st}) 
fits these same numerical simulations nearly as well.

The exponential sensitivity to mass and redshift evident in these expressions
for the cluster mass function is both a blessing and a curse.  On the one
hand, it makes cluster counts and their evolution with redshift a very powerful 
probe of cosmological parameters.   Figure~\ref{fig:mfunc}, showing the cluster
mass function and its evolution with time for five different cosmologies,
illustrates how sensitive mass-function evolution is to the matter density.  
On the other hand, any systematic errors in the measurement of cluster mass, 
including inconsistencies in the definition of cluster mass, are also exponentially 
amplified by the steepness of the mass function.

\subsubsection{Cluster Bias}
\label{sec:bias}

Another observable feature of the cluster population, closely related to the
mass function, is the tendency of galaxy
clusters to cluster with one another.  Fluctuations in the number density 
of clusters on large scales are observed to be more pronounced than 
the fluctuations of the underlying matter density \citep[e.g.,][]{bs83, kk83,
phg92, Collins00, Bahcall03_Sloan_corr}.  In other words, 
the fractional deviation of $dn_M/d \ln \sigma^{-1}$ from its mean value within 
a given volume of the universe is observed to be larger than 
$\delta \rho / \rho$ in that same volume.  The ratio $b(M)$ between
the perturbation in the number density of clusters of mass $M$ and
the perturbation amplitude of the matter density is known as the
bias parameter, and it is taken to be independent of length scale, 
as long as that length scale is much larger than a cluster.  

Cluster bias can be interpreted as a modulation of the collapse
threshold by long-wavelength density modes \citep{Kaiser84, wfde87, ck89}. 
The idea here is that a long-wavelength
density enhancement of amplitude $\delta \rho / \rho = \epsilon$ 
lowers the effective collapse threshold for smaller-scale structures to
$\delta_c - \epsilon$, thereby inducing an offset in $dn_M/d \ln \sigma^{-1}$
from its mean value on mass scale $M$.  This contribution adds to the 
perturbation $\epsilon$ in cluster number density owing to the amplitude 
of the large-scale mode.  Dividing the sum of these two offsets by
$\epsilon$ leads to an expression relating the bias parameter to
the mass function \citep{mw96, st99}: 
\begin{equation}
 b(M) = 1 - \frac {d} {d \delta_c}  \left[ \ln \left( \frac {d n_M} {d \ln \sigma^{-1}} 
 							\right) \right] 
                                                \; \; .
\end{equation}
Plugging in the Sheth-Tormen mass function of equation (\ref{eq:mfunc_st})
produces
\begin{equation}
 b(M) = 1 + \frac {1} {\delta_c} \left[ \frac {a_{\rm s} \delta_c^2} {\sigma^2} -1 
                       + \frac {2p_{\rm s}} {1 + (a_{\rm s} \delta_c^2 / \sigma^2)^{p_{\rm s}}} \right] \; \; .
\end{equation}
\citet{hk03} show that the parameter values $a_{\rm s} = 0.75$ 
and $p_{\rm s} = 0.3$ accurately reproduce the bias of cluster-sized 
halos seen in large-scale numerical simulations, when cluster mass 
is taken to be $M_{\rm 180m}$.  Notice that small values of $\sigma(M)$ lead
to large values of $b(M)$, meaning that rare, high-mass objects are
much more likely to be found in regions of the universe where the
surrounding matter density is higher than average. 

\subsection{Measuring the Cluster Mass Function}
\label{sec:obsmassfcn}

Equations (\ref{eq:mfunc_ps}), (\ref{eq:mfunc_st}), and 
(\ref{eq:mfunc_jenk}) illustrate why cosmologists are so
enthusiastic about the cluster mass function.  Dividing an 
accurate measurement of the mass function by $\Omat \rhocro$ 
directly leads to an accurate measurement of 
the primordial power spectrum $\sigma(M)$ on mass scales
$\sim 10^{14} - 10^{15} \, M_\odot$.   Furthermore, any uncertainty
in $\rhocro$ scales out of the power spectrum's normalization, 
because measured values of cluster number density scale as
$h^{3}$, making the quantity $M_8 \rhocro^{-1} (dn_M / d \ln \sigma^{-1})$
independent of $h$.  One is left only with a degeneracy between
$\sigma_8$ and $\Omat$.  Taking the logarithmic derivative of
(\ref{eq:mfunc_ps}) with respect to $\sigma$ at constant $M$ shows that the 
mass function is roughly $\propto \sigma^{2}$ 
in the region where $\sigma \approx 1$.  Hence, the measured
level of that normalization in the local universe reflects the 
parameter combination $\sigma_8 \Omat^\alpha$, with 
$\alpha \approx 0.5$.

This degeneracy can be broken in three ways.  First, one can simply
measure $\Omat$ or $\sigma_8$ in some other way.  Second, one can measure
the cluster mass function over a range of masses and rely on a
precise measurement on the mass function's shape to break the
degeneracy, assuming that the CDM power spectrum (\S~\ref{sec:cdm}) is valid.  
Or, third, one can measure the evolution of the cluster mass function, 
which is highly sensitive to $\Omat$.
We will explore that option in more depth in \S~\ref{sec:mfuncev}, but first
we need to examine some of the obstacles to accurate mass-function
measurements.

\subsubsection{Linking Mass with Observables}
\label{sec:massobs}

In order to measure the mass function using a large sample of clusters, 
we need to relate cluster mass to an easily observable quantity.
Doing this properly requires a consistent definition of mass (\S~\ref{sec:massdef}) 
and a well-calibrated relation linking that mass to some observable, 
but which mass definition works best?  Expressions like the Jenkins 
mass function (equation \ref{eq:mfunc_jenk}) appear to be cosmologically 
invariant when cluster masses are defined with respect to the background 
mass density (e.g., $M_{\rm 180m}$, see \S~\ref{sec:massfcn}).  On the other hand, the structure of
a cluster, as reflected in its dark-matter velocity dispersion, seems to be cosmologically 
invariant when cluster masses are defined with respect to the critical mass 
density (e.g., $M_{200}$, see \S~\ref{sec:massdef}).  To paraphrase \citet{Evrard04}, 
Nature appears to do {\em accounting} relative to the mean mass density
and {\em dynamics} relative to the critical density.  

Because simulations suggest that dynamical quantities like the galaxy velocity dispersion 
and the X-ray determined gas temperature should be more tightly correlated 
with $M_{200}$ than with $M_{\rm 180m}$, we will take $M_{200}$ to be the
primary definition of cluster mass.  Methods like those outlined in \S~\ref{sec:massdef}
can then be used to convert a mass function in $M_{200}$ to one in
$M_{\rm 180m}$.  Alternatively, one can fit the results of large-scale simulations
to determine a cosmology-dependent correction to the Jenkins mass function
for use with the mass definition $M_{200}$.  \citet{Evrard02} have done
that, finding that substituting $A_{\rm J} = 0.27 - 0.07 (1 - \Omat)$, 
$B_{\rm J} = 0.65 + 0.11 (1 - \Omat)$, and $\epsilon_{\rm J} = 3.8$ into equation
(\ref{eq:mfunc_jenk}) reproduces the $M_{200}$ mass function in
simulations at $z=0$.  Despite the tight relationship between $M_{200}$ and the 
dark-matter velocity dispersion in simulations, the link between $M_{200}$ and 
observable quantities is still a potentially large source of systematic
error.  Even if the galaxy velocity dispersion were identical to that of
the dark matter, accurately measuring that dispersion within a {\em sphere}
of radius $r_{200}$ requires an enormous amount of data to minimize
projection effects \citep[e.g.,][]{rgkd03}.  

To see how systematic errors corrupt the mass-function measurement, 
consider the general case for a generic observable $X$.  Suppose a 
cluster survey determines the comoving number density distribution 
$dn_M / d \ln X$ within logarithmic bins of the observable.  Converting
this distribution to a mass function $d n_M / d \ln \sigma^{-1}$ via the chain 
rule requires, at minimum, knowledge of the normalization and effective 
power-law index $\alpha_X \equiv d \ln X / d \ln M$ of the $M_{200}$-$X$ 
relation over the observed range in $X$, as well as the effective power-law
index $\alpha_M \equiv d \ln \sigma^{-1} / d \ln M$ of the mass fluctuations.
Fitting a semi-analytic expression for the mass function 
like equation (\ref{eq:mfunc_jenk}) to the observations 
for a fixed value of $\Omat$ then determines $\sigma_{\rm fit} = \sigma(M_{\rm fit})$ 
on a particular mass scale $M_{\rm fit}$, and consequently determines 
$\sigma_8 \approx (M_{\rm fit}/M_8)^{\alpha_M} \sigma_{\rm fit}(M_{\rm fit})$.

Any systematic offset $\Delta M/M$ in the normalization of the $M_{200}$-$X$
relation produces a corresponding offset in the measured power-spectrum 
normalization:
\begin{eqnarray}
  \frac {\Delta \sigma_8} {\sigma_8} & = & 
        \left( \alpha_M + \frac {d \ln \sigma_{\rm fit}} {d \ln M_{\rm fit}} \right) \frac {\Delta M} {M} 
            \\
     ~ & = &  \alpha_X \left[ \alpha_M + 
                  \frac {1} {\epsilon_{\rm J} ( \ln \sigma^{-1} + B_{\rm J} )^{\epsilon_{\rm J} -1} } \right] \frac 
                         {\Delta X} {X}  \nonumber
    \label{eq-sig8err}
\end{eqnarray}
\citep[e.g.,][]{v00, Evrard02, Seljak02}.   The second line of this equation assumes that
$\sigma_{\rm fit}$ has been determined using the Jenkins mass function of
equation (\ref{eq:mfunc_jenk}).  On the mass scale of rich clusters 
($\gtrsim 10^{14.5} \, M_\odot$), \citet{Evrard02} find that the factor in 
parentheses is $\approx 0.4$. implying that a systematic 25\% error in 
mass would lead to a 10\% error in the measurement of $\sigma_8$.  Below this
mass scale the factor in parentheses increases, leading to an even larger
error in the power-spectrum normalization for a given mass offset 
\citep{hw02}.

Dispersion in the value of the mass-tracing observable for clusters of fixed mass is 
another important source of uncertainty that must be dealt with carefully because 
of the exponential slope of the mass function \citep[e.g.,][]{pbsw03}.   
Significant scatter boosts the normalization of $d n_M / d \ln X$ over the expectation 
for the no-scatter case, as the overall number of lower-mass clusters 
scattering to higher values of $X$ far exceeds the number of higher-mass 
clusters scattering in the opposite direction.  Underestimating this scatter
leads to an overestimate of $\sigma_8$ that can be particularly severe
if the scatter has a long non-gaussian tail to large values of $X$. 
Situations in which such a tail could arise include merger shocks that
substantially boost the temperature and luminosity in a significant subset
of X-ray selected clusters \citep[e.g.,][]{rs01, rsr02} and superpositions
of galaxies that boost the apparent richness and velocity dispersion in
an optically-selected sample \citep{vfw97}.

Surveys that probe deep into the universe for clusters must also cope with
redshift evolution in the mass-observable relation.  That is, if $M_{200} \propto
X^{\alpha_X} (1+z)^{b_X}$, then one needs to know the value of $b_X$.  
This source of uncertainty affects both the mapping of $X$ onto mass
for individual clusters and the number density one infers for clusters of a 
given mass from a survey based on the observable $X$. 

A sufficiently large cluster survey can circumvent many of these systematic problems
through self-calibration \citep{lsw02, Hu03, mm03a, mm03b}.   
This procedure treats all parameters describing the
systematic uncertainties in such things as the scatter, normalization, and evolution 
of the mass-observable relation as free parameters in the overall cosmological model.  
Fitting a large number of clusters spanning a wide range in redshift to this overall
model, one can then determine not only the global cosmological parameters but
also the most likely values of the free parameters in the mass-observable relation.
However, treating the systematic uncertainties in this way has a cost.  Each free
parameter added weakens the statistical constraints on the cosmological 
measurements \citep{mm03a}.

\subsubsection{Mass-Temperature Relation}
\label{sec:mtrel}

Among the observables that trace cluster mass, X-ray temperature has
received considerable recent attention because it is closely related to the
depth of a cluster's potential well and can be readily observed to $z \sim 1$ 
with current X-ray telescopes (\S~\ref{sec:tx}).    \citet{ha91} pioneered
the technique of measuring the cluster mass function with X-ray temperatures,
using cluster temperatures determined at $z \approx 0$ with 
the {\em Einstein}, {\em Exosat}, and {\em HEAO/OSO} satellites.  
Cluster temperatures measured
with the {\em ASCA} satellite improved the precision of this measurement
\citep{Ikebe02}, and temperatures measured with the {\em Chandra} and 
{\em XMM-Newton} telescopes should improve that precision even more.
Because the data are now of such high quality, systematic uncertainty
in the link between mass and temperature is the main factor limiting 
this technique.

Mass and temperature ought to be simply related for a cluster in hydrostatic equilibrium.
The gas temperature of a singular isothermal sphere with mass $M_{200}$ 
inside radius $r_{200}$ is
\begin{eqnarray}
  \kB T_{200} & =  & \frac {G M_{200} \mu m_p} {2 r_{200}} 
               \;  \;   =  \; \;  \frac {\mu m_p} {2} [10 \, G M_{200} H(z)]^{2/3}   \nonumber \\
             ~  & = & (8.2 \, \keV) \left( \frac {M_{200}} {10^{15} \, h^{-1} \, M_\odot} \right)^{2/3}
                                \left[ \frac {H(z)} {H_0} \right]^{2/3} \, .
\end{eqnarray}
Realistic departures from hydrostatic equilibrium can be assessed with simulations 
of structure formation that include hydrodynamics, but they do not have a large
effect on the mass-temperature relation.  These models indeed find that gas temperature 
scales with $M_{200}^{2/3}$, so that $T \approx T_{200}$, with
a scatter of only 10-15\% \citep[e.g.,][]{emn96, f99_sb}.
However, to calibrate the mass-temperature relation more precisely, 
we need a more specific definition of temperature.

Clusters are not perfectly isothermal, so any single number specifying
a cluster's gas temperature is some sort of weighted mean.  The luminosity-weighted
mean temperature $\Tlum$ obtained by weighting each gas parcel's temperature by
$\propto \rho_g^2$ is a popular choice for comparing theory with observations because 
each temperature component contributes in proportion to its photon flux in the cluster's
overall spectrum (\S~\ref{sec:tx}).  The spectral-fit temperature $\Tsp$ has not yet received
much attention in theoretical work because it depends somewhat on the procedure used to
fit the spectrum, but \citet{mrmt04} have recently developed a temperature
weighting scheme for theoretical models that appears to track $\Tsp$ quite closely.

A bewildering variety of parameters has been used in the literature to express
the mass-temperature relation's normalization.  Here we will express
the normalization of the $M_{200}$-$\Tlum$ relation in terms of $\Tlum/T_{200}$.
This choice has two advantages:  It does not link the normalization to any 
particular mass or temperature scale, and $\Tlum/T_{200} = 1$ for both 
a singular isothermal sphere and an isothermal beta model with $\beta = 2/3$.  

Uncertainty in this normalization factor is currently the single most important
issue afflicting cluster mass-function measurements with X-ray observations.
Table~\ref{tab:mtnorm} provides some recent observational and theoretical
calibrations of this relation, in three different groups.  The first group gives
calibrations from hydrodynamical simulations that do not account for galaxy
formation, which generally fall into the range $\Tlum/T_{200} = 0.8-1.0$.  In some cases,
an $M_{500}$-$T_{\rm lum}$ relation has been converted to $M_{200}$
assuming $M_{200} = 1.4 M_{500}$, appropriate for halo concentration
$c = 5$.  The second group gives calibrations inferred from observations, 
which fall into the range $\Tlum/T_{200} = 1.1-1.4$.  In other words, 
clusters of a given temperature seem to be 30\% to 60\% less massive than one 
would expect from the simulations.  Apparently galaxy formation changes the
normalization of the $M_{200}$-$\Tlum$ relation, for reasons discussed
in detail in \S~\ref{sec:baryons}, although some of this discrepancy may also stem
from systematic offsets in the observational interpretation.  
The third group of normalization factors, which tend
to lie in between the first two groups, come from simulations that attempt 
to account for the effects of galaxy formation.
Given this uncertainty in the mass-temperature normalization, the
systematic uncertainty in $\sigma_8$ values derived from cluster temperatures
is about 25\%, because $\sigma_8 \propto (\Tlum/T_{200})^{3/5}$ according 
to equation (\ref{eq-sig8err}) for rich clusters \citep{Evrard02}.  

\begin{table}
\caption{\label{tab:mtnorm} Normalization of the Mass-Temperature 
Relation}

\begin{ruledtabular}
\begin{tabular}{lc}
 
 Models without Radiative Cooling & $\Tlum/T_{200}$\footnote{
 $T_6 \equiv \kB \Tlum/6\,\keV$}  \\
 
  \hline 
 
 \citet{nfw95} & 0.99 \\
 \citet{emn96}\footnote{Conversion from $M_{500}$ assumes $M_{200} = 1.4 M_{500}$.}   & 0.91 \\
 \citet{bn98}     & 0.80 \\
 \citet{thomas01} & 0.98 \\
 \citet{mtkp02}\footnote{$\Tlum$ computation includes gas with cooling time $< 6$~Gyr} & 0.57 \\
 \citet{mtkp02}\footnote{$\Tlum$ computation excludes gas with cooling time $< 6$~Gyr} & 0.90 \\ 
 
 \hline \hline
 Observations  & $\Tlum/T_{200}$$^a$  \\ 
   \hline 

 \citet{hms99}\footnote{Masses estimated using isothermal beta model}  & $(1.08 \pm 0.04)T_6^{-0.19}$  \\  
 \citet{hms99}\footnote{Masses estimated using polytropic beta model}  & $(1.40 \pm 0.16)T_6^{-0.02}$  \\ 
 \citet{nmf00}\footnote{Conversion from $M_{1000}$ assumes $M_{200} = 2.0 M_{1000}$.}   
 									& ($1.20 \pm 0.12) \, T_6^{-0.20}$ \\ 
 \citet{frb01}$^{b,}$\footnote{Full sample, masses from isothermal beta model}  & ($1.18 \pm 0.10)T_6^{-0.33}$ \\
 \citet{frb01}$^{b,}$\footnote{Full sample, masses from polytropic beta model}  & $(1.26 \pm 0.11)T_6^{-0.19}$ \\
 \citet{frb01}$^{b,}$\footnote{Subset with $\kB \Tlum$, polytropic beta model}  &  
 					$(1.33 \pm 0.18)T_6^{-0.02}$ \\ 

  \hline \hline
 Models with Radiative Cooling  & $\Tlum/T_{200}$$^a$  \\ 
 \hline 
 
 \citet{mtkp02} & $0.79 \, T_6^{-0.31}$  \\
 \citet{mtkp02}\footnote{$\Tlum$ computed with cooling cores removed} & $0.88\, T_6^{-0.09}$  \\
 \citet{Borg04}$^{b}$ & $(1.03 \pm 0.03) T_6^{-0.06}$ \\
 \citet{Borg04}$^{b,}$\footnote{Masses inferred from polytropic beta-model fits.} 
 							& $(1.24 \pm 0.03) T_6^{-0.06}$ \\
\end{tabular} 

\end{ruledtabular}
\end{table}

Efforts are underway to reconcile the observed normalization of the
$M_{200}$-$\Tlum$ relation with theoretical expectations.  Some of
the discrepancy probably arises from systematic errors in the masses
derived from X-ray observations.  Hydrostatic equilibrium is usually
assumed, but the turbulent velocities in simulated clusters 
can sometimes be $\sim$20-30\% of the sound speed, in which case the
hydrostatic assumption would lead to masses underestimated
by 10-15\% \citep{rs01, rtm03}.  In addition, the beta model formalism often
used to derive cluster mass may have systematic problems.
Applying this model to simulated clusters tends to underestimate
their masses \citep[][see the last line of Table~\ref{tab:mtnorm}]{mtkp02, Borg04},  
and recent {\em XMM-Newton} observations suggest that
the correction for temperature gradients may be excessive
\citep{pa02, pa03, Mush04}.  

Alternative mass measurements would be very helpful in solving 
these problems.  Calibration of the mass-temperature relation with
lensing observations has met with mixed success.  Weak-lensing
measurements of massive, relaxed clusters agree with the masses
derived from X-ray data under the assumption of hydrostatic equilibrium 
\citep{asf01}.  However, that agreement seems to 
be poorer for less relaxed clusters \citep{Smith03}.
Measurements of cluster mass from the galaxy velocity field in 
and around a few very well observed clusters also tend to support 
the X-ray derived masses \citep{rgkd03}.  Because the calibration may depend
systematically on how clusters are selected, self-calibration of a large
cluster survey may ultimately be the best way of calibrating the
$M_{200}$-$\Tlum$ relation.  A thorough understanding
of  how galaxy formation affects that relation would help reduce the 
number of free parameters that need to be calibrated, thereby reducing
the statistical uncertainties achievable with self-calibration.

\subsubsection{Mass-Luminosity Relation}
\label{sec:mlrel}

X-ray luminosity also correlates well with cluster mass and is easier
to measure than X-ray temperature, allowing for mass-function measurements 
using much larger cluster samples.  However, the correlation between mass and 
luminosity is not as tight as that between mass and temperature,
having a scatter $\sim$50\% \citep{ReiBohr02}.  Additionally, the normalization and slope
of the relation depend heavily on the physics of galaxy formation (\S~\ref{sec:scaling}).
Because our understanding of the connection between galaxy formation
and a cluster's X-ray luminosity is not yet mature enough to calibrate
the mass-luminosity relation with simulations, we need to rely
solely on observational calibrations.

One common way to calibrate the mass-luminosity relation is to 
combine the mass-temperature relation with the observed
luminosity-temperature relation \citep[e.g.,][]{borg99}.  On cluster scales,
the relation between the total (bolometric) X-ray luminosity
and $\Tlum$ is approximately a power law.  Normalizing
the relation at 6~keV, in the heart of the temperature range
for rich clusters, leads to the expression $\Lx = L_6 (\Tlum / 6 \, \keV )^{\alpha_{LT}}$,
and Table~\ref{tab:ltrel} gives some representative values of $L_6$ 
and $\alpha_{LT}$.   Excising the central regions of clusters, out to
about 100~kpc in radius, reduces the scatter in the relation
because cooling and non-gravitational heating processes 
affect the temperature and luminosity of these regions differently
from cluster to cluster \citep{fcem94, mark98, af98, vbbb02}.

\begin{table}
\caption{\label{tab:ltrel} Luminosity-Temperature Relation at $z \approx 0$}

\begin{ruledtabular}
\begin{tabular}{lcc}
 
 Source & $L_6$\footnote{Bolometric X-ray luminosity is 
 $\Lx = L_6 (\Tlum/ 6 \, \keV)^{\alpha_{LT}}$ with $L_6$ in units of 
 $10^{44} \, h_{70}^{-2} \, {\rm erg \, s^{-1}}$.} & $\alpha_{LT}$  \\
 
  \hline 
 
 \citet{es91}         &  $6.3 \pm 1.3$  &  $2.62 \pm 0.10$  \\
 \citet{David93}   &  $5.6 \pm 0.9$  &  $3.37 \pm 0.05$  \\
 \citet{mark98}\footnote{Cores of clusters excised to avoid cool cores.}
                              &  $6.4 \pm 0.6$  &  $2.64 \pm 0.27$  \\
 \citet{af98}\footnote{Clusters without cool cores.}
 		          &   $5.7 \pm 3.4$  &  $2.92 \pm 0.45$ \\
 \citet{af98}\footnote{Clusters with cool cores.}
 		          &   $14.6 \pm 7.3$  &  $3.08 \pm 0.58$  \\
 \citet{ae99}\footnote{Sample avoids clusters with cool cores.}         
                              &  $5.9 \pm 0.4$  &  $2.88 \pm 0.15$  \\
\citet{xw00}         &  $7.6 \pm 1.2$  &  $2.79 \pm 0.08$  \\
\citet{nsh02}        &  $6.0 \pm 4.2$  &  $2.82 \pm 0.43$  \\
\citet{edm02}         &  $7.3 \pm 1.8$  &  $2.54 \pm 0.42$  \\
                             
\end{tabular} 

\end{ruledtabular}
\end{table}

The power-law index of the $\Lx$-$\Tlum$ relation clearly indicates 
that galaxy formation has affected the $\Lx$-$\Tlum$ relation.  If the density 
distribution of intracluster gas within $r_{200}$ were self-similar, 
independent of cluster mass, then one would expect bremsstrahlung 
emission to give $\Lx \propto \rho_g M_{200} \Tlum^{1/2}
\propto \Tlum^2$ \citep{Kaiser86}.   The steepness of the observed
power-law index indicates that non-gravitational processes have
raised the entropy of the intracluster gas, making it harder to compress,
particularly in the shallower potential wells of cool clusters.  
This excess entropy therefore lowers the luminosities 
of all clusters by lowering the mean gas density and steepens
the $\Lx$-$\Tlum$ relation because the impact of excess entropy
decreases as cluster temperature rises \citep[][\S~\ref{sec:scaling}]{eh91, 
Kaiser91}. 

Calibrating the mass-luminosity relation by coupling the mass-temperature
relation with the observed luminosity-temperature relation leads, not surprisingly,
to values of $\sigma_8$ that are similar to those derived from the
mass-temperature relation alone and are subject to the same systematic
uncertainties that plague the mass-temperature calibration.  
There is, however, a route to the mass-luminosity calibration that circumvents
the middle step involving the mass-temperature relation.

The mass-luminosity relation can be calibrated more directly with
high-quality X-ray imaging and temperature data on a complete sample 
of clusters.  \citet{ReiBohr02} have done this with {\em ROSAT}
imaging data and {\em ASCA} temperatures, finding
\begin{equation}
 \Lx = 10^{45.0 \pm 0.3} \, h_{70}^{-2} \, {\rm erg \, s^{-1}}
                   \left( \frac {M_{200} }  {10^{15} \, h_{70}^{-1} \, M_\odot} \right)^{1.8} \; \; .
\end{equation}
Their mass calibration assumes that the cluster gas is in hydrostatic equilibrium
and obeys an {\em isothermal} beta model.  The masses they derive are therefore
higher than those one would find after correcting for a possible negative 
temperature gradient at large radii but do not account for any 
turbulent pressure support.  With this mass-luminosity
relation, they find a cluster mass function whose normalization corresponds to
$\sigma_8 = 0.68 (\Omat / 0.3)^{-0.38}$.  

Furthermore, because their observed cluster sample extends over two 
decades in mass, \citet{ReiBohr02} attempted to break the 
$\sigma_8$-$\Omat$ degeneracy by fitting the mass-function's 
shape with a CDM power spectrum, finding a best fit of $\Omat = 0.12^{+0.06}_{-0.04}$ 
and $\sigma_8 = 0.96^{+0.15}_{-0.12}$, with $\Omat < 0.31$ at the 3$\sigma$ level.  
The unusually low best-fit value of $\Omat$ arises because their derived mass function 
is shallower than that expected for $\Omat = 0.3$.  However, \citet{pbsw03}  
have applied that same $L_X$-$M_{200}$ relation to the larger {\em REFLEX} cluster 
sample \citep{Bohringer02}, finding $\sigma_8 = 0.86^{+0.12}_{-0.16}$ and 
$\Omat = 0.23^{+0.10}_{-0.06}$.  Results similar to these latter values are also
found from the mass-luminosity relation when cluster evolution is used to
break the $\sigma_8$-$\Omat$ degeneracy (\S~\ref{sec:mfuncev}).

\subsubsection{Mass-Richness Relation}
\label{sec:mrich}

Optical telescopes have gathered much larger cluster samples than have X-ray
telescopes, but deriving a mass function from these samples is not so straightforward.
Projection effects complicate both the measurement of cluster mass and
the computation of the sample volume associated with a given mass.
Clusters in optical surveys are selected on the basis of richness, which
depends on the number of galaxies observed within a certain projected
radius from the center of the cluster (\S~\ref{sec:richness}).  Thus, even if optical luminosity
traces mass exactly, galaxy concentrations lying outside $r_{200}$ but
projected along the same line of sight can boost the apparent mass,
introducing non-gaussian uncertainties in the mass-richness relation.
Likewise, the effective volume associated with a given cluster mass in
an richness-selected survey is harder to quantify than in a survey with
a definite X-ray flux cutoff.    Nevertheless, when richness is rigorously
defined, it correlates well with a cluster's X-ray properties \citep{ye03}. 
However, the scatter between optical richness and X-ray luminosity
is still large compared with the accuracy to which one would like to
derive cosmological parameters \citep{d01_rox1, d02_rox2, 
Kochanek03, Gilbank03}.

Measuring cluster masses purely on the basis of galaxy-count richness 
necessitates a different approach to defining a cluster's radius and 
therefore its mass.  Because traditional measures of richness depend 
on the radius within which galaxies are counted, they are defined with 
respect to a fixed physical radius, independent of mass, at each redshift.
For this reason, observations of cluster richness are sometimes compared 
with simulations on the basis of cluster masses measured within a constant
physical or comoving radius \citep{Bode01}.   Deriving a mass-richness relation from
simulations of galaxy formation also involves an observational calibration
of the cluster mass-to-light ratio, which according to equation (\ref{eq-sig8err}) 
introduces a systematic uncertainty in $\sigma_8$ that is $\sim$40\% of the uncertainty
in the mass-to-light conversion.

Making such a comparison with the early clusters from the Sloan Digital 
Sky Survey, \citet{bahcall_maxBCG03} find a mass-function normalization implying 
$\sigma_8 = 0.69 \pm 0.07 (\Omat / 0.3)^{-0.6}$.
Adding mass-function shape information to break the degeneracy
leads to $\Omat = 0.19^{+0.08}_{-0.07}$ and $\sigma_8 = 0.9^{+0.3}_{-0.2}$, 
in reasonably good agreement with the X-ray derived values.  
Unfortunately, because there is as yet no simple parametric form, analogous
to the Jenkins mass function, giving the mass function defined with respect
to a fixed radius, it is not clear how to self-calibrate the associated 
mass-richness relation to high accuracy with a large survey.

\subsubsection{Velocity Dispersion and Mass}
\label{sec:vdmass}

Velocity dispersion is the optical analog to X-ray temperature. Thus, one
would expect a mass function defined on the basis of velocity dispersion
to coincide with those defined with respect to cluster temperature.
Measurements of rich clusters indicate that $\sigv^2 = (1.0 \pm 0.1) \kB \Tlum
/ \mu m_p$ \citep[e.g.,][\S~\ref{sec:tx}]{xw00}, which reassuringly suggests
that both quantities accurately trace mass.  On the other hand, \citet{Evrard02}
have pointed out that combining this relation with the observational
calibration of the $M_{200}$-$\Tlum$ relation ($\Tlum \approx 1.2 \, T_{200}$)
leads to a puzzle.  While it might be possible for non-gravitational effects
associated with galaxy formation to boost $\Tlum$, it is more difficult to
imagine why non-gravitational effects would boost the galaxy velocity
dispersion by a similar factor.  

Mass functions derived from velocity dispersion measurements
also suggest that the masses derived from those measurements
are larger than those derived from X-ray data.   Using the virial theorem
with a pressure correction term (\S~\ref{sec:vdisp}), \citet{Girardi98} derive a cluster mass 
function from velocity dispersions whose normalization indicates 
$\sigma_8 = (1.01 \pm 0.07) (\Omat / 0.3)^{-0.43}$, implying an overall 
number density at a given cluster mass about two times larger than 
the X-ray measurements.  A discrepancy in $\sigma_8$ as large as 
30\% could arise if the optically determined masses
were over 50\% larger, but the actual mass discrepancies appear not to be quite 
so large.  \citet{ReiBohr02} find that in the 42 clusters they have in 
common with \citet{Girardi98} the virial masses are 25\% larger, on average, 
than the X-ray masses.  Another factor that could contribute to this discrepancy
is scatter in the $M_{200}$-$\sigv$ relation.  An underestimate of the scatter would
drive up the inferred mass-function amplitude, raising the best-fitting value
of $\sigma_8$.  

Part of the discrepancy between the optical and X-ray determined masses
may stem from how velocity dispersions are observed.  
Because $\sigv$ declines with projected radius, its observed value will depend 
on the cutoff radius.  Also, any foreground or background interlopers projected 
onto the cluster can contaminate the velocity-dispersion measurement.
Ideally, one would like to cut off the measurement at a spherical boundary with radius
$r_{200}$, inside of which the relation between dark-matter velocity dispersion 
and mass is well-calibrated, but the large number of galaxy velocities needed
to accurately measure the mass profile near the virial radius make this approach impractical 
for large cluster samples.  In a small sample of eight rich clusters, \citet{rgkd03}  
have used an average of almost 200 galaxy velocities per cluster, extending
to well beyond $r_{200}$, to measure the mass $M_{200}$ within $r_{200}$.
The masses they find are consistent with both the X-ray determined masses
and with the virial theorem including a surface-pressure correction.

\subsubsection{Weak Lensing and Mass}
\label{sec:wlmass}

Weak lensing is a very promising method for measuring cluster masses that
is independent of a cluster's baryon content, dynamical state, and mass-to-light ratio.
The main systematic problem in weak-lensing mass measurements comes from 
the lensing done by excess mass outside the virial radius 
but along the line of sight through the cluster \citep{mwnl99, mwl01, Hoekstra01}.  
So far, weak-lensing's main contribution to cluster studies has been to assist 
in the calibration of other mass estimators \citep[e.g.,][]{asf01}.  

Techniques for compiling cluster samples selected on 
the basis of weak lensing are still in their infancy.  
Only a few clusters with confirmed spectroscopic redshifts have been 
detected in weak lensing surveys \citep[e.g.,][]{Wittman01, Dahle03, 
Schirmer03, Wittman03}.  However, deep optical surveys
covering wide patches of the sky should turn up many more such clusters
in the coming decade.  In the meantime, smaller weak-lensing surveys sensitive
to large-scale structure are complementing the cluster work because they provide values of
$\sigma_8$ that are independent of the cluster measurements.  Numbers
currently in the literature span approximately the same range as those
derived from clusters, going from $\sigma_8 = (0.72 \pm 0.08)(\Omat/0.3)^{-0.57}$ 
\citep{Jarvis03} on the low end to $\sigma_8 = (0.97 \pm 0.14)(\Omat /0.3)^{-0.44}$ 
\citep{Bacon03} on the high end.

\subsubsection{Baryons and Mass}
\label{sec:mbaryon}

Yet another technique for measuring the cluster mass function relies on
the constancy of the ratio of baryons to dark matter in massive clusters.
X-ray observations from {\em Chandra} indicate that the ratio of hot 
baryonic gas to total gravitating matter within a given radius asymptotically
approaches $(0.113 \pm 0.005) h_{70}^{-3/2}$ in relaxed, high-mass
clusters \citep{asf02}.   Correcting for the baryons
in stars, whose mass is approximately $0.16 h_{70}^{1/2}$ times that 
of the hot gas (\S~\ref{sec:condensation}, raises the overall ratio of 
baryons to dark matter in clusters to $f_b  = 0.13$ for $h_{70} = 1.0$.  

This ratio is itself one of the best tools for measuring $\Omat$ \citep{wnef93, Evrard97,
asf02, djf95}.  No known hydrodynamic process can 
drive a large proportion of a rich cluster's baryons out of the cluster's deep 
potential well.  Thus, the ratio of baryons to dark matter in a cluster is expected 
to be similar to the global ratio in the universe.  Dividing the mean baryon density
$\Obar = 0.045 \, h_{70}^{-2}$ consistent with both the abundances of light elements
\citep{bnt01} and microwave background fluctuations \citep{WMAP03} by the baryon fraction
$f_b$ implies $\Omat \approx 0.3$.  \citet{asf02} find $\Omat = 0.30^{+0.04}_{-0.03}$
after marginalizing over the uncertainties in $\Obar$ and Hubble's constant. 

One can also use ratio of baryons to dark matter to constrain dark energy 
\citep{Sasaki96, Pen97}.  Measurements of this ratio in clusters depend on the 
relationship between transverse size and redshift, which depends on both $\Omat$ 
and $\Olam$ (\S~\ref{sec:glogeo}).  If the actual ratio remains constant with redshift, 
then the measured ratios will be independent of redshift only if the correct values of 
$\Omat$ and $\Olam$ are used in the measurement.  \citet{Allen04} have 
recently shown that the measured baryon to dark matter ratio in a sample of
26 clusters ranging up to $z \approx 0.9$ is consistent with the low-redshift ratio for
$\Olam = 0.94_{-0.23}^{+0.21}$ \citep[see also][]{asf02, etp03}.  However, 
the degree to which the actual ratio is redshift-independent is not yet known.

If the ratio of baryons to dark matter were completely independent of cluster
mass and radius, then measurements of the baryon mass inside a radius
containing a mean baryon density of $200 f_b \rhocr$ would directly give
$M_{200}$.  The cluster mass function could then be determined by
measuring the baryon masses within a given scale radius
\citep{vv03, Vikh03}.  In fact, the baryon
fraction is not quite constant in clusters, probably owing to the same
galaxy-formation effects that shift the $M$-$T$ and $L$-$T$ relations
(\S~\ref{sec:scaling}).  For example, \citet{mme99} find that 
the ratio of gas mass to dark matter is $\propto \Tlum^{0.36  \pm 0.22}$
in clusters cooler than about 6~keV and is statistically
inconsistent with a constant value at the 99\% level.  Other studies concur 
that the proportion of hot gas in low-mass clusters is smaller than that in 
high-mass clusters \citep{na01, spflm03}.  

After correcting for this effect, \citet{vv03} infer a cluster
mass function from the baryon mass function signifying $\sigma_8 = 0.72 \pm 0.04$
for the assumed cosmology ($\Omega = 0.3$, $\Olam = 0.7$, and $h=0.71$).  
Furthermore, the shape parameter $\Gamma = 0.13 \pm 0.07$ of the mass 
function is consistent with the CDM power spectrum given the assumed 
values of $\Omat$ and $h$.   Notice that this value of $\sigma_8$ agrees
with those derived from the observationally calibrated $M_{200}$-$\Tlum$ 
and $M_{200}$-$\Lx$ relations, even though it does not explicitly rely 
on those calibrations.
   
\subsection{Evolution of the Mass Function}
\label{sec:mfuncev}

Measurements of evolution in the cluster mass function can considerably tighten
all these constraints on cosmological parameters.  What we actually
observe, of course, is the dependence on redshift of the observables that
trace the cluster mass function.  For a given cluster sample we can measure the 
number of clusters $dN$ within a given solid angle $d \Omega$ and redshift
interval $[z,z+dz]$ that fall into the range $[X,X+dX]$ of the observable $X$.
With full knowledge of the mass-observable relation $M(X,z)$ and its scatter
as a function of redshift, we could then derive the redshift distribution
\begin{equation}
  \frac {d^3N} {dM \, d\Omega \, dz} (M,z) = \frac {dn_M} {dM} (M,z) \cdot
  						\frac {d^2 V_{\rm co}} {dz \, d\Omega} (z)
\end{equation} 
for clusters of mass $M$ directly from the observations.  This distribution of clusters with 
redshift would then provide strong constraints on cosmological models through both
the mass-function evolution factor $dn_M/dM$ and the comoving volume factor
$d^2 V_{\rm co}/d\Omega \, dz $ from equation (\ref{eq:vco}).

As the reader probably suspects by now, our ability to constrain cosmological
parameters through the redshift distribution of clusters is currently limited
by our understanding of evolution in the mass-observable relations.  However,
this problem is not as severe as one might expect because the evolution in the
mass function itself is so dramatic, especially for $\Omat \approx 1$.  This part
of the review discusses what we have learned about structure formation and
cosmological parameters by observing cluster evolution.  It begins with a description
of how mass-function evolution depends on cosmological parameters and then
considers the complications arising from evolution of the observables themselves.  
It concludes with a summary of current constraints on $\Omat$ from cluster evolution and 
a look at the prospects for measuring $\Olam$ and $w$ with large cluster surveys.

\subsubsection{Dependence on Cosmology}
\label{sec:depcos}

Evolution of the mass function is highly sensitive to cosmology because the matter 
density controls the rate at which structure grows.  When the mass function
can be expressed in terms of formulae like (\ref{eq:mfunc_ps}), (\ref{eq:mfunc_st}),
or (\ref{eq:mfunc_jenk}), its evolution is controlled entirely by the growth
function $D(z)$, which is a well defined function of $\Omat$, $\Olam$, 
and $w$ (\S~\ref{sec:massfcn}).  Small-amplitude 
density perturbations grow as $D(z) = (1+z)^{-1}$ 
when $\Omat (z) \approx 1$, but perturbation growth stalls when $\Omat (z) \ll 1$.  
This effect manifests itself most strongly in high-mass clusters because 
they are the latest objects to form in a hierarchical cosmology with a 
CDM-like power spectrum \citep{Evrard89, pdj89, ob92, ecf96, vl96}.  
The exponential dependence of the 
mass function on $\sigma(M,z) = D(z) \sigma(M,0)$ makes the effect
quite dramatic for objects sufficiently massive that $\sigma(M,0) < 1$.

Dependence of the mass function on $\Olam$ and $w$ is a little more
subtle.  These parameters affect mass-function evolution by altering
the redshift at which $\Omat (z) $ departs significantly from unity for
a given value of $\Omat$ at $z=0$ \citep{hmh01}. 
The time at which dark energy begins to dominate the dynamics
of the universe is later for both larger values of $\Olam$ and smaller (more 
negative) values of $w$ (see Figure~\ref{fig:omegas}), leading 
to greater evolution of the mass function between $z \sim 1$ 
and the present \citep{ws98, wbk02, bw03}. 

Measurements of how the mass function changes with redshift can 
provide additional information about $\Olam$ and $w$ through the 
expansion rate of the universe.  If the mass function of clusters is
precisely known, then number counts of clusters exceeding a given
mass in each redshift interval $dz$ reveal the volume associated
with that redshift interval and can be used to determine the
dynamics of the universe's expansion.  The number of clusters
with mass $>M$ on the celestial sphere in the redshift interval $dz$ 
is given by
\begin{equation}
  \frac {dN} {dz} (M)  = \frac {4 \pi r_\kappa^2 (z) c} {H(z)}  n_M(M,z)   \; \; .
\end{equation}
Figure~\ref{fig:dNdz} shows this number-redshift distribution for several different
cosmological models.  
Notice that the statistical power of cluster surveys is ultimately limited
by the total number of massive clusters in the observable universe, which
is of order $10^5$.

\begin{figure}
\includegraphics[width=3.4in , trim = 1.5in 0.8in 2.0in 0.7in , clip]
{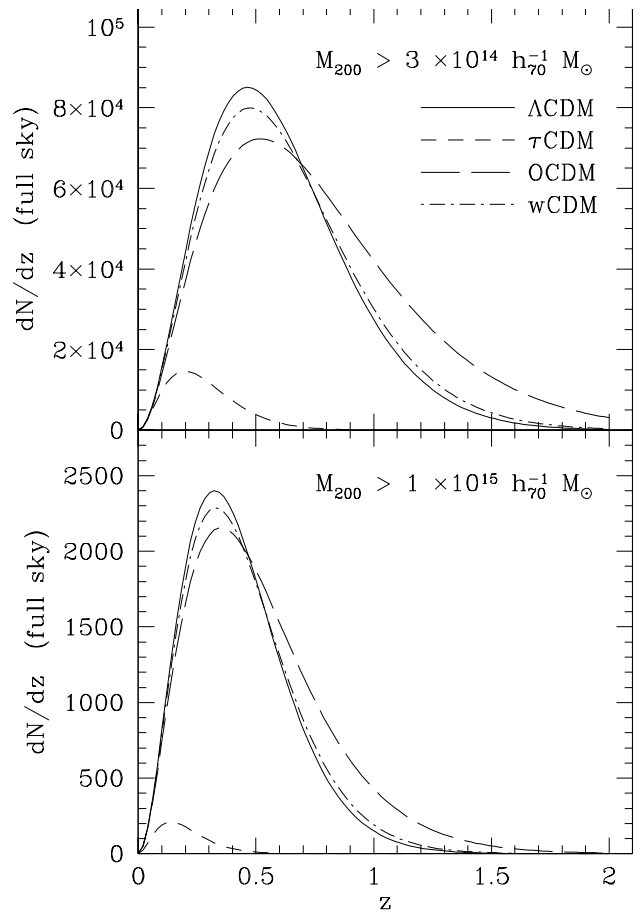}
\caption{Predicted number of clusters on the sky as a function of redshift 
in different cosmologies. The upper panel shows the number of clusters
per unit redshift with $M_{200} > 3 \times 10^{14} \, h_{70} \, M_\odot$
over the entire sky.  Notice that there are a few tens of thousands of such 
clusters on the sky in models with $\Omat = 0.3$, most of them at $z < 1$.  
There are many fewer massive clusters at $z > 0.5$ in the $\tau$CDM model with
$\Omat =1$ because cluster evolution is so rapid in that case.  The lower panel 
shows the numbers of clusters with $M_{200} > 1 \times 10^{15} \, h_{70} \, M_\odot$.  
Differences between models with $\Omat \approx 0.3$ but differing values 
of $\Olam$ and $w$ should be detectable in large cluster surveys containing
$\sim 10^4$ clusters and extending to $z \sim 1$.}
\label{fig:dNdz}
\end{figure}

\subsubsection{Evolution of the Observables}
\label{sec:obsevol}

All of the mass-observable relations discussed in \S~\ref{sec:obsmassfcn} evolve with
redshift, partly because the definition of $M_{200}$ is pinned to the
critical density and partly because of galaxy-formation physics.
Clusters of a given mass are hotter earlier in time because their
matter density is larger;  both $T_{200}$ and the square of the 
dark-matter velocity dispersion for a fixed value of $M_{200}$
vary with redshift as $H^{2/3}(z)$ (\S~\ref{sec:mtrel}).  
One therefore expects $\Tlum$ and the square of the galaxy velocity 
dispersion to depend on redshift in the same way, but it is possible 
that the physics of galaxy formation adds additional redshift evolution 
that must be accounted for in precise cosmological measurements.  
Galaxy formation plays a more explicit role in the mass-richness and 
$M_{200}$-$\Lx$ relations, because the optical luminosities of galaxies
evolve with time and because the physics of galaxy formation alters
the $\Lx$-$\Tx$ relation (\S~\ref{sec:scaling}).  Scatter in the mass-observable
relation might also be larger at higher redshifts, given that the proportion 
of relaxed clusters may well be smaller earlier in time.

As an example of how mass-observable evolution affects observations of mass-function 
evolution, consider its effects on X-ray surveys.  The upper left of 
Figure~\ref{fig:mtlfunc} shows how the cluster 
mass function evolves for two different cosmologies, a standard $\Lambda$CDM
model ($\Omat = 0.3$, $\Olam = 0.7$, $w =-1$, $\sigma_8 = 0.9$) and a $\tau$CDM model 
($\Omat = 1.0$, $\Olam = 0.0$, $\sigma_8 = 0.5$, $\Gamma = 0.21$) whose power 
spectrum has been adjusted by hand so that its shape is similar to that of the 
$\Lambda$CDM model, as required by observations 
of large-scale structure (\S~\ref{sec:obsmassfcn}).
Mass-function evolution is quite pronounced in both models but is far stronger in 
the $\tau$CDM model.

\begin{figure*}
\includegraphics[width=6.5in , trim = 0.4in 0.3in 0.1in 0.5in , clip]
{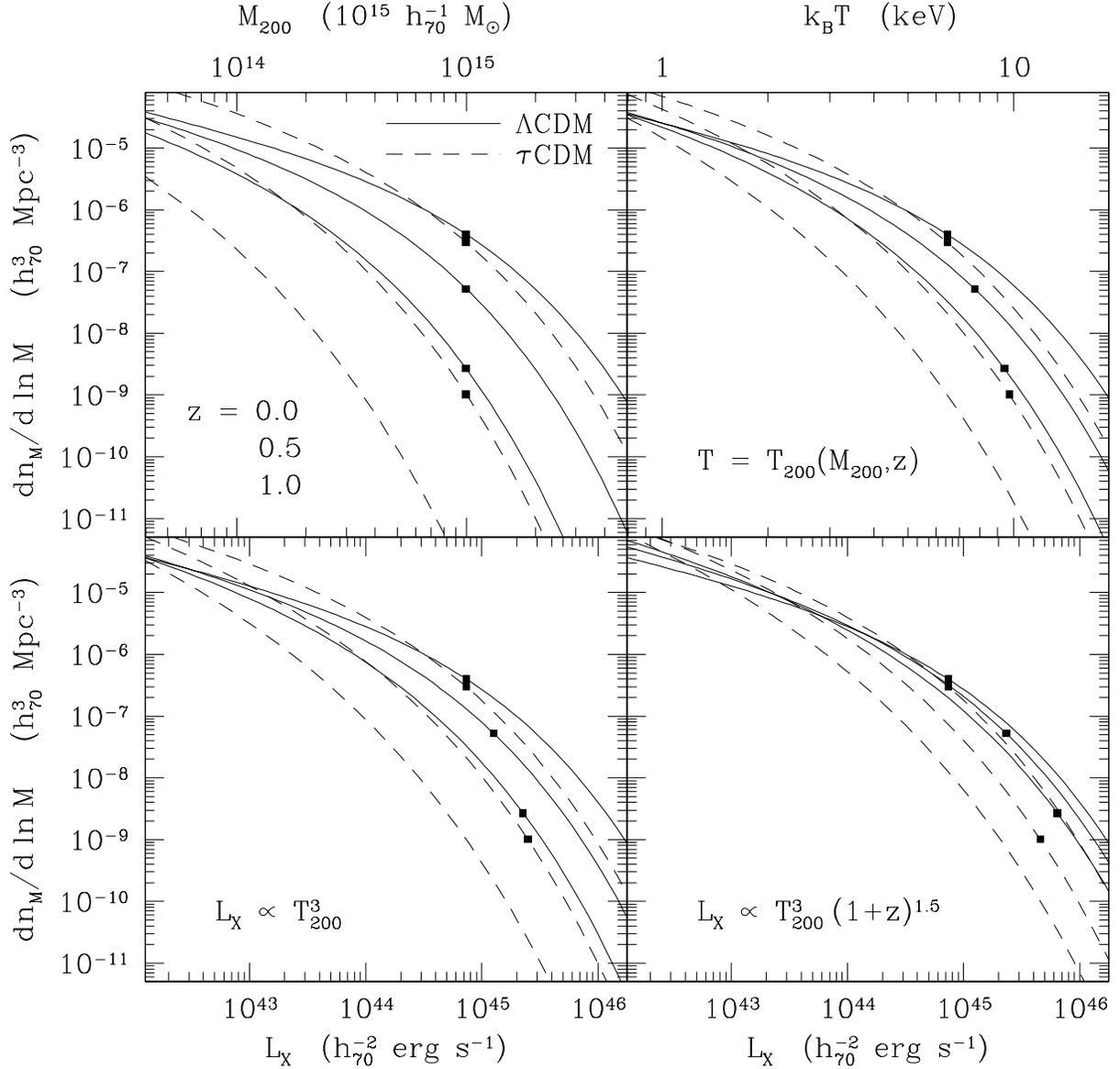}
\caption{Evolution of the cluster mass function and its manifestations in temperature 
and luminosity.  The two models in this comparison are $\Lambda$CDM ($\Omat = 0.3$,
$\Olam = 0.7$, $w = -1$, $\sigma_8 = 0.9$) and $\tau$CDM ($\Omat = 1.0$, $\Gamma = 0.21$,
$\sigma_8 = 0.5$).  Evolution of the mass function, shown at the upper left, is far more
pronounced in $\tau$CDM (dashed lines) than in $\Lambda$CDM (solid lines)
because it is so sensitive to the current matter density.  Each set
of three lines shows the differential mass function $dn_M / d \ln M$ at $z = 0$, 0.5, and 1.0,
from top to bottom, and black squares show the value of the mass function at a fiducial
mass of $10^{15} \, h_{70}^{-1} \, M_\odot$.  The upper right panel shows the same mass
functions plotted against temperature, assuming $T = T_{200}(M_{200},z)$.  Notice that
the higher-redshift curves have shifted to the right, weakening the evolution in temperature
space, because clusters of a given mass have higher temperatures at higher redshifts. 
In order to convert these curves to temperature functions, one would need to convolve
them with the scatter in the mass-temperature relation and multiply by $d \ln M / d \ln T 
\approx 1.5$.  The lower two panels show these same curves as a function of luminosity,
assuming $L_X = (6 \times 10^{44} \, h_{70}^{-2} \, {\rm erg \, s^{-1}}) T_{200}^3$ at $z = 0$ and two
different redshift dependences of the $\Lx$-$T$ relation.  In the case without $\Lx$-$T$ evolution
on the left-hand side, the curves are just relabeled versions of the ones in the upper-right panel.
However, the strong $\Lx$-$T$ evolution in the right-hand panel ($\Lx \propto T^3 (1+z)^{1.5}$)
shifts the three curves in the $\Lambda$CDM case nearly on top of one another at 
$\Lx \approx 10^{44} \, h_{70}^{-1} \, {\rm erg \, s^{-1}}$.  Convolving these curves with the
dispersion in the mass-luminosity relation and multiplying by $d \ln M / d \ln L \approx 0.5$
converts them to luminosity functions.}
\label{fig:mtlfunc}
\end{figure*}

Evolution in the mass-temperature relation weakens the observed amount of
cluster evolution when cluster number density is plotted as a 
function of temperature.  The upper right of   
Figure~\ref{fig:mtlfunc} shows the result of using a
$M_{200}$-$\Tlum$ relation with $\Tlum/T_{200} = 1$ and zero dispersion.  
Because clusters of a given mass are hotter at earlier times,
the higher redshift curves have translated to higher temperatures, 
compared with their positions in the upper-left panel.  Additional
mass-temperature evolution exceeding that predicted by the virial theorem and 
corresponding to values of $\Tlum/T_{200}$ that increase with redshift, would
further reduce the evolution, but there is currently no evidence for such 
evolution.

Redshift-dependent changes in the luminosity-temperature relation can have 
additional evolution-softening effects.   The $\Lx \propto \Tlum^3$ power-law form
of the relation appears to remain the same with redshift, but the amount of evolution 
in the normalization is uncertain.  Early assessments suggested no evolution
in the normalization \citep{ms97, Don99, borg99}.  The lower right of 
Figure~\ref{fig:mtlfunc} shows the evolution of cluster number density plotted
against luminosity for a non-evolving normalization 
and $\Lx = 6 \times 10^{44} \, h_{70}^{-2} \, {\rm erg \, s^{-1}} 
(\Tlum/ 6 \, \keV)^3$, again with no dispersion.
These curves differ from the temperature-function curves only in the labeling
of the horizontal axis.  More recent results indicate that higher-redshift 
clusters of a given temperature are more luminous, with an evolving relation  
$\Lx \propto \Tlum (1+z)^{b_{LT}}$ where $0.5 \lesssim b_{LT} \lesssim 1.5$ 
\citep[][see \S~\ref{sec:sc-evol}]{Vikh02, Lumb03, Ettori03}.  
Figure~\ref{fig:mtlfunc} shows the same distribution functions for $b_{LT} = 1.5$, 
at the high end of the suggested range.  The extra redshift dependence in this case 
slides the high-redshift curves even further to the right, roughly compensating 
for all of the evolution in the underlying mass function.

These examples underscore the importance of constraining evolution in the
mass-observable relations, even if the observables could be perfectly
measured.  In addition, one must bear in mind that the observations themselves can
introduce spurious redshift dependences in the mass-observable relations, 
largely because distant clusters are more difficult to observe than nearby ones.  
Optical projection effects become progressively harder to deal with at high
redshift, complicating observations of richness and velocity dispersion,
observations of weak lensing have fewer background galaxies to measure,
and the decline in X-ray surface brightness makes cluster temperature measurements
more difficult.  In many ways, the Sunyaev-Zeldovich effect is the most promising
observable for characterizing high-redshift clusters because its magnitude does
not depend on redshift (\S~\ref{sec:clsz}).  

There are three basic ways to deal with evolution in the normalization and
perhaps the scatter of a mass-observable relation:
\begin{itemize}
\item{Assume a model for the evolution of the relation.  Numerical simulations
          can be very helpful in providing a model for evolution of the normalization
          and scatter of mass-observable relations but give accurate results only
          if they include all the relevant physics.}
\item{Assume a parametric form for the mass-observable relation inspired 
          by theoretical models and try to calibrate it directly with 
          observations.  The normalization of the relation is usually assumed 
          to be proportional to either $(1+z)$ or $H(z)$ raised to 
          a power determined by a fit to observations.  In practice, however,
          the mass-observable relations for distant clusters are not directly calibrated. 
          What we have instead are relations that link one easily observed quantity, 
          such as X-ray luminosity, to another that is more closely related to mass, 
          like X-ray temperature or the weak-lensing distortion.}
\item{Assume a parametric form for the mass-observable relation and apply 
          self-calibration techniques to a large cluster survey to find the most
          likely parameters describing mass-observable evolution \citep{lsw02, Hu03,
          mm03a, mm03b}.  
          Parameters involving redshift-dependent scatter in the relation can
          also be included in such an analysis.  This technique is very promising 
          but requires large surveys of distant clusters which are not yet in hand.  
          Its accuracy is limited by the number of free parameters 
          needed to describe the mass-observable relations---the fewer, the 
          better.  Having a realistic physical model for 
          mass-observable evolution helps boost the accuracy achievable with
          self-calibration by reducing the number of unknown parameters.}
\end{itemize}
A decade from now, when much larger cluster samples will be available, 
self-calibration will probably be the best way to calibrate the
mass-observable relations.  In the meantime, it would be wise to spend
some effort on direct observational calibrations through cross-comparisons of
multiple mass-tracing observables.

\subsubsection{Constraints on Dark Matter}
\label{sec:omat}

Surveys of distant clusters find modest evolution in their comoving number
density fully consistent with cosmological models in which $\Omat \approx 0.3$.  
Because the rate of mass-function evolution at moderate redshifts ($z \lesssim 0.5$) 
is governed primarily by the overall matter density, this conclusion does not depend
strongly on the value of $\Olam$.  Here we focus on the constraints on
$\Omat$ derived from X-ray surveys, whose observables---$\Lx$, $\Tx$, and
baryonic mass---are related to to the spherical mass $M_{200}$
through simple parametric relations.  

Evolution in the X-ray temperature function was first observed by \citet{Henry97}, 
who showed that the comoving number density of $\sim 5 \, \keV$ clusters at 
$z \sim 0.35$ was only slightly smaller than it is today.  Assuming standard
evolution of the mass-temperature relation,
\citet{ecfh98} derived matter-density constraints $\Omat = 0.38 \pm 0.2$ for
$\Olam = 1 - \Omat$ and $\Omat = 0.44 \pm 0.2$ for $\Olam = 0$ from these data
using a maximum likelihood analysis to take full advantage of the sparse
temperature data.  More conservative analyses that simply counted clusters
hotter than a given temperature found weaker constraints \citep{vl99}.
\citet{Henry00} provides a complete discussion of the cluster temperature
data and the maximum-likelihood analysis technique.

\begin{figure}
\includegraphics[width=3.2in , trim = 0in 0in 0in 0in , clip]
{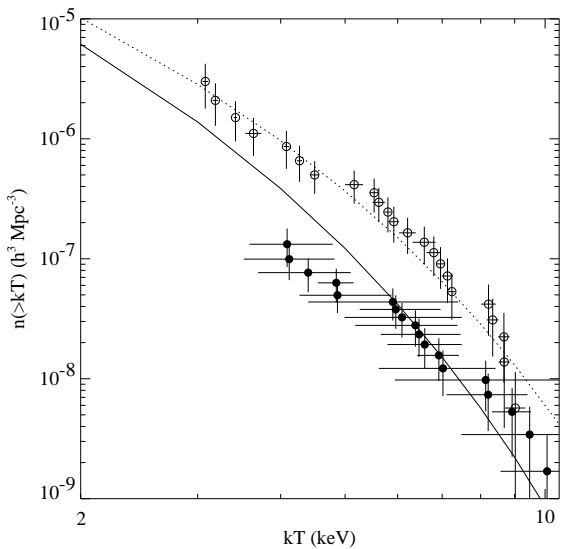}
\caption{Observed evolution in the integrated cluster temperature function $n(>kT)$
giving the comoving number density of clusters with temperatures exceeding $kT$. 
Open circles give the low-redshift temperature function from a sample of clusters
with mean redshift $z = 0.051$.  Filled circles give the observed temperature function
of clusters with a mean redshift of $z = 0.429$.  The data points in each case are 
correlated because $n(> kT)$ at a given temperature is a cumulative function 
depending on all data points at higher temperatures.  Lines give the predicted
temperature functions at $z = 0.051$ (dotted line) and $z = 0.429$ (solid line)
for the best-fitting model: $\Omat = 0.28$, $\Olam = 0.98$, $\sigma_8 = 0.68$.  
(Figure courtesy of Pat Henry.)}
\label{fig:tfunc_obs}
\end{figure}

Temperature measurements of a handful of hot clusters at higher redshifts have 
shown that the rate of cluster temperature evolution remains modest at higher 
redshifts \citep{Donahue96, dvglhs98, Don99, Henry00}.  The comoving number
density of $\gtrsim 8 \, \keV$ clusters at $z \sim 0.5 - 0.8$ is no less than about one
tenth of its current value, in strong disagreement with the standard expectation
in an $\Omat = 1$ universe (see Figure~\ref{fig:mtlfunc}).  Including these hot, distant clusters in 
the analysis further strengthens the constraints on the matter density, ruling out
$\Omat = 1$ at the 3$\sigma$ level for standard mass-temperature evolution
\citep{dvglhs98, bf98, dv99, Evrard02}.  In order
for such hot clusters to exist in a flat, matter-dominated universe, the mass-temperature
relation would have to evolve in a non-standard way, with an increase in either
the scatter or the normalization at $z \gtrsim 0.5$.  \citet{Evrard02} have
shown that the $\tau$CDM mass function of Figure~\ref{fig:mtlfunc} is consistent with the
temperature function observations only if the mass-temperature normalization
factor $\Tlum/T_{200}$ is $\sim$1.5 times higher at $z \sim 0.5$ than at present.
Such a big change seems unlikely in light of alternative observations of these 
hot, high-redshift clusters that agree with the large masses inferred from the standard 
normalization \citep{lg95, dvglhs98, Tran99}. 

Observations of evolution in the X-ray luminosity function have greater
statistical power because many more clusters have known luminosities than
have known temperatures, but uncertainties in luminosity-temperature evolution
dilute the constraints they place on $\Omat$.  Many X-ray surveys have shown 
that the comoving number density of clusters at a given luminosity changes
very little from redshift $z \sim 0.8$ to the present for $\Lx \lesssim 10^{44} \,
{\rm erg \, s^{-1}}$;  significant evolution is seen only for clusters with 
$\Lx \gtrsim 10^{45} \, {\rm erg \, s^{-1}}$ \citep{rbn02, Mullis04}.
Evolution this mild is generally expected in cosmological models with 
$\Omat \approx 0.3$.  Strong evolution in the luminosity-temperature relation
must occur in models with $\Omat = 1$ for the observed evolution in the luminosity function
to be so weak (see Figure~\ref{fig:mtlfunc}).
An extensive analysis by \citet{borg01} of luminosity-function evolution in the 
{\em ROSAT} Deep Cluster Survey, which extends to $z \sim 1$, indicates that 
$\Omat = 0.35^{+0.13}_{-0.10}$, where the error bars signify the 1$\sigma$
confidence interval.  Models with $\Omat = 1$ fall outside the 3$\sigma$
confidence interval, even when the normalization of the luminosity temperature
relation is allowed to vary with redshift as $\Lx \propto \Tlum (1+z)$.  


\begin{figure}
\includegraphics[width=3.4in , trim = 0.1in 0.1in 0.1in 0.1in , clip]
{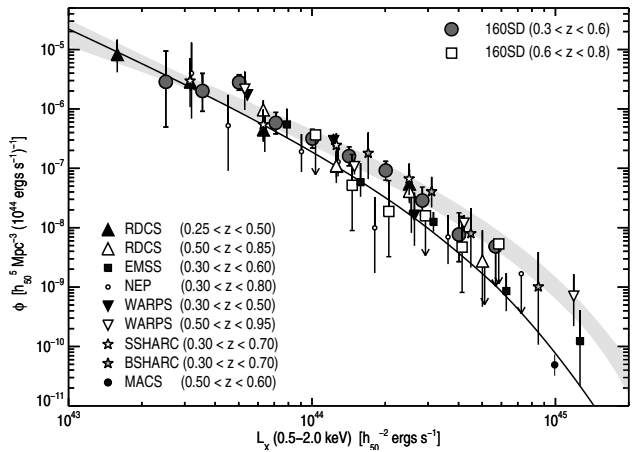}
\caption{Observed evolution in the cluster luminosity function.  Many different
cluster surveys spanning the range $0 \lesssim z \lesssim 1$ are shown on this
figure.  The vertical axis gives the luminosity functions derived from these surveys
in terms of $\phi \equiv dn/d \Lx$,and the shaded region shows the luminosity
function at $z \approx 0$.  Significant evolution is seen only at $\Lx \gtrsim
10^{45} \, {\rm erg \, s^{-1}}$, consistent with $\Lambda$CDM models with
a moderate amount of evolution in the $\Lx$-$T$ relation (see Figure~\ref{fig:mtlfunc}.)
 (Figure from \citet{Mullis04}.)}
\label{fig:lfunc_obs}
\end{figure}

The evolution of the baryon mass function observed with X-ray telescopes
agrees with the conclusions drawn from the luminosity and temperature
functions.  \citet{Vikh03} have measured the baryon mass function
in a sample of clusters at $z \sim 0.5$, finding that the comoving number
density of massive clusters at that redshift is roughly one
tenth of the current value.  This result implies $\Omat = 0.25 \pm 0.1$
(1$\sigma$ confidence interval) for $\Olam = 1 - \Omat$.

Optical studies concur that cluster evolution has been relatively modest since
$z \sim 0.5$, buttressing the conclusion that $\Omat < 1$.  In fact, the 
evolution of optically selected clusters appears even milder than the evolution 
in X-ray selected clusters \citep[e.g.,][]{Postman02}, which would imply an 
even smaller value of $\Omat$.  However, it is not yet clear how much of
this discrepancy arises from differences between the projected masses 
measured by optical surveys and the spherical masses measured by
X-ray surveys.

\subsubsection{Constraints on Dark Energy}
\label{sec:olam}

Existing cluster surveys, taken by themselves, do not yet place strong constraints
on dark energy, but that situation is likely to change in the coming decade, with the
advent of large, deep cluster surveys in the optical, X-ray, and microwave bands.
Currently, the most interesting information that clusters provide about dark energy 
comes from combining the results of cluster surveys with other information.
If the overall geometry of the universe is indeed flat, as seems quite evident 
from the temperature patterns in the cosmic microwave background \citep[e.g.,][]{WMAP03}, 
then the matter density inferred from clusters implies $\Olam = 1 - \Omat = 
0.7 \pm 0.1$, in agreement with measurements of $\Olam$ from the supernova 
magnitude-redshift relation \citep{Riess98, Perl99}.   Geometrical arguments involving clusters
provide weaker support for this conclusion.  If the baryon fraction of clusters at a given 
temperature is assumed to remain constant with time, then the transverse sizes of 
clusters as a function of redshift can be used to constrain the geometry of the universe.  
Studies using such methods disfavor $\Olam = 0$ \citep{Mohr00, aan02}.

Large cluster surveys extending to $z \sim 1$ have the potential to place much stronger
constraints on the dark-energy parameters $\Olam$ and $w$, independent of
other information, as long as these surveys are large enough to permit self-calibration 
of the mass-observable relationships \citep{hhm01, lsw02, wbk02}. 
The accuracy achievable with self-calibration depends
critically on the nature of cluster evolution, because the self-calibration
procedure requires evolution of the relevant mass-observable relation to be 
expressed in a parametric form.  Constraints on the cosmological parameters
are considerably weaker if the actual evolution does not follow the assumed 
parametric form.   However, cross-calibration of mass-observable 
evolution through intensive supplementary observations of a small subset 
of the large survey restores much of the potential inherent in self-calibration 
\citep{mm03a}. 

Including information on cluster bias inherent in a large cluster survey
further tightens the constraints on dark energy.  Because
the tendency of clusters to cluster with one another depends in a simple
way on the cosmological model (\S~\ref{sec:bias}), folding this information into the
self-calibration procedure improves the accuracy with which cosmological
parameters can be measured \citep{mm03b}.  Figure~\ref{fig:mm04} shows
how the estimated constraints on $\Omat$ and $w$ tighten when 
information about cluster bias is added.  It assumes that the universe
is flat ($\Olam = 1 - \Omat$) and considers three different planned cluster
surveys: two large Sunyaev-Zeldovich surveys (SPT and {\em Planck}) and
a large X-ray survey (DUET), each of which will find 20,000 to 30,000 cluster
to $z \gtrsim 1$ \citep[see][for details]{mm03b}.  In the most optimistic
cases, the parameters $\Omat$, $\Olam$, and $w$ will be measured with $\sim$5\% accuracy. 

\begin{figure}
\includegraphics[width=3.4in , trim = 2.3in 0.15in 0.3in 0.4in , clip]
{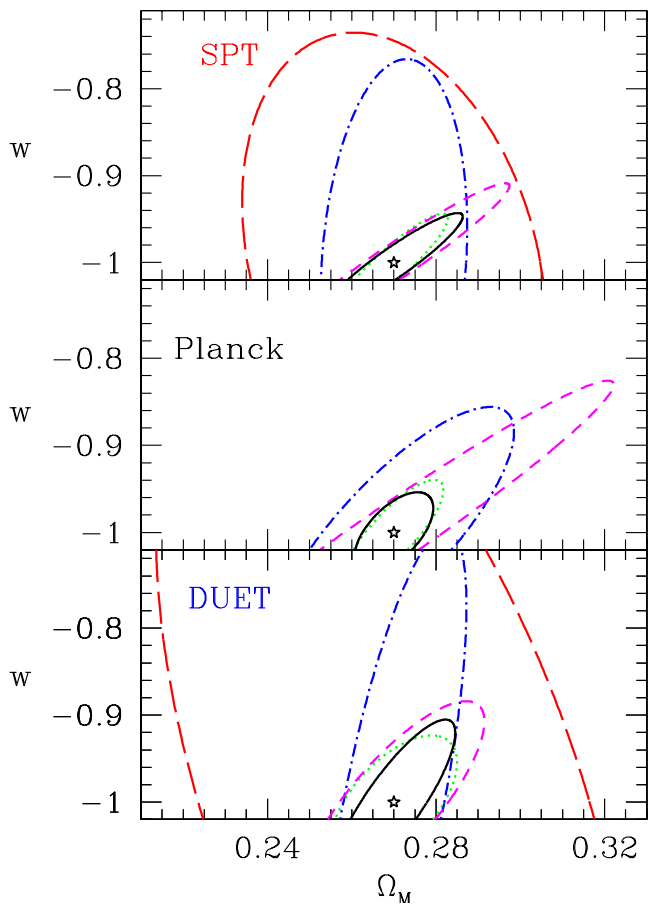}
\caption{Expected constraints on cosmological parameters from self-calibrated surveys.
The SPT and {\em Planck} surveys will find 20,000 to 30,000 clusters through the S-Z
effect.  DUET is a proposed X-ray survey designed to find $\sim 20,000$ clusters.
Dotted contours show the expected constraints on $w$ and $\Omega_M$ from 
self-calibration if redshift evolution of the cluster observables behaves exactly 
according to the standard scaling relations, allowing the redshift dependences
of those scaling relations to be fixed.  Long-dashed contours show how the 
constraints loosen when redshift evolution is determined as part of the self-calibration.
Dot-dashed contours show how the contraints begin to tighten when information
about cluster bias is included in the calibration.  Solid contours show the best-case
scenario in which the self-cailbration includes both information about cluster bias
and supplementary follow-up calibrations of a small subset of the survey.  (Figure from
\citet{mm03b}.) } 
\label{fig:mm04}
\end{figure}

A large survey also minimizes the sample variance that arises from 
cluster bias \citep{Evrard02, hk03}.   Because
clusters tend to be clustered, the variance in the number of clusters within small
sample volumes is larger than the gaussian expectation, adding systematic
uncertainty to the measured mass function.  This effect is generally not large for current 
cluster surveys but should be taken into account if one is designing a cluster survey 
for making high-precision cosmological measurements.

In summary, observations of cluster evolution already constrain the density
of gravitating matter to be $\Omat \approx 0.3 \pm 0.1$, meaning that 
$\Olam \approx 0.7 \pm 0.1$ if the universe is flat.  Using this value of
$\Omat$ to break the $\Omat$-$\sigma_8$ degeneracy leads to $\sigma_8
\approx 0.7-1.0$, depending on the mass-temperature calibration. The major
source of uncertainty in all these cosmological parameters comes not from 
the statistics of the survey but rather from uncertainties in the normalization
and rate of evolution in the mass-observable relations.  In order to better
understand these relations and how they evolve, we need to know how
galaxy formation affects the evolution of the stuff we can observe---the baryons 
in clusters.  That is where we turn our attention next.

\section{Evolution of the Baryonic Component}
\label{sec:baryons}

Those who are cosmologists at heart are interested in how galaxy formation
affects the intracluster medium primarily because they would like to know how 
to measure cluster masses more accurately.  Those who are astronomers at heart 
are interested in the intracluster medium as well, but for them its main attraction 
is that the hot gas contains valuable information about the physical 
processes that govern galaxy formation.   Clarifying the connections between
galaxy formation and the mass-observable relations is therefore important
to both of these lines of research.

One of the nagging mysteries in our current picture of the universe is why so few 
of the universe's baryons have turned into stars \citep{wr78, wf91, Cole91}.  
Numerical simulations of cosmological structure formation 
that include baryonic hydrodynamics and the radiative cooling processes
that lead to galaxy formation predict that $\gtrsim 20$\%
of the baryons should have condensed into galaxies, but $\lesssim 10$\%
have been found in the form of stars \citep[e.g.,][]{bpbk01}.   
Some form of feedback, involving supernovae and perhaps outflows from
active galactic nuclei, seems to have stymied condensation, but we are 
still largely ignorant about how this feedback works.  

Galactic winds like those observed from nearby starburst galaxies, in which
multiple clustered supernovae are driving the powerful outflows, are likely 
to be important in regulating early star formation, but observational constraints 
on the mass and energy flux in such winds are sketchy at best 
\citep{Martin99, Heckman02},  particularly at high 
redshift \citep{Pettini00, Pettini01, Adel03}. 
These galactic winds presumably had a dramatic impact on the 
intergalactic medium and subsequent galaxy formation, 
with effects that may have persisted until the present day
\citep[e.g.,][]{ob03, BenMad03}. 
Likewise, quasars and other forms of activity driven by 
black-hole growth in the nuclei of young galaxies may
also have produced powerful outflows with lasting consequences
for the intergalactic gas, but the energy input from these objects
is still highly uncertain \citep{v94, v96, is01, nr02, so04}. 

Unfortunately, the low-redshift intergalactic medium, where most of the universe's 
baryons are thought to reside, is notoriously hard to observe.  Because the majority 
of this gaseous matter remains undetected, it is sometimes referred 
to as the ``missing baryons'' \citep[e.g.][]{co99}.
A handful of quasars are bright enough beacons for probing the missing baryons
via absorption-line studies with the ultraviolet spectrographs on the
Hubble Space Telescope \citep[e.g.,][]{ssp96, pss02}, and that
number will increase if the Cosmic Origins Spectrograph 
is installed on Hubble. However, the inferences drawn from 
such studies depend critically on the uncertain heavy-element abundance
and ionization state of these intergalactic clouds \citep[e.g.,][]{tsj00, stg03}. 

Clusters of galaxies are still the only places in the 
universe where we have anything approaching a complete 
accounting of intergalactic baryons, their thermal state, 
and their heavy-element enrichment.  Thus, observations of 
the intracluster medium can provide unique insights into the cooling
and feedback processes that govern galaxy formation.  
In order to interpret the signatures of galaxy
formation in the intracluster medium we need to understand how 
the thermodynamic properties of today's clusters are linked to 
the physics of the intergalactic baryons at $z \gtrsim 2$, the epoch 
of galaxy formation.

This section of the review discusses the current understanding
of the interactions between galaxy formation and the intracluster
medium, focusing in particular on how those interactions affect
the mass-observable relations so crucial to cosmology.
It begins by outlining the properties that clusters would have
if radiative cooling of the universe's baryons and subsequent galaxy formation
were suppressed.  Because these properties do not agree with
observations, radiative cooling and galaxy formation must
somehow have altered the structure of the intracluster medium, 
with important consequences for the mass-observable relations.  The middle of this
section summarizes some of the recent progress that has been
made in understanding the role of galaxy formation and its
impact on the observable properties of clusters.  It then concludes
with a brief discussion of the existing constraints on baryon
condensation in clusters.

\subsection{Structure Formation and Gravitational Heating}
\label{sec:gravheat}

People who study clusters of galaxies are sometimes asked
how the X-ray emitting gas gets so hot.  The answer to that
question is simple.  If radiative cooling is negligible,
then gravitationally driven processes will heat diffuse gas to the 
virial temperature of the potential well that confines it.
A tougher question would be to ask why the intracluster
medium has the density that it does.  In order to answer
that question, one needs to know what produces the entropy
of the X-ray emitting gas.  Without galaxy formation in the
picture, shocks driven by hierarchical structure formation are
the only source of entropy for the intracluster medium, and 
this mode of entropy production leads to clusters whose
density and temperature structures are nearly self-similar.

\subsubsection{Intracluster Entropy}
\label{sec:entropy}

Entropy is of fundamental importance for two reasons:
it determines the structure of the intracluster 
medium and it records the thermodynamic history of the cluster's 
gas.  Entropy determines structure because high-entropy gas floats 
and low-entropy gas sinks.  A cluster's intergalactic gas therefore 
convects until its isentropic surfaces coincide with the equipotential 
surfaces of the dark-matter potential.  Thus, the entropy 
distribution of a cluster's gas and the shape of the dark-matter 
potential well in which that gas sits completely determine the 
large-scale X-ray properties of a relaxed cluster of galaxies.  
The gas density profile $\rho_g(r)$ and temperature profile $T(r)$ 
of the intracluster medium in this state of convective and 
hydrostatic equilibrium are just manifestations of its entropy 
distribution.
  
This review adopts the approach of other work in this field and
defines ``entropy'' to be
\begin{equation}
  K \equiv \frac {\kB T} { \mu m_p \rho_g^{2/3} }  \; \; .  
\end{equation}
The quantity $K$ is the constant of proportionality in the equation 
of state $P = K \rho_g^{5/3}$ for an adiabatic monatomic gas, 
and is directly related to the standard thermodynamic entropy 
per particle, $s = \kB \ln K^{3/2} + s_0$, where $s_0$ is a constant
that depends only on fundamental constants and the mixture of
particle masses.
Another quantity frequently called ``entropy" in the cluster literature 
is $S = \kB T n_e^{-2/3}$.  In order to avoid confusion with the classical
definition of entropy, we will call this quantity $K_e$.  For the
typical elemental abundances in the intracluster medium, one
can convert between these definitions using the relation 
\begin{eqnarray}
  K_e & = & \kB T n_e^{-2/3}  \\
 ~      &  = & 960 \, \keV \, {\rm cm^2} \, 
        \left( \frac {K} {10^{34} \, {\rm erg \, cm^2 \, g^{-5/3}}} \right) \; \; . \nonumber
\end{eqnarray}
A cluster achieves convective equilibrium when $dK/dr \geq 0$ everywhere, 
and the entropy distribution that determines the gas configuration in
this state can be expressed as $K(M_g)$, where the inverse
relation $M_g(K)$ is the mass of gas with entropy $< K$.

Comparisons between the entropy distributions of clusters that differ 
in mass can be simplified by casting those distributions into
dimensionless form \citep[e.g.,][]{vbbb02}.  Combining the mean
density of dark matter within the scale radius $r_{200}$, the 
global baryon fraction $f_b = \Obar / \Omat$, and the characteristic
halo temperature $T_{200}$ gives the characteristic entropy
scale
\begin{eqnarray}
  K_{200} & = & \frac { \kB T_{200} } { \mu m_p (200 f_b \rhocr)^{2/3} }   \\
       ~        & = & \frac {1} {2} \left[  \frac {2 \pi} {15} 
                                  \frac {G^2 M_{200}} {f_b H(z)} \right]^{2/3}  \; \; . \nonumber 
\end{eqnarray}
For $f_b = 0.022 h^{-2}$, this entropy scale reduces to
\begin{eqnarray}
  K_{e,200} & =  & 362 \, \kB \Tlum \, {\rm cm^2}  \left( \frac {T_{200}} {\Tlum}  \right)
                    \nonumber \\
           ~ & ~ & \; \; \; \; \; \; \times
                           \left[ \frac {H(z)} {H_0} \right]^{-4/3} \left(  \frac {\Omat} {0.3} \right)^{-4/3} \; \; .
  \label{eq-ke200}
\end{eqnarray}
Writing the entropy scale in this way makes explicit the fact that
the observed temperature of a cluster is not necessarily a reliable
guide to the characteristic entropy $K_{200}$ of its halo.  If the intracluster
medium of a real cluster is either hotter or cooler than $T_{200}$,
then one must apply the correction factor $T_{200} / \Tlum$ when
computing the cluster's value of $K_{200}$.

\subsubsection{Entropy Generation by Smooth Accretion}
\label{sec:smoothacc}

One way to approach the problem of gravitationally driven
entropy generation is through spherically symmetric
models of smooth accretion, in which gas passes through 
an accretion shock as it enters the cluster \citep[e.g.,][]{kp97,
tn01, vbblb03}.
If the incoming gas is cold, then the accretion shock is the
sole source of intracluster entropy.  If instead the incoming gas has
been heated before passing through the accretion shock,
then the Mach number of the shock is smaller and the intracluster
entropy level reflects both the amount of preheating and the production
of entropy at the accretion shock. 

Let us first consider the case of cold accreting gas,
in which the pressure and entropy of the incoming gas are
negligible.  Suppose that mass accretes in a series of
concentric shells, each with baryon fraction $f_b$,
that initially comove with the Hubble flow as in the
spherical collapse model of \S~\ref{sec:sphercoll}.
In this simple model, a shell that initially encloses 
total mass $M$ reaches zero velocity at the turnaround 
radius $r_{\rm ta}$ and falls back through an accretion
shock at a radius $r_{\rm ac}$ in the neighborhood of the virial
radius $r_{\rm ta}/2$.  

Because the cold accreting gas is effectively pressureless, 
the equations that determine the postshock entropy are
\begin{eqnarray}
  \dot{M}_g & = & 4 \pi r_{\rm ac}^2 \rho_1 v_{\rm ac}  \\
  v_{\rm ac}^2 & = & \frac {2GM} {r_{\rm ta}} \label{eq-vac} \\
 \kB  T_2 & = & \frac {1} {3} \mu m_p v_{\rm ac}^2  \label{eq-tps} \\
  \rho_2 & = & 4 \rho_1 \; \; \label{eq-rhops} ,
\end{eqnarray}
where $\dot{M_g} = f_b \dot{M}$ is the gas accretion rate, $\rho_1$ is the 
preshock gas density,  $T_2$ and $\rho_2$ are postshock quantities,
and the accretion radius has been set to $r_{\rm ac} = r_{\rm ta}/2$.
Equations (\ref{eq-tps}) and (\ref{eq-rhops}) are restatements
of the jump conditions for strong shocks, assuming that the
postshock velocity is negligible in the cluster rest frame
\citep[e.g.,][]{LanLif59, cmt97}, 
and equation (\ref{eq-vac}) is exact only for 
cosmologies with $\Olam = 0$.

The postshock entropy produced by smooth accretion of
cold gas at time $t$ is therefore
\begin{eqnarray}
 K_{\rm sm} & = & \frac {v_{\rm ac}^2} {3 (4 \rho_1)^{2/3}} \\ 
      ~     & = & \frac {1} {3} 
 		\left( \frac {\pi G^2} {f_b} \right)^{2/3}
                \left[ \frac {d \ln M} {d \ln t} \right]^{-2/3} 
                (Mt)^{2/3}
              \nonumber \; \; .
  \label{eq-kcold}
\end{eqnarray}
Because the entropy generated at the shock front increases monotonically
with time, such an idealized cluster never convects but rather accretes shells of baryons
as if they were onion skins.  The resulting entropy distribution in dimensionless form is 
\begin{eqnarray}
 \frac {K_{\rm sm}(M_g)} {K_{200}} & = & \frac {2} {3} \left( \frac {15} {2} \right)^{2/3}
                      (H_0 t_0)^{2/3}   \; \; \; \nonumber \\
 ~ & ~ & \; \; \, \times     
                 \left[ \frac {d \ln \eta} {d \ln t} \right]^{-2/3}
                 \left[ \frac {\eta t(\eta)} {t_0} \right]^{2/3} \, ,
\label{eq-khatcold}
\end{eqnarray}
where $\eta \equiv M_g(t)/f_b M_{200}(t_0)$ is effectively a radial
coordinate corresponding to the amount of gas accreted by
time $t$ divided by the amount accreted by the present time $t_0$.
Given these assumptions, the entropy profile arising from smooth 
accretion of cold gas  depends entirely on the mass accretion history 
$M(t)$, and the profiles of objects with similar accretion
histories should be self-similar with respect to $K_{200}$.

This simple model yields entropy distributions whose overall shape
agrees with cluster observations \citep{vbblb03}.   The rate at which 
a cluster accretes matter through hierarchical structure formation 
depends on the growth function $D(t)$ and the power-law slope 
$\alpha_M \equiv d \ln \sigma^{-1} / d \ln M$ of the perturbation spectrum 
on the mass scale of the cluster:  $M(t) \propto [D(t)]^{1/2 \alpha_M}$ 
\citep{lc93, vd98}.    Clusters ranging in mass from 
$10^{14} M_\odot$ to $10^{15} M_\odot$ grow roughly as $M(t) \propto t$ to
$M(t) \propto t^2$ in the concordance model \citep{tn01, vbblb03}. 
Plugging these growth rates into equation (\ref{eq-kcold}) leads 
to entropy distributions between $K \propto M_g$ and $K \propto M_g^{4/3}$. 
Throughout much of a cluster, the gas mass encompassed within a given
radius rises approximately linearly with radius (\S~\ref{sec:sx}), meaning that
the $K(r)$ relation should be slightly steeper than linear.  Numerical models 
of smooth accretion by \citet{tn01} find $K(r) \propto r^{1.1}$.
The entropy profile observed outside the core regions of clusters also obey 
$K(r) \propto r^{1.1}$ \citep{pa02, pa03}, but the extent to which this agreement 
is coincidental is not clear.

If the accreting gas is not cold, then the intracluster entropy profile
produced by smooth accretion has an isentropic core with an entropy
level similar to the preshock entropy \citep{bbp99, tn01}. 
A non-zero initial entropy level $K_1$ changes the cold-accretion model outlined
above by altering the jump conditions represented by equations (\ref{eq-tps}) 
and (\ref{eq-rhops}).   When $K_1$ is no larger than the entropy generated at the
accretion shock, then the entropy profile created by smooth accretion of warm gas
can be closely approximated by adding $0.84 K_1$ to the entropy profile 
$K_{\rm sm}(M_g)$ from the cold-accretion case \citep{dd02, vbblb03}. 
If $K_1$ is large compared with $K_{\rm sm}$, then the accretion
shock is weak or non-existent, and accretion is nearly adiabatic, leading to
an isentropic entropy profile with the constant value $K_1$.

\subsubsection{Entropy Generation by Hierarchical Merging}
\label{sec:entmerg}

In real clusters the accreting gas is lumpy, not smooth, which transforms
the nature of entropy generation.  Incoming gas associated with accreting 
sublumps of matter enters the cluster with a wide range of densities.  
There is no well-defined accretion shock but rather a complex network of 
shocks as different lumps of infalling gas mix with the intracluster medium 
of the main halo.   Numerical simulations of this process beginning 
with cosmological initial conditions produce clusters that 
have nearly self-similar entropy structure \citep[e.g.,][]{nfw95}, 
as expected from the scaling properties of hierarchical 
structure formation \citep{Kaiser86}.

Figure~\ref{fig:kr_sph} shows entropy profiles of 30 clusters generated with a
numerical simulation of a $\Lambda$CDM cosmology including 
hydrodynamics but not radiative cooling \citep{Kay04, vkb04}. 
The masses of these clusters span more than a factor of 10, but
when their entropy profiles are divided by the appropriate value 
of $K_{200}$, they lie nearly on top of one another, at least 
outside the approximate core radius $0.1 r_{200}$.
This result is not unique to the simulation method.  All codes with
sufficiently high resolution find that non-radiative clusters have
approximately self-similar entropy structure, and consequently
self-similar density and temperature structure \citep{f99_sb, vkb04}.

\begin{figure}
\includegraphics[width=3.4in , trim = 1in 5in 2.0in 0.2in , clip]
{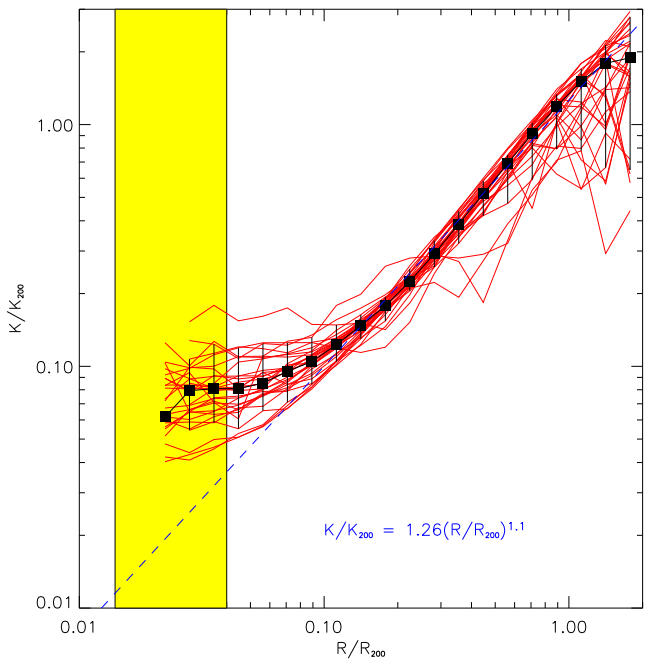}
\caption{ \footnotesize
Dimensionless entropy $K/K_{200}$ as a function of scale radius $r/r_{200}$ for 30 clusters
simulated without radiative cooling or feedback.  Black squares show the median profile, and the 
dashed line illustrates the power-law relation $K/K_{200} = 1.26 (r/r_{200})^{1.1}$.  Most 
of the entropy profiles shown lie close to this relation in the radial range 
$0.1 \lesssim r/r_{200} \lesssim 1.0$.  At smaller radii, the entropy profiles generally
flatten, and their dispersion increases. This flattening is likely to be a real effect, as
it sets in well outside the shaded box showing the gravitational softening
length of the simulation.
\label{fig:kr_sph}}
\end{figure}

The self-similarity of the entropy profile in non-radiative clusters is
a very useful point of comparison for sleuthing the effects of galaxy
formation.  Deviations from this baseline profile are likely to be
due to a combination of radiative cooling and the feedback processes that ensue.
\citet{vkb04} find that a good representation of the baseline 
entropy profile produced outside the cores of clusters by hierarchical 
structure formation is given by the power law 
$K_{\rm PL} = (1.35 \pm 0.2) K_{200} (r/r_{200})^{1.1}$.
Specifying the baseline entropy profile within the cluster core ($< 0.1 r_{200}$)
is more difficult both because there is more dispersion in that region among
simulated clusters and because the results there depend somewhat on the
hydrodynamical method used in the simulations.

Another notable aspect of self-similarity in non-radiative clusters is
that the gas density profile and the dark-matter density profile outside
$0.1 r_{200}$ have virtually identical shapes \citep{nfw95, f99_sb}.  
This feature leads to another useful approximation
to the entropy profiles of non-radiative clusters \citep{Bryan00}.  
One can specify the gas density profile by assuming
that it obeys an NFW density profile (\S~\ref{sec:massprof}) with the same concentration
as the dark matter and a total baryon mass $f_b M_{200}$ within $r_{200}$ 
and then compute the temperature profile that would keep the gas in 
hydrostatic equilibrium.  The temperature and density profiles in this 
kind of model approximately obey the polytropic relation
$T(r) \propto [\rho(r)]^{\gammaeff -1}$, with $\gammaeff \approx 1.1-1.2$ 
\citep{ks01, vbbb02}.  Combining them
produces an alternative baseline entropy profile that depends on the 
concentration parameter $c_{200}$ of the dark-matter halo and that
this review will denote as $K_{\rm NFW} (r)$.   

Despite the complexity of the shock structure in hierarchical accretion, 
the numerically simulated entropy profiles are similar in shape to those
created by smooth accretion models \citep{borg01, borg02}.  
However, these profiles have lower overall entropy levels 
than the smooth-accretion profiles \citep{vbblb03}. 
Figure~\ref{fig:kcomp} demonstrates this point by comparing the two approximations, 
$K_{\rm PL}$ and $K_{\rm NFW}$, 
of simulated non-radiative clusters with two entropy profiles drawn 
from smooth accretion models,  one from the numerical computations 
of \citet{tn01} and the other from equation (\ref{eq-khatcold}) 
assuming $M \propto t^{3/2}$ and $H_0 t_0 = 1$, which 
are reasonable assumptions for $\Lambda$CDM models.

\begin{figure}
\includegraphics[width=3.5in , trim = 1.1in 1.0in 0.9in 1.0in , clip]
{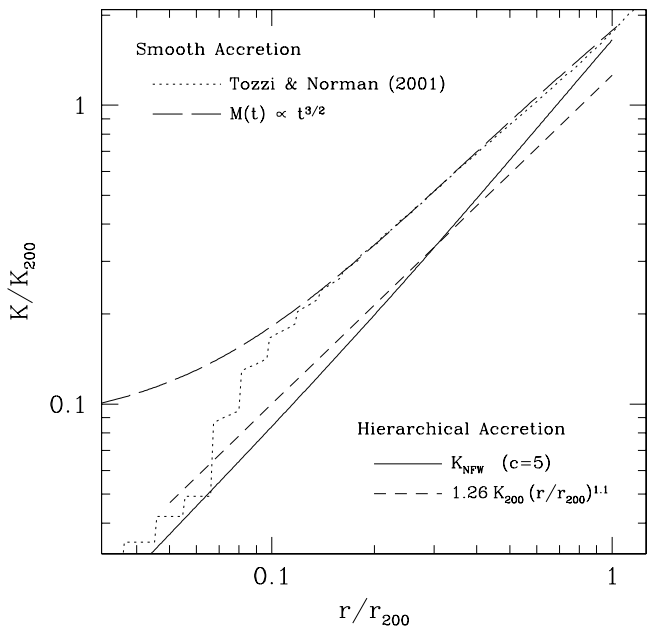}
\caption{Entropy profiles from smooth accretion and hierarchical accretion.
Smoothing of the gas accreting onto a cluster boosts entropy production while
maintaining the characteristic $K(r) \propto r^{1.1}$ entropy profile.  The two
lower lines illustrate approximate entropy profiles produced by hierarchical
accretion, including the power-law expression from Figure~\ref{fig:kr_sph}
and the $K_{\rm NFW}$ model described in the text, with $c_{200} = 5$. 
The two upper lines illustrate entropy profiles resulting from smooth accretion models,
including profile computed by \citet{tn01} for a $10^{15} \, h^{-1} \, M_\odot$ 
cluster with $300 \, {\rm keV \, cm^2}$ of preheating and radiative cooling
implemented (dotted line) and a profile computed from equation~(\ref{eq-khatcold})
and preheating amounting to $0.1 K_{200}$.  The two smooth models run
roughly parallel to the hierarchical accretion models but their normalizations
are $\sim 1.5$ times higher.
\label{fig:kcomp}}
\end{figure}

The likely reason for this discrepancy is that smooth accretion maximizes
entropy production because it minimizes the mean mass-weighted density
of accreting gas \citep{vbblb03, psf03}.   Smoothing the accreting 
gas does not change the accretion velocity but does reduce the 
mean density of accreting gas lumps.  Because postshock entropy 
scales as $v_{\rm ac}^2 \rho_1^{-2/3}$, the mean entropy of lumpy accreted gas 
is therefore less than in the smooth-accretion case.  This effect of smoothing
might not be entirely academic, because the observed entropy profiles of 
low-temperature clusters show a similar entropy boost relative to the 
baseline profile \citep{vp03}.

\subsubsection{Observed Entropy Profiles}
\label{sec:obsent}

Astronomers have known for more than a decade that the structure of the
intracluster medium in real clusters cannot be self-similar because the 
luminosity-temperature relation of clusters does not agree with self-similar
scaling \citep[][\S~\ref{sec:mlrel}]{es91, Kaiser91, eh91}.
Only within the last couple of years has the nature of that deviation from self-similarity
become clear.  High-quality cluster observations with the {\em XMM-Newton} 
satellite are showing that intracluster entropy profiles have the $K(r) \propto
r^{1.1}$ shape characteristic of gravitational structure formation outside of
the core, but the overall normalization of these profiles scales as $\Tlum^{2/3}$
instead of $\Tlum$, as in the baseline profiles \citep{pa03}. 
Analyses of much larger cluster samples observed with earlier
X-ray telescopes have arrived at the same conclusion.   Instead of self-similarity 
with $K(r/r_{200}) \propto \Tlum$,  \citet{psf03} find altered similarity with
$K(r/r_{200}) \propto \Tlum^{2/3}$ at both the core radius $0.1 r_{200}$ and
farther out in clusters, at the scale radius $r_{500} \approx 0.66 r_{200}$.
The question to be answered is therefore how galaxy formation and feedback
manage to produce such a shift in the overall normalization of cluster entropy
profiles without substantially changing their shape. 

\subsection{Galaxy Formation and Feedback}
\label{sec:feedback}

In the decade since astronomers became aware of similarity breaking in clusters
there have been many numerical simulations devoted to understanding it.   
Our understanding of this problem remains incomplete because including
galaxy formation in cosmological models of cluster formation is a formidable
computational challenge, requiring codes that simulate three-dimensional hydrodynamics 
spanning an enormous dynamical range in length scales and that track 
a large number of physical processes.  The volume required to model a cosmologically
significant sample of clusters is of order $10^{26}$~cm in linear scale, individual galaxies have
sizes $\sim 10^{23}$~cm, star-forming regions within those galaxies can be as small
as $10^{19}$~cm, and the stars themselves are only $\sim 10^{11}$~cm in size.
Sophisticated hydrodynamical techniques are now able to model the formation
of the first stars from cosmological initial conditions (Abel, Bryan, \& Norman 2001),
but are far from being able to track in detail the formation of an entire galaxy's worth
of stars, let alone all the feedback processes that can occur.

For the time being, the difficulty of solving this problem from first principles means 
that modelers have to be selective about the physical processes and conditions 
that merit modeling.  Important clues to what the essential processes 
are can be gleaned from the observed characteristics of clusters.  This part of
the review sifts through some of those clues, showing that radiative cooling is likely
to be the process that sets the entropy scale of similarity breaking but that radiative
cooling cannot act alone.  Otherwise, too much baryonic matter would condense
into stars and cold gas clouds.

\subsubsection{Preheating}
\label{sec:preheating}

Early approaches to the problem of similarity breaking in clusters
postulated that some sort of heating process imposed 
a universal minimum entropy---an ``entropy floor''---on the intergalactic
gas before it collected into clusters \citep{eh91, Kaiser91}.  
Imposing a global entropy floor helps to bring the theoretical $\Lx$-$\Tlum$
relation into better agreement with observations because this extra entropy
makes the gas harder to compress in cluster cores, where entropy is smallest, 
particularly in the shallower potential wells of low-temperature clusters.
This resistance to compression breaks cluster similarity by lowering 
the core density and therefore the X-ray emissivity in low-$T$ clusters 
more than in high-$T$ clusters, thereby steepening the $\Lx$-$\Tlum$ relation.

According to this preheating picture, the core entropy level and scaling relations
of clusters should reflect the global entropy floor produced at early times. 
Initial measurements of entropy at the core radius $r_{0.1}$ demonstrated
that low-temperature clusters had greater amounts of entropy than
expected from self-similarity and suggested that the level of the entropy 
floor was $\sim 135 \, \keV \, {\rm cm^2}$ \citep{pcn99, lpc00}. 
This result matched well with numerical simulations 
of cluster formation with preheating levels of $50 - 100 \, \keV \, {\rm cm^2}$
that produced clusters with approximately the right $\Lx$-$\Tlum$ relation
\citep{bem01}. 

However, simple preheating now appears to be too crude an explanation
for similarity breaking.  In the preheating picture, low-temperature clusters
should have large isentropic cores \citep{bbp99, tn01}, 
but this prediction disagrees with the observations showing that the shapes
of cluster entropy profiles do not depend significantly on temperature (\S~\ref{sec:obsent}).
In addition, the abundant evidence for intergalactic gas at $\lesssim 10^5$~K
from quasar absorption line studies clearly shows that preheating cannot
be global at $z \gtrsim 2$, and the preheating models themselves do not
explain why the level of the entropy floor should be 
$\sim 135 \, \keV \, {\rm cm^2}$.

\subsubsection{Radiative Cooling}
\label{sec:cooling}

In contrast, the observed entropy scale of similarity breaking emerges naturally
from the process of radiative cooling.  Intergalactic gas both inside and outside of
clusters radiates thermal energy at a rate given by the cooling function
$\Lambda_c (T)$, described in more detail in \S~\ref{sec:sx}.  Cooling that radiates
an energy $\Delta q$ per particle reduces the entropy by $\Delta \ln K^{3/2} = 
\Delta q / \kB T$.  Thus, the equation expressing these radiative losses 
can be written 
\begin{equation}
  \frac {d K^{3/2}} {dt} = - \frac {3} {2} \frac {K_c^{3/2}(T)} {t_0} \; ,
  \label{eq-dkdt}
\end{equation}
where
\begin{equation}
  K_c(T) = \left[ \frac {2} {3} 
                        \left( \frac {n_e n_p} {\rho^2} \right)
                        \frac {(\kB T)^{1/2} \Lambda_c(T)} {(\mu m_p)^{1/2}}
                          \right]^{2/3} t_0^{2/3}
\label{eq-kc}
\end{equation}
is the entropy level at which constant-density gas at temperature $T$ 
radiates an energy equivalent to its thermal energy in the time $t_0$. 
The latter formula reduces to 
\begin{equation}
  K_c(T) \approx 81 \, \keV \, {\rm cm^2} \left( \frac {t_0} {14 \, {\rm Gyr}} \right)^{2/3}
                   \left( \frac {T} {1 \, \keV} \right)^{2/3} \; \; 
 \label{eq-kcbrems}
\end{equation}
when pure bremsstrahlung cooling is assumed. 

The fact that the entropy threshold below which gas cools within the universe's 
lifetime is close to the entropy floor inferred from clusters with $\sim2$~keV  temperatures 
suggests that radiative cooling sets the entropy scale for similarity breaking
\citep{vb01}.   \citet{vp03} further quantify this point.  Figure~\ref{fig:kcool}
shows how entropy measurements at $0.1 r_{200}$ in a large sample of clusters
\citep{psf03} compare with the cooling threshold $K_c(T)$ for gas with
heavy-element abundances equal to 30\% of their solar values relative to hydrogen. 
Both the measured core entropies and the entropy threshold for cooling
scale as $T^{2/3}$, and they are approximately equal, although
the scatter in the data is quite significant.

\begin{figure}
\includegraphics[width=3.4in , trim = 0in 0in 2.4in 5.45in , clip]
{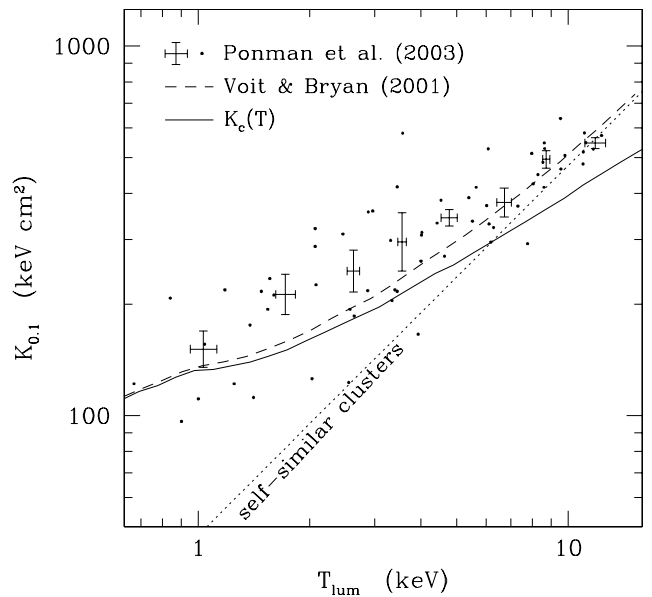}
\caption{ 
Comparison between entropy measured at $0.1 r_{200}$ and the cooling
threshold in a large sample of clusters.  Small points show the entropy $K_{0.1}$
measured at $0.1 r_{200}$ in a sample of 64 clusters, and points with error
bars show the mean entropy measurement in temperature bins of eight clusters
each.  The dotted line gives the mean entropy predicted by simulations of clusters
without radiative cooling of feedback.  The solid line shows the value of the 
cooling threshold $K_c(T)$ computed for heavy-element abundances 0.3 times
their solar values and $t_0 = 14$~Gyr.  The dashed line shows the entropy predicted
at $0.1 r_{200}$ by the simple analytical model of \citet{vb01}.
\label{fig:kcool}}
\end{figure}

Section~\ref{sec:sc-cool} below shows that radiative cooling also accounts well for the
scaling relations of global X-ray properties like $\Lx$ and $\Tlum$ with
mass.  However, casting equation (\ref{eq-kcbrems}) in dimensionless
form illustrates why at least some feedback must compensate for cooling:
\begin{equation}
  \frac {K_c(T)} {K_{200}} \approx 0.2 \, (Ht)^{2/3} \left[ \frac {H(z)} {H_0} \right]^{2/3}
               \left( \frac {T} {1 \, \keV} \right)^{-1/3} \; \; .
\end{equation}
The cooling threshold in low-temperature clusters at the present time is $\sim$20\%
of the characteristic entropy $K_{200}$ and greater than that if emission-line
cooling from heavy elements is included.  At earlier times, the dimensionless
cooling threshold is even higher, meaning that a large proportion of the baryons
belonging to the progenitor objects that ultimately assembled into present-day clusters
should have condensed into stars or cold gas clouds.  This is one of the 
manifestations of the classic overcooling problem of hierarchical galaxy formation 
\citep{wr78, Cole91, bvm92}. 
Because the observed mass ratio of stars to hot gas in clusters is only about
10\% (\S~\ref{sec:condensation}), wholesale baryon condensation doesn't 
seem to have happened.  

Recognition of this overcooling problem led \citet{vb01} to propose
a way for radiative cooling to determine the entropy scale of similarity breaking 
without acting alone.  The basic idea is that gas with entropy less than $K_c(T)$
cannot persist indefinitely.  It must either cool and condense or be heated until its entropy exceeds 
$K_c(T)$.  At any given time, feedback is triggered by condensing gas parcels with entropy 
less than the cooling threshold and acts until those parcels are eliminated by either
cooling, heating, or some combination of the two.  Thus, the joint action of cooling
and feedback imprint an entropy scale corresponding to the cooling threshold, regardless
of how strong the feedback is.  This kind of effect has now been seen in a number of
numerical simulations that include cooling and differing forms of feedback 
\cite{dkw02, borg02, Vald03, ktt03, Borg04}.

The fact that similarity breaking is not very sensitive to the efficiency of feedback 
is good news for cosmologists but bad news for astrophysicists.  It offers hope that
we can understand the mass-observable relations of clusters without solving all 
the messy astrophysical problems of feedback.  Yet, it also implies that the 
mass-observable relations alone do not tell us much about the nature 
of that feedback.  Instead, we must look to the spatially resolved entropy profiles 
of clusters \citep{vp03, Kay04} and the ratio of condensed baryons 
to hot gas in clusters \citep{bpbk01, borg02, ktt03, Borg04}. 

\subsubsection{Feedback from Supernovae}
\label{sec:SNe}

Supernovae are the most obvious candidate for supplying the feedback that
suppresses condensation, but it is not clear that supernova heating and
the galactic winds it drives can provide enough entropy to keep the fraction 
of condensed baryons below about 10\%.  Heavy-element abundances
in clusters imply that the total amount of supernova energy released during
a cluster's history amounts to $\sim 0.3 - 1 \, \keV$ per gas particle in the intracluster
medium \citep{fad01, pmbb02}.   The amount of energy
input needed to explain the mass-observable relations while avoiding overcooling
is $\sim 1$~keV \citep{wfn01, vbbb02, tbsmmm03}, at the upper end of the range inferred
from heavy elements, but the transfer of supernova energy to the intracluster 
medium must be highly efficient, which seems unlikely \citep{ky00}.    
Supernova energy would have to be converted to almost entirely
to thermal energy with very little radiated away.  

In order to avoid radiative losses, supernova heating must raise the entropy
of the gas it heats to at least $100 \, {\rm keV \, cm^2}$.  An evenly distributed
thermal energy input of order $1 \, {\keV}$ would therefore have to go into
gas significantly less dense than $10^{-3} \, {\rm cm^{-3}}$ to avoid such
losses.  Gas near the centers of present-day clusters, not to mention the
galaxies where supernovae occur, is denser than that, particularly at 
earlier times when most of the star formation happened.  Simulations
that spread supernova feedback evenly therefore produce too many
condensed baryons in clusters \citep{borg02}.  Artificial algorithms 
that target supernova feedback at gas parcels that would otherwise cool 
are more successful at preventing overcooling \cite{ktt03}.   However,
efforts to implement a more realistic version of targeted feedback in the
form of galactic winds are still not entirely successful at preventing overcooling
\citep{Borg04}. 

It remains to be seen whether supernova feedback alone can account for the
observed entropy profiles of clusters.  \citet{vbblb03} and \citet{psf03} 
have proposed that entropy input from galactic winds preceding the accretion 
of gas onto clusters could lead to a form of entropy amplification 
that would explain the observations.  If galactic winds are strong enough 
to significantly smooth out the lumpiness of the local intergalactic gas, 
then the mode of accretion of this gas onto clusters will be closer to smooth accretion
than to hierarchical accretion, thereby boosting the entropy generated through accretion
shocks without changing the profile's characteristic shape.  This effect is a 
plausible explanation for the altered similarity of the observed entropy profiles,
but it has not yet been thoroughly tested in simulations.  Intriguing results by
\citet{Kay04} show that an extremely targeted feedback model, in which supernovae 
heat the local gas to $1000 \, {\rm \keV \, cm^2}$, successfully reproduces both
the normalization and shape of the observed entropy profiles.

\subsubsection{Feedback from Active Galactic Nuclei}
\label{sec:AGNs}

If supernovae cannot prevent overcooling, then perhaps supermassive black
holes in the nuclei of galaxies are what stop it \citep{vs99, wfn01, clm02}. 
The omnipresence of supermassive black holes at the centers of galaxies 
\citep{Magorrian98} and the excellent correlation of their masses 
with the bulge and halo properties of the host galaxy \citep{fm00, Gebhardt00} 
strongly suggest that the growth of black holes in 
the nuclei of galaxies goes hand-in-hand with galaxy formation.  Furthermore, 
the centers of many clusters with low-entropy gas whose cooling time is 
less than the age of the universe also contain active galactic nuclei that are 
ejecting streams of relativistic plasma into the intracluster medium \citep{Burns90}.  
It is therefore plausible that supermassive black holes at the centers of clusters 
provide feedback that suppresses further cooling whenever condensing 
intracluster gas accretes onto the central black hole.

Such a feedback loop is attractive and consistent with the circumstantial 
evidence, but the precise mechanism of heating remains unclear.
The bubbles of relativistic plasma being inflated by the active galactic
nuclei in clusters appear not to be expanding fast enough to shock heat the 
intracluster medium because the rims of the bubbles are no hotter
than their surroundings \citep{McNamara00, Fabian00}.  
Also, if active galactic nuclei simply injected
heat energy into the center of a cluster, then one would expect to
see a flat or reversed entropy gradient in clusters with strong nuclear
activity, indicating that convection is carrying heat outward.  
Instead, the entropy gradients in these cluster cores increase
monotonically outward \citep{David01, Horner04}.  
One possibility is that heating is episodic
\citep{kb03} and that we have not yet found a cluster
in the midst of an intense heating episode.  Another is that heating
is somehow spread evenly throughout the cluster core in a way that
maintains the entropy gradient \citep{bk02, rb02}.  
Yet another possibility is that
bursts of relativistic plasma drive sound waves into the intracluster
medium that eventually dissipate into heat \cite{Fabian03}. 

Unfortunately, the none of these heating mechanisms have yet
been tested in the context of cosmological structure formation, so
we do not know their overall impact on either baryon condensation
or the global entropy profiles of clusters.  Also, many aspects of 
the relationship between cosmology and nuclear activity in galaxies 
remain highly uncertain.  A major role for quasar feedback is plausible.
However, the connection between 
the growth of central black holes in galaxies and galaxy formation 
itself is not well understood, and the efficiency with which black holes 
convert accretion energy into outflows is unknown. 

\subsubsection{Transport Processes}
\label{sec:cond}

Heat transport processes like thermal conduction and turbulent mixing
may also mitigate radiative cooling because gas that condenses
sets up a temperature gradient along which heat energy can flow.  In gas without
magnetic fields, electrons conduct heat along temperature gradients
giving a heat flux $\kappa_s \nabla T$, with $\kappa_s \approx 6 \times 10^{-7} 
\, T^{5/2} \, {\rm erg \, cm^{-1} \, s^{-1} \, K^{-7/2}}$ \citep{Spitzer62}, the so-called
Spitzer rate, valid when the scale length of the temperature gradient is
longer than the electron mean free path.  Clusters with central cooling times
less than $H_0^{-1}$ indeed tend to have positive temperature gradients
within the central $\sim 100$~kpc, raising the possibility that heat conduction 
at least partially balances radiative losses.  Many models for conduction in 
cluster cores have been developed \citep[e.g.,][]{tr83, bm86, bd88, rt89, Sparks92}, 
but conduction does not satisfactorily balance radiative cooling.   
Temperature-gradient observations are inconsistent with steady-state balance 
between cooling and conduction in a number of cluster cores 
\citep{Horner04, vf04}.   However, mixing of hot gas with cooler gas 
facilitated by intracluster turbulence \citep{kn03b} or AGN activity \citep{bk02}
could enhance the effectiveness of heat conduction. 

It is possible that cooling, conduction, feedback, and perhaps mixing as well are all 
needed for a complete solution that explains the observed core temperature gradients 
without overcooling.  Conduction that balances cooling in a steady state 
has often been dismissed on the grounds that it is not stable enough to 
preserve the observed temperature and density gradients 
for periods of order $\gtrsim 1$~Gyr \citep{cb77, Fabian94}. 
Because of conduction's extreme sensitivity to temperature, it is difficult for 
radiative cooling and conduction to achieve precise thermal balance 
with a globally stable temperature gradient \citep{bd88, Soker03}. 
On the other hand, conduction would have to be suppressed by at least two
orders of magnitude for radiative cooling to produce the observed gradients
\citep{bc81, fhcg81}.  Recent theoretical analyses 
of conduction have concluded that this level of suppression is unrealistically 
high \citep{Maly01, mk01, nm01}.  
Combining cooling, conduction, and feedback offers a
way out of this dilemma.  Hybrid models in which conduction compensates
for cooling in the outer parts of the core while feedback from an active galactic
nucleus compensates for it in the inner parts have had some success 
in reproducing the observations \citep{rb02, bm03}.

\subsection{Galaxy Formation and Cluster Observables}
\label{sec:scaling}

Earlier we saw that the X-ray properties of the self-similar clusters produced 
by purely gravitational structure formation do not agree with observations (\S~\ref{sec:gravheat}).  
Observed clusters of a given mass appear to be hotter than their theoretical 
counterparts and also less luminous, especially at the cool end of the 
cluster temperature range.  Such disagreements have been worrisome 
to cosmologists who would like to understand what governs
the cluster observables used to measure mass, but these problems 
are on their way to being solved.  Both analytical work and 
hydrodynamical simulations performed during the last several 
years are showing that the observed $\Lx$-$\Tlum$, $M_{200}$-$\Tlum$, 
and $\Lx$-$M_{200}$ relations are natural outcomes of galaxy formation.  
Significant uncertainties remain, but the theoretical foundation for the mass-observable 
relations essential for probing cosmology with clusters is growing firmer.

\begin{figure*}
\includegraphics[width=7.0in , trim = 0.4in 1.7in 0.5in 2.0in , clip]
{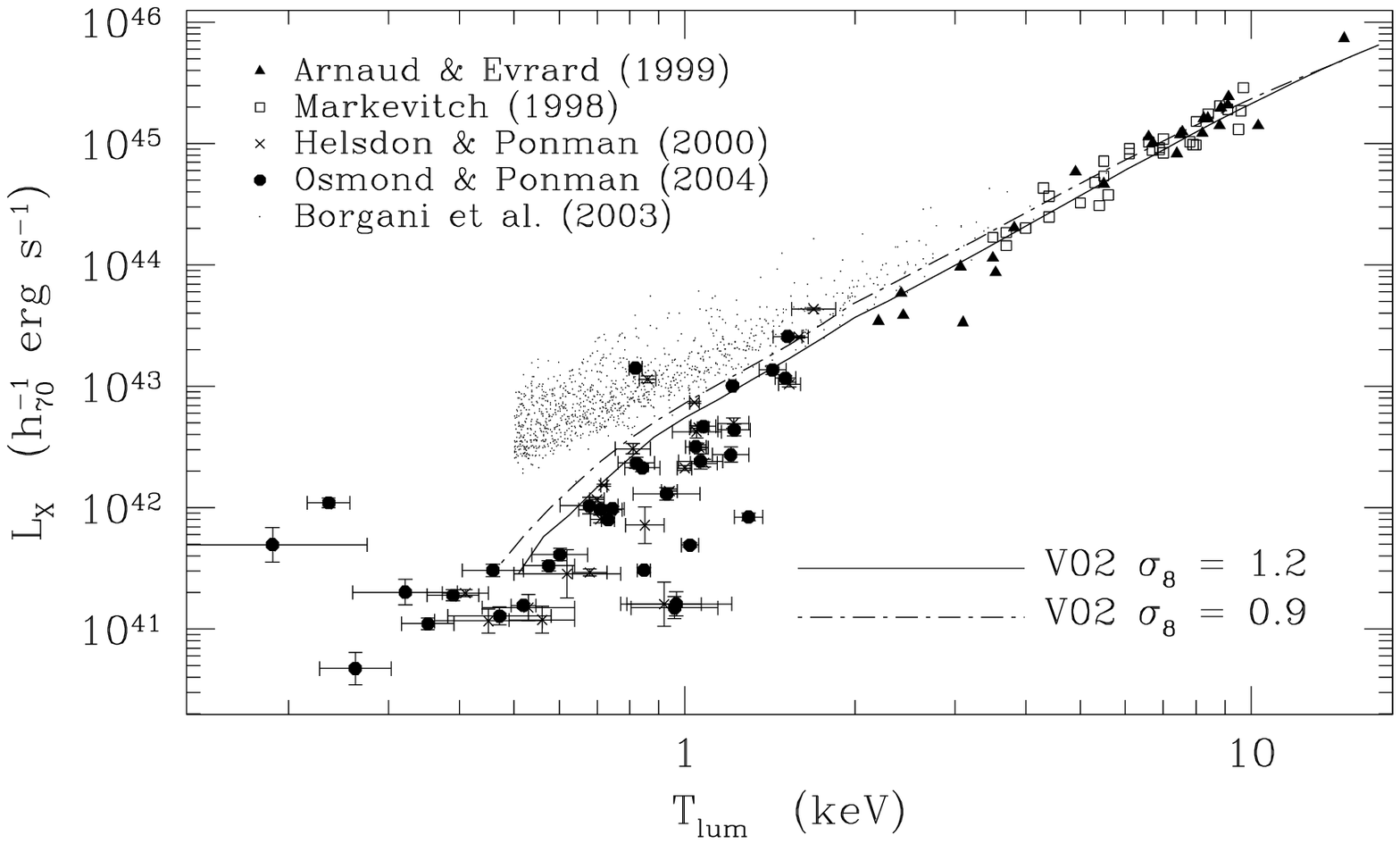}
\caption{ 
Luminosity-temperature relation.  Points show cluster data from \citet{ae99} (solid
triangles) who avoided clusters with cool cores, cluster data from \citet{mark98} (open
squares) with cool cores excised, two sets of group data from \citet{hp00} (crosses) and
\citet{op04} (solid octagons) that were not corrected for cool cores, and simulated 
clusters from \citet{Borg04} (small points).  These 
simulations implement radiative cooling and supernova feedback in the form of galactic winds.
Lines show modified-entropy models from \citet{vbbb02} with entropy truncated at the cooling
threshold.  There is a slight dependence on $\sigma_8$ in these models because
higher values of $\sigma_8$ lead to dark-matter halos with more concentrated cores.
Both the analytical and numerical models agree well with the data at $k_B \Tlum \gtrsim
2 \, \keV$.  Agreement is less good at lower temperatures, but the reasons for the disagreements
are unclear.  More feedback may be needed in the numerical models to suppress the
luminosities, and the large scatter in the observations at $\lesssim 1$~keV may reflect 
a wide range in the effectiveness of feedback.
\label{fig:ltrel}}
\end{figure*}

\begin{figure*}
\includegraphics[width=7.0in , trim = 0.2in 1.2in 0.5in 2.5in , clip]
{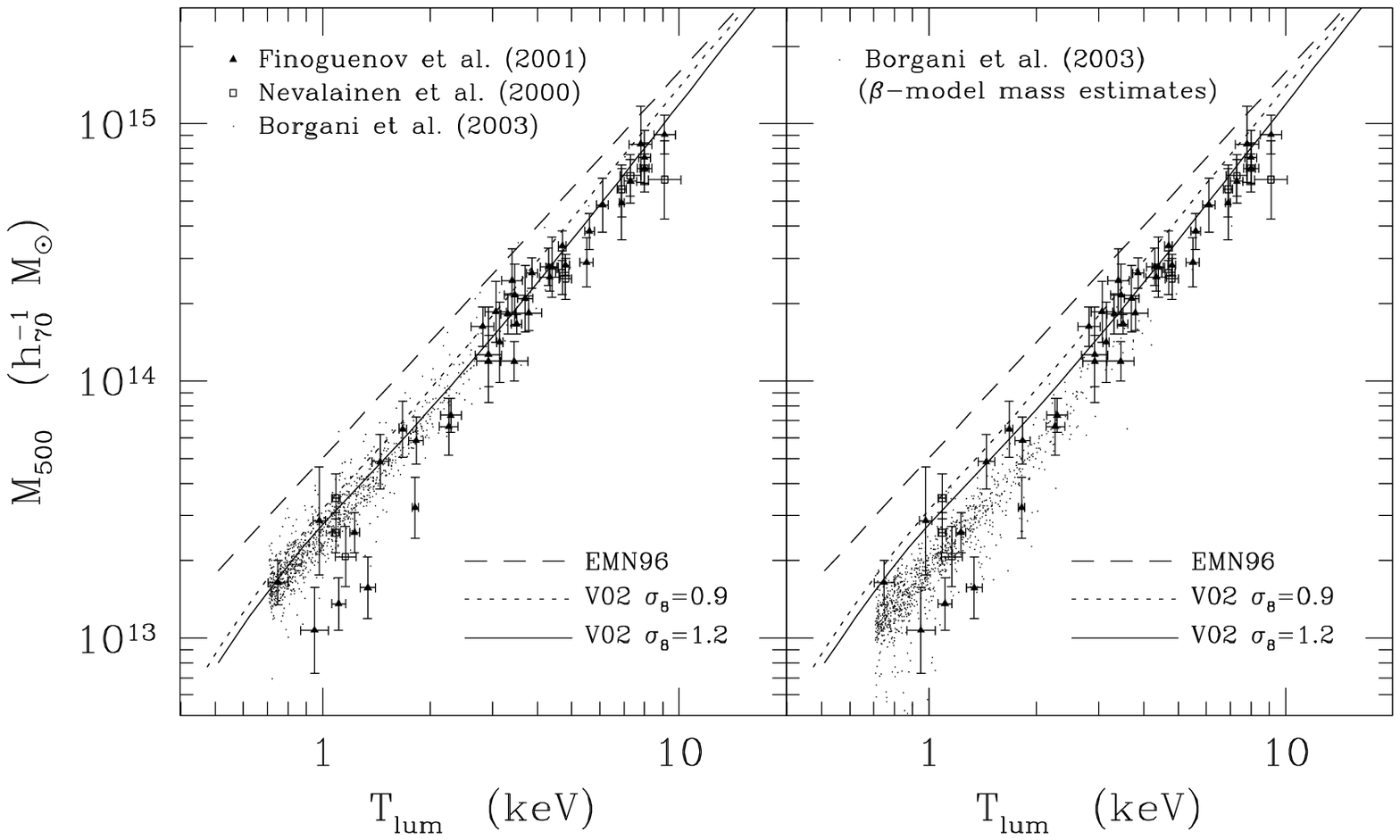}
\caption{ 
Mass-temperature relation.  Large points show cluster data from \citet{frb01} (solid triangles) 
and \citet{nmf00} (open squares), in which cluster masses were inferred from fitting polytropic
beta models (see \S~{sec:tx}).  Dashed lines illustrate the $M_{500}$-$\Tlum$ relation measured
in clustered simulated without cooling and feedback by \citet{emn96}, which clearly disagree
with the data points.  The other lines show the $M_{500}$-$\Tlum$ relations predicted
by the analytical models of \citet{vbbb02}, which agree much better with the data.  
There is a slight difference between models with $\sigma_8 = 0.9$ (dotted lines) 
and $\sigma_8 = 1.2$ because higher values of $\sigma_8$ lead to clusters with 
higher halo concentrations that produce slightly higher temperatures.  Tiny points
show data for clusters simulated by \citet{Borg04} with radiative cooling and feedback
in the form of supernova-driven galactic winds.  The left-hand panel uses the actual
values of $M_{500}$, which agree with the analytical models.  The right-hand panel
uses values of $M_{500}$ inferred from fitting polytropic beta models to the observations,
which underestimate true cluster masses, especially at low temperature, suggesting
there may be a systematic observational bias in this method of mass measurement.
\label{fig:mtrel}}
\end{figure*}

\subsubsection{Role of Cooling}
\label{sec:sc-cool}

Radiative cooling turns out to the most important process to include.  While
it might seem paradoxical, allowing the intracluster medium to radiate thermal
energy actually causes its luminosity-weighted temperature to rise.  The reason
for this behavior is that cooling selectively removes low-entropy gas from the
intracluster medium, raising the mean entropy of what remains \citep{kp97, ptce00, Bryan00}.
In non-radiative cluster simulations, the entropy of 
gas in the vicinity of the cluster core is below the cooling threshold $K_c$.  This
aspect of non-radiative models is unphysical, because gas with entropy less 
than $K_c$ would radiate an amount of energy greater than its total thermal 
energy content over the course of the simulations.   When cooling is allowed 
to occur, this low-entropy core gas condenses out of the intracluster medium 
and is replaced by higher entropy core gas having a higher temperature, 
a lower density, and therefore a lower luminosity.

A simple analytical model illustrates the effect of the cooling threshold on the 
$\Lx$-$\Tlum$ and $M_{200}$-$\Tlum$ relations \citep{vb01, vbbb02, wx02}. 
The model assumes that the intracluster entropy distribution in
the absence of galaxy formation would be the $K_{\rm NFW}(M_g)$ distribution derived
from the density profile of the dark matter.  Because condensation and feedback both act to
eliminate gas below the cooling threshold, the model simply truncates the entropy
distribution at $K_c(T_{200})$ and discards all the gas with lower entropy. 
One can interpret this gas removal either as condensation 
or as extreme feedback that heats the sub-threshold
gas to a much higher entropy level.  This cooling and feedback need not
occur at the center of the cluster.  In a hierarchical cosmology, much of the low-entropy
gas cools, condenses into galaxies, and produces feedback long before the
cluster is finally assembled.

Computing the hydrostatic configuration of the modified entropy distribution in 
the original dark-matter potential gives $\Lx$ and $\Tlum$ as a function of the mass 
$M_{200}$ and concentration $c_{200}$ of the dark-matter halo.  
Figures~\ref{fig:ltrel} and \ref{fig:mtrel} show that the resulting $\Lx$-$\Tlum$ and 
$M_{200}$-$\Tlum$ relations generally agree well with observations but may slightly 
overpredict $\Lx$ for objects cooler than $\sim 2 \, \keV$ and do not account for the
large scatter at low temperatures.
There are no free parameters in this model, other than the cosmological parameters, 
because the $M_{200}$-$c_{200}$ relation and the age of the universe used to 
compute $K_c$ depend only on cosmology, and the heavy-element abundance 
used to compute the cooling threshold is taken from observations.  

Numerical simulations in which feedback is either weak or non-existent produce 
clusters whose properties are quite similar to the ones in this simple analytical model. 
Early numerical investigations of cooling in individual clusters
gave inconclusive results \citep{so98, Lewis00}, but 
simulations by \citet{mtkpc01} showed that adding cooling to a large-scale 
cluster simulation could give an $\Lx$-$\Tlum$ like the observed one.  Subsequent
numerical work has confirmed that result \citep[e.g.,][]{dkw02, borg02, Vald03, ktt03}.  
Adding radiative cooling to the cosmological model produces good agreement with 
observations at all cluster temperatures $\gtrsim 2 \, \keV$.

Even when the simulations implement strong feedback, the X-ray scaling relations 
change remarkably little from the cooling-only case \citep{borg02, ktt03}. 
The main effect on the $\Lx$-$\Tlum$ relation of adding strong feedback to simulations
that already include cooling is to slightly reduce the luminosity 
of cool ($\lesssim 2 \, \keV$) clusters, bringing them into
better agreement with observations.  This insensitivity to the efficiency 
of feedback is another strong indication that the cooling threshold governs
the entropy scale for similarity breaking.  

One point of disagreement between the analytical models, the simulations, and the
observations concerns the central temperature gradient.  Many observed clusters have a
relatively small amount of gas in their cores whose cooling time is 
less than the age of the universe, and in 
those clusters the core temperature gradient is generally positive ($d T /dr > 0$).
In the simple analytical models outlined above, no gas is allowed to be below the cooling
threshold, resulting in a core that is nearly isentropic and thus has a negative temperature
gradient ($dT/dr < 0$).  Likewise, simulations with cooling and feedback also tend
to have flat or negative temperature gradients in the neighborhood of the core radius
($\sim 100$~kpc).  

This problem deserves attention because elevated core temperatures in models
with cooling are what bring the theoretical $M_{200}$-$\Tlum$ relation into agreement
with observations.  Making the analytical model slightly more realistic brings the 
predicted temperature gradient into better agreement with observations.  
The discontinuous cooling threshold applied by the simplest models
is overly crude because it completely removes gas just below the 
threshold while gas just above the threshold does not
cool at all.  Instead, cooling acts upon the entropy distribution as described by equation
(\ref{eq-dkdt}).  \citet{vbbb02} show that modifying the baseline profile $K_{\rm NFW}$ using
this equation with $T = T_{200}$ for a time $t_0$ leads to an entropy distribution
that reproduces the observed temperature gradients.  

Simulations involving pure cooling do not agree with this result.  The temperature-gradient 
discrepancy between analytical models and simulations in the pure-cooling case is still 
not understood, but may have something to do with the implicit stability of the cooling process in
the analytical model.  In that model the present-day intracluster medium is spherically
symmetric with a positive entropy gradient, by definition, whereas thermal instabilities
in the simulations that lead to a more heterogenous entropy pattern at each radius, may
be at the root of the negative temperature gradient.  Perhaps the observations are
telling us that a stabilizing influence like conduction erases small-scale thermal
instabilities without shutting off global cooling.

\subsubsection{Role of Feedback}
\label{sec:sc-feedback}

The primary role of feedback is to regulate how many baryons condense into stars
and cold gas clouds.  As mentioned in the discussion of cooling, strong feedback 
does not have a large effect on the $\Lx$-$\Tlum$ relation, aside from a slight
decrease in the luminosity of low-temperature clusters, as long as it is strong enough
to shut off cooling in the gas parcels that it affects.  However, moderate feedback
that heats gas to $\lesssim 100 \, \keV \, {\rm cm^2}$ can boost $\Lx$ because it does 
not allow the core gas to cool but rather maintains it in an entropy state that allows
it to radiate considerable thermal energy \citet{ktt03}. 

Some preheating and feedback models adequately explain the scaling relations
without explicitly including cooling \citep[e.g.,][]{bbp99, tn01, bem01, bblp02}.  
In these models, the minimum entropy
level introduced by heating is typically a free parameter that is adjusted to give 
the best-fitting $\Lx$-$\Tlum$ relation.  The value of this best-fitting entropy
level turns out to be $100-400 \, {\rm \keV \, cm^2}$, approximately corresponding
to the level of the cooling threshold.  This correspondence is consistent with the
idea that the amount of heating needed to explain the mass-observable relation 
is determined by the need to shut cooling, in which case cooling still sets the
entropy scale of similarity breaking, even when it is not explicitly included in
the model \cite{vbbb02}.

From the standpoint of the mass-observable relations, the most important effect
of feedback itself has to do with cluster richness.  In both the simulations and the 
analytical models, pure cooling leads to a larger fraction of condensed baryons
in cool clusters \citep{mtkpc01, dkw02, vbbb02, borg02, Borg04}, 
implying that these objects might have a higher star-to-baryon
ratio and therefore a lower mass-to-light ratio.  There are some observational
indications that the ratio of stellar luminosity to mass in clusters is a function of
mass \citep{lms03}, but not all such studies agree.  This issue will need to be settled
in order for optical richness measurements to deliver high-precision mass 
functions (\S~\ref{sec:condensation}).

\subsubsection{Role of Smoothing}
\label{sec:sc-smooth}

A full understanding of the $\Lx$-$\Tlum$ relation may involve feedback indirectly, 
through its smoothing effects on the intergalactic medium (\S~\ref{sec:entmerg}).  If the 
observed preservation of $K(r) \propto r^{1.1}$ entropy profiles is indeed due to
smoothing of the intergalactic medium followed by accretion onto clusters, then the 
present-day entropy profiles of clusters are evidence that galactic winds were widespread 
prior to the accretion of gas into today's clusters.  Rather than just affecting the core 
entropy of clusters, a modest amount of entropy produced  by early winds may have 
been amplified by smooth accretion, boosting the entire entropy profile by a
common factor determined by the cooling threshold \citep{vp03}.    
If that is indeed what happens, then it would explain the observed alteration
of cluster similarity such that $K(r/r_{200}) \propto \Tlum^{2/3}$ \citep{pa03, psf03}, 
which leads directly to the relation $\Lx \propto \Tlum^3 (T_{200} / \Tlum)^{1.5}$ for pure
bremsstrahlung emission, in agreement with the observations. 

\subsubsection{Predictions for Evolution}
\label{sec:sc-evol}

Preheating, the cooling threshold, and the altered similarity indicative of 
smoothing affect the time-dependent behavior of the $\Lx$-$\Tlum$ relation
differently, offering a way to gather further information about their
relative influence on cluster structure.   Defining
\begin{equation}
 \hat{L} =  \int_0^{r/r_{200}} \left( \frac {\rho_g} {200 f_b \rhocr} \right)^2 
                                          \hat{r}^2 d \hat{r} \; \; , 
\end{equation}
one can express the scaling of a cluster's integrated X-ray luminosity
as
\begin{eqnarray}
  \Lx & \propto & \Lambda_c(\Tlum) M_{200} \rhocr  \hat{L} \\
    ~ & \propto & \Tlum^2 \left( \frac {T_{200}} {\Tlum} \right)^{3/2} H(z)  \hat{L}
         \; \; ,       
\end{eqnarray}
where the first line assumes the cluster is approximately isothermal and the
the second line is an approximation that assumes pure bremsstrahlung emission.  
The self-similar case, 
\begin{equation}
   \Lx \propto \Tlum^2 H(z)
\end{equation}
is well known to be a poor description of the data because its power-law
slope at $z \approx 0$ is too shallow.  

The modified-entropy models of \citet{vbbb02} show that enforcing 
a minimum core entropy level $K_{\rm min}$ breaks self-similarity in such a way that 
$\hat{L} \propto K_{\rm min}^{-3/2} T_{200}^{3/2} H^{-2}$, if $K_{\rm min}$ is a 
significant fraction of the cluster's characteristic entropy $K_{200}$.  In the pure
preheating case, $K_{\rm min}$ is assumed to be independent of both cluster
mass and of redshift, leading to
\begin{equation}
   \Lx \propto \Tlum^{3.5} \left( \frac {T_{200}} {\Tlum} \right)^3 \frac {1} {H(z)} \; \; .
\end{equation}
In other words, pure preheating steepens the $\Lx$-$\Tlum$ a little more than
necessary but causes high-redshift clusters to be less luminous than one
would expect from their temperatures because the entropy floor $K_{\rm min}$
is a larger proportion of $K_{200}$ earlier in time.  This prediction appears
to conflict with recent observations indicating evolution in the opposite 
direction \citep[e.g.,][]{Vikh02}.  Tying the minimum entropy scale to the cooling threshold 
$K_c \propto \Tlum^{2/3} t^{2/3}$ helps to solve this problem because it leads to
\begin{equation}
   \Lx \propto \Tlum^{2.5} \left( \frac {T_{200}} {\Tlum} \right)^3 \frac {1} {H(z) t(z)} \; \; .
\end{equation}
In this case, a little bit of tilt in the $\Tlum/T_{200}$ relation, consistent with
observations (see Table~\ref{tab:mtnorm}), is needed to sufficiently steepen the $\Lx$-$\Tlum$
relation, and the sense of the evolution agrees with observations.  
In a $\Lambda$CDM universe, the redshift dependence 
of the luminosity normalization is $H^{-1}t^{-1} 
\sim (1+z)^{0.5} \sim H^{0.75}$ out to $z \sim 0.5$.
Altered similarity linked to the cooling threshold is in better agreement
with the slope but produces less evolution.  
Assuming density profiles that scale as 
$\rho_g(r/r_{200}) \propto (\Tlum / K_c)^{3/2}$ yields
\begin{equation}
   \Lx \propto \Tlum^{3} \left( \frac {T_{200}} {\Tlum} \right)^3 \frac {1} {H^3(z) t^2(z)} \; \; .
\end{equation}
The normalization of luminosity in this relation varies as $H^{-3} t^{-2} \sim (1+z)^{0.3}
\sim H^{0.5}$ to $z \sim 0.5$.

Observations of evolution in the luminosity-temperature relation are not
yet precise enough to distinguish between these latter two possibilities.
The usual procedure is to compare the $\Lx$-$\Tlum$ relation
measured in a significantly redshifted cluster sample to the relation 
measured at $z \approx 0$.  \citet{Vikh02} were the first to detect 
evolution, finding $\Lx(\Tlum) \propto (1+z)^{b_{LT}}$ with $b_{LT}
= 1.5 \pm 0.3$, assuming a $\Lambda$CDM cosmology.   These authors
compared the low-redshift sample of \citet{mark98} to a collection of 22 clusters
in the redshift range $0.4 < z < 0.8$.  \citet{Lumb03} found a similar 
amount of evolution, $b_{LT} = 1.52_{-0.27}^{+0.24}$, using a smaller
sample of eight clusters at $z \approx 0.4$, but not all studies find such 
strong evolution, which exceeds the predictions of the basic models 
outlined above.  For example, \citet{Ettori03} find $b_{LT} = 0.62 \pm 0.28$
for a sample of 28 clusters at $z > 0.4$ using the Markevitch (1998) sample as
the low-redshift baseline and $b_{LT} = 0.98 \pm 0.20$ relative to the
\citet{ae99} low-redshift baseline.  Furthermore, the strength
of the evolution found by \citet{Ettori03} becomes {\em smaller} for high-redshift
clusters, consistent with no evolution at all ($b_{LT} = 0.04 \pm 0.33$) 
when they include only their 16 clusters with $z > 0.6$ in the comparison
with the Markevitch sample.  Apparently, there are some systematic 
uncertainties in these evolution measurements that need to
be accounted for.

\subsection{Constraints on Baryon Condensation}
\label{sec:condensation}

The ultimate test for feedback models is that they must account for both the
proportion of condensed baryons to hot gas in clusters and any dependence
of that proportion on cluster mass.  In order to apply that test, we would like
to have firm numbers for the amount of condensed baryons in clusters, but
such measurements can be difficult.   Even if the amount of starlight were
perfectly measured, converting integrated starlight to stellar mass involves 
uncertain assumptions about both the star-formation history of a cluster 
and the distribution function of stellar masses at birth, a quantity known as the initial
mass function.  Any variation in the star-formation history or initial mass function 
with cluster mass can lead to spurious systematic trends in the cluster mass 
function inferred from cluster richness.

Baryons contained in cold clouds are even harder to constrain because 
gaseous matter in this form can be nearly invisible, if it is sufficiently cold
\citep{ffj94, ffj02}.  However, it seems unlikely that large amounts of baryons exist
in such a form, at least in rich clusters.  Adding the amount of baryons inferred
from starlight to the amount of hot gas observed in rich clusters accounts for 
nearly all the baryons expected from the global ratio of baryons to dark matter,
leaving little room left in the baryon budget for cold gas clouds.

The situation is less clear in lower-mass clusters and groups of galaxies,
in which the proportion of hot gas to dark matter is significantly smaller.
Summing the masses of stars and hot gas accounts for only about half
the expected number of baryons in some cases, yet there is no observational 
evidence for large quantities of cold gas \citep[e.g.,][]{Waugh02}.
Circumstantial evidence argues against there being large reservoirs
of cold baryons in groups.  Presumably, the rich clusters in which we now
see virtually all the baryons were hierarchically assembled from objects
like the baryon-poor groups of galaxies we observe today.  If large amounts
of baryons in their higher-redshift counterparts were locked away in some
cold, condensed form, then how were they released when these groups
of galaxies merged to form large clusters?

A more complete accounting of intracluster baryons, especially in low-mass
systems, is sorely needed in order test the various feedback models described
in \S~\ref{sec:feedback}.  The rest of this section summarizes some of the recent
work on constraining the amount of condensed intracluster baryons in the form of stars,
the prospects for measuring baryon condensation through the S-Z effect, 
and X-ray observations of nearby clusters that may help solve the puzzles 
surrounding condensation and feedback.

\subsubsection{Mass and Light in Clusters}
\label{sec:m2l}

Inferences of stellar mass from the observed starlight are generally based on a
mass-to-light ratio expressed in solar units.  That is, the mass-to-light ratio 
of the Sun in all wavebands equals unity.  Because young stellar populations
tend to emit large amounts of blue light that quickly dies out as the population
ages, most recent assessments of the stellar mass in clusters have concentrated 
on measurements of infrared starlight in the $K$ band at roughly 2 microns.
Observing starlight in this band minimizes the uncertainties owing to a cluster's 
star formation history.  The old stellar populations characteristic of elliptical
galaxies tend to have a $K$-band mass-to-light ratio $\Upsilon_K \approx 0.8 
\, h_{70}^{-1}$, and mass-to-light ratios in spiral and irregular galaxies can
be up to a factor of two smaller \citep{bd00}.  
For the mix of galaxies seen in clusters, \citet{lms03} estimate that the mean 
mass-to-light ratio ranges from $\Upsilon_K = 0.7 \, h_{70}^{-1}$ to $0.8 \, h_{70}^{-1}$ 
as cluster temperature climbs from 2~keV to 10~keV.  From this mass-to-light
ratio, they infer that the fraction of intracluster baryons in stellar form is 
$f_* \approx 0.1$ for rich clusters \citep[see also][]{bpbk01}.   Notice that
this value is about half that predicted by current simulations of cluster formation
including strong feedback, a discrepancy that could become even larger with
higher-resolution simulations \citep{Borg04}.

Many studies, but not all of them, suggest that the fraction of condensed baryons
in stars may be a function of cluster mass.  The ratio of $K$-band light to {\em total} 
cluster mass within $r_{500}$ found by \citet{lms03} is $\Upsilon_K  = 
(47 \pm 3) \,  h_{70} \, (M_{500} / 3 \times 10^{14} \, h_{70}^{-1} M_\odot )^{0.31}$, which
translates to a temperature dependence $\Upsilon_K \propto \Tlum^{0.5 \pm 0.1}$.
The ratio of stellar mass to total mass in this study therefore ranges from $\sim$2.2\%
at $10^{14} \, h_{70}^{-1} \, M_\odot$ to $\sim$1.2\% at $10^{15} \, h_{70}^{-1} , M_\odot$.
Similar trends with shallower slopes are seen at other wavelengths.  \citet{bc02}
find the ratio of total mass to starlight in the heart of the visible spectrum ($V$-band)
is $\Upsilon_V \propto \Tlum^{0.3 \pm 0.1}$. In blue light ($B$-band), 
\citet{Girardi02} find $\Upsilon_B \propto M^{0.25}$.  However, other studies
have found no significant dependence on mass.  According to \citet{Kochanek03} 
the $K$-band mass-to-light ratio inside $r_{200}$ scales as $\Upsilon_K
\propto M_{200}^{-0.10 \pm 0.09}$.

\subsubsection{Intergalactic Stars}
\label{sec:igstars}

Measurements of the total stellar luminosity in clusters generally focus on the light
from galaxies, but what about stars that are not in galaxies?  At least some
of a cluster's stars float unmoored in the spaces between a cluster's galaxies
\citep{ftv98}.  These stars are thought to have originated in galaxies but were 
later stripped from their homes by tidal forces during a close encounter 
with another galaxy.   Current observational limits, however, indicate that
no more than 10-20\% of a cluster's stars are outside of galaxies \citep{Durrell02}, 
implying that failing to account for intergalactic stars does not lead to large
errors in measured mass-to-light ratios. 

\subsubsection{Global S-Z Effect}
\label{sec:globalsz}
  
 If the baryons missing in low-mass clusters are not in condensed form, 
 then they must be in the form of hot gas beyond the regions detectable
 with X-ray telescopes.  If that is indeed the case, then the best way of
 finding them may be through the Sunyaev-Zeldovich effect.  Section
 \ref{sec:sze} showed that the integrated microwave distortion from a cluster
 scales with the electron temperature of the cluster and the overall mass
 in hot electrons.  If a significant proportion of baryons have condensed, 
 then the associated electrons are also locked away in cold clouds, where
 they don't contribute to the S-Z signal.
 
 Simulations of cluster formation that include cooling indicate how the
 mean value of the $y$-distortion owing to clusters depends on cooling
 and feedback processes.  Models by \citet{da_silva01} produce
 $y = 3.2 \times 10^{-6}$ in the non-radiative case, dropping to
 $y = 2.3 \times 10^{-6}$ in the case of radiative cooling without
 feedback.   The difference between the radiative case and non-radiative
 case is somewhat smaller when feedback is implemented.  \citet{whs02} find 
 $y = 2.5 \times 10^{-6}$ in the non-radiative case and $y = 2.1 \times 10^{-6}$
 when both cooling and feedback are turned on.  Testing for baryon
 condensation in this way may eventually be possible, but the mean
 value of the S-Z distortion is also very sensitive to other cosmological
 parameters, such as $\sigma_8$, which will have to be very well
 constrained before we can use the global $y$ parameter to test
 feedback models.

\subsubsection{Cooling Flows in Clusters}
\label{sec:cflows}

Cores of present-day clusters are among the best places in the universe
to observe the interplay between condensation and feedback.  Gas at 
the centers of many clusters can radiate an amount of energy equal to its
thermal energy in less than a billion years, yet the majority of that gas
is not condensing \citep[see][for a recent review]{dv04}.  Early interpretations
of clusters with central cooling times less than the age of the universe
suggested that the core gas should gradually condense and be replaced
by the surrounding material in an orderly flow of cooling gas \citep{fn77, cb77, mb78}.
The mass condensation rates inferred from X-ray imaging ranged as high 
as $\sim 10^2$ to $10^3 \, M_\odot \, {\rm yr}^{-1}$ implying that the cores of these
``cooling-flow'' clusters should contain $\gtrsim 10^{12} \, M_\odot$ in the form
of condensed baryons.  However, exhaustive searches for this mass
sink generally have not found stars forming at such a high rate \citep{om89, mo89},
nor have they found sufficiently large collections of cold baryonic clouds to
account for the deposited mass
\citep{bd94, mj94, odea94, opk98, vd95}.

Now X-ray spectroscopy itself is showing that condensation proceeds at 
a considerably slower rate, if it happens at all.  The central gas in clusters with
short cooling times appears to reach temperatures $\sim \Tlum / 2$, but very
little X-ray line emission is seen from gas at $\lesssim \Tlum / 3$ \citep{pet_etal_01,
pet_etal_03}.  Some sort of heating mechanism seems to be inhibiting condensation
below this temperature.  There are plenty of candidates for resupplying the
radiated heat energy---supernovae, outflows from active galactic nuclei, electron
thermal conduction, and turbulent mixing have all been suggested 
(see \S~\ref{sec:feedback})---but there is still no consensus on the relative importance of
these mechanisms.

A reduced amount of condensation still appears to be occurring.   For 
example, plenty of circumstantial evidence links short central cooling times with star
formation at the centers of clusters.  Objects whose central cooling time is
less than the age of the universe frequently contain emission-line nebulae
whose properties suggest that they are energized primarily by hot, young
stars \citep{jfn87, vd97}.  Nebulae like these are never seen in clusters
where the central cooling time is greater than the universe's age \citep{hcw85}. 
Also, objects with prominent nebulae tend to have abundant cool molecular 
hydrogen gas, the seed material for star formation \citep{Donahue00, Edge01,
ef03}.  Efforts to estimate the star formation rate from the ultraviolet light emanating from 
the centers of clusters indicate that it may be consistent with the current
upper limits on the condensation rate drawn from X-ray spectroscopy
\citep{mwm04}.

An understanding of what regulates condensation and star formation at the
centers of present-day clusters will help to solve more than just the overcooling 
problem of galaxy formation.  It is also relevant to an aspect of bright galaxies 
that remains difficult to understand.  The luminosity distribution function of galaxies
cuts off very sharply at the high-luminosity end, far more sharply than called for
in standard models of galaxy formation.  Extremely powerful feedback can produce 
a sharp cutoff, but the amount of energy input required seems to implicate active 
galactic nuclei as the primary feedback source \citep[e.g.,][]{bbflbc03, so04}.
Alternatively, thermal conduction might produce a sharp cutoff because its efficiency
rises so rapidly with temperature \citep{bbflbc03, fvm02}.  As the halo of a massive
galaxy grows and its characteristic temperature rises through a critical threshold 
$\sim 10^7$~K, conduction can strongly suppress further cooling and star formation, 
if it is not inhibited by magnetic fields.   Detailed studies of cluster cores will be needed 
to test these possibilities.  Early efforts are indicating that conduction might not be
efficient enough to prevent overcooling \citep{Dolag04, jsd04}.

\section{Concluding Remarks}
\label{sec:concl}

The next decade of research into cluster evolution promises to be
very exciting.  Large optical surveys like the Sloan Digital Sky Survey
are greatly increasing the number of well-studied clusters of galaxies
in the low-redshift universe.  Deep surveys looking for the Sunyaev-Zeldovich
effect will be finding thousands of clusters to distances well beyond a
redshift of $z =1$.  The {\em Chandra} and {\em XMM-Newton} X-ray
observatories are providing our most detailed look yet at the intracluster
medium, its thermodynamical state, and some of the feedback processes
that regulate condensation of intergalactic gas into galaxies and stars.
Also, dedicated X-ray satellite missions to survey a large fraction of the
sky for distant clusters are currently being planned.

Making the most of these opportunities will require cooperation between
observers in those different wavebands, and theoretical modeling that
closely links those cluster observables to cosmological parameters.
Optical and infrared followup of S-Z surveys will be critical in order to 
determine the redshifts of the cluster candidates.  X-ray followup of
a subset of the S-Z clusters will also be necessary to establish how
the thermodynamics of galaxy formation affects evolution of the
mass-observable relations in the microwave band.  Concentrated
efforts to observe a calibration set of clusters in all of these wavebands
will be very valuable in helping to establish how well the various
observables trace mass and the scatter in each of these observables
at a given mass.

If the $\Lambda$CDM concordance model is indeed a good description
of the overall architecture of the universe and its initial perturbation spectrum, 
then the parameters describing the cosmological context in which 
galaxy formation happens ought to be quite precisely established 
within this decade.  Studies of cluster evolution will be just one part 
of this overall effort, which also includes distance determinations 
to high-redshift supernovae, increasingly sensitive observations 
of the cosmic microwave background, and the mapping of large-scale
structure.   However, consistency between the dynamics of cluster evolution
and the geometry of the universe, as measured with supernova and
microwave observations, will stand as a particularly critical test of the
overall model.  With success, most of the remaining secrets about galaxy 
formation---other than what dark matter and dark energy actually 
are---will concern baryons and their complex cooling and feedback 
processes.  

Our understanding of what baryons do is rapidly progressing, thanks
in large part to large-scale cosmological simulations on massively 
parallel computers.  Clusters and their evolution place unique constraints
on those models because clusters are the only places in the universe 
where the majority of the baryons emit detectable radiation, revealing
their thermodynamic state and elemental abundances.  Galaxy formation 
has clearly left its mark in the intracluster medium, but we are just beginning 
to decipher what it has written there in the gases between the galaxies.
Perhaps in ten more years there will be as much optimism about understanding
the baryonic side of galaxy formation as there is now about the understanding
the darker side.

\section*{Acknowledgments}

Many colleagues have helped me to understand the issues covered in this
review, particularly M. Donahue, G. Bryan, A. Evrard, J. Mohr, E. Ellingson, 
S. Borgani, T. Ponman, M. Balogh, R. Bower, C. Lacey, S. Kay, M. Postman, 
H. Ebeling, A. Fabian, C. Scharf, P. Tozzi, A. Finoguenov, S. Allen, 
B. McNamara, R. Mushotzky, C. Sarazin, R. Wechsler, H. Hoekstra, and M. White.  
Thanks also to A. Vikhlinin, P. Henry, and C. Mullis for providing some of the
figures.  Parts of this article were written at the very hospitable Aspen Center 
for Physics.

\bibliographystyle{apsrmp}
\bibliography{clusters}

\end{document}